\documentclass[a4paper,12pt]{paper}
\usepackage[margin=2.5cm]{geometry}
\usepackage[usenames,dvipsnames]{xcolor}
\usepackage{booktabs}
\usepackage{apalike}

\usepackage{natbib}
\sectionfont{\large\sf\bfseries\color{black!70!blue}} 
\usepackage{subcaption}
\usepackage[colorlinks,allcolors=MidnightBlue]{hyperref}
\hypersetup{
    colorlinks=true,
    linkcolor=MidnightBlue,
    filecolor=magenta,      
    urlcolor=cyan, 
    citecolor=purple,
    }
    
\usepackage[utf8]{inputenc}
\usepackage{booktabs}
\usepackage{multirow}
\usepackage{array}
\usepackage[pdftex]{graphicx} 
\usepackage{enumitem}
\usepackage[raggedright]{titlesec}

\titleformat{\paragraph}[hang]{\normalfont\normalsize\bfseries}{\theparagraph}{1em}{}
\titlespacing*{\paragraph}{0pt}{1.25ex plus 0.1ex minus .0ex}{0.05em}

\setlist{noitemsep} 

\usepackage{enumitem}
\usepackage[raggedright]{titlesec}
\usepackage{blindtext}
\usepackage{url}

\usepackage{ragged2e}
\justifying

\usepackage{authblk}

\usepackage{booktabs}
\usepackage{multirow}
\usepackage{array}
\newcolumntype{L}{>{$}l<{$}} 

\RequirePackage{mathptmx} 

\RequirePackage[T1]{fontenc} \RequirePackage[tt=false, type1=true]{libertine} \RequirePackage[varqu]{zi4} \RequirePackage[libertine]{newtxmath} 

\title{User-Centered Course Reengineering: An Analytical Approach to Enhancing Reading Comprehension in Educational Content}
\author{Madjid Sadallah}
\date{}

\begin{document}
\begin{center}
    \Huge
    \textbf{User-Centered Course Reengineering}\\
    \vspace{0.5cm}
    \LARGE
    An Analytical Approach to Enhancing Reading Comprehension in Educational Content\\

    \vspace{1cm}
    \Huge Madjid Sadallah \\ \vspace{1cm}
    \large January 2022 \\  \vspace{1cm}

    \normalfont Research Report \\
    Derived from  PhD thesis of Madjid Sadallah  titled \\ ``\textit{Models and Tools for Usage-based e-Learning Document Reengineering}''\\
    Defended in April 2019 at University Abderrahmane Mira of Bejaia \\
    \vspace{.5cm}
    \Large \textit{Advisors} \\ \large Pr. Yannick Prié \\ \large Dr. Maredj Azze-Eddine \\ \large Dr. Benoît Encelle \\ \vspace{1cm}
\end{center}

\begin{abstract}
Delivering high-quality content is crucial for effective reading comprehension and successful learning. Ensuring educational materials are interpreted as intended by their authors is a persistent challenge, especially with the added complexity of multimedia and interactivity in the digital age. Authors must continuously revise their materials to meet learners' evolving needs. Detecting comprehension barriers and identifying actionable improvements within documents is complex, particularly in education where reading is fundamental. This study presents an analytical framework to help course designers enhance educational content to better support learning outcomes. Grounded in a robust theoretical foundation integrating learning analytics, reading comprehension, and content revision, our approach introduces usage-based document reengineering. This methodology adapts document content and structure based on insights from analyzing digital reading traces-interactions between readers and content. We define reading sessions to capture these interactions and develop indicators to detect comprehension challenges. Our framework enables authors to receive tailored content revision recommendations through an interactive dashboard, presenting actionable insights from reading activity. The proposed approach was implemented and evaluated using data from a European e-learning platform. Evaluations validate the framework's effectiveness, demonstrating its capacity to empower authors with data-driven insights for targeted revisions. The findings highlight the framework's ability to enhance educational content quality, making it more responsive to learners' needs. This research significantly contributes to learning analytics and content optimization, offering practical tools to improve educational outcomes and inform future developments in e-learning.
\end{abstract}

\newpage
\section{Introduction}

The evolution of digital technologies represents one of the most profound shifts in the production and dissemination of knowledge since the invention of the printing press. With the proliferation of digital content, readers increasingly prefer engaging with materials in electronic formats, driven by the convenience, accessibility, and interactive features these formats offer. These digital documents often integrate rich media, such as graphics, images, and videos, which not only enhance engagement but also enable nonlinear, interactive reading experiences. This shift has significantly impacted the realm of education, where traditional pedagogical methods are continuously evolving to incorporate information technologies both inside and outside the classroom. As a result, online and hybrid learning modalities have gained tremendous popularity, contributing to the rapid expansion of educational platforms and the surge in enrollments for online courses.

In this context, vast amounts of instructional resources are being published on a daily basis by a growing number of course creators, providing millions of learners with access to knowledge for quick reference or in-depth study. A key challenge for content creators, particularly in educational settings, is ensuring that their materials effectively support learners' reading, comprehension, and knowledge assimilation. While the technological capabilities of digital platforms have revolutionized the way content is produced and consumed, they have also introduced new complexities, such as the integration of multimedia elements, which can obscure or distort the transmission of the intended message. Furthermore, these challenges are amplified by the intrinsic difficulty of structuring complex ideas in a way that is easily digestible and engaging for a diverse audience, a task made all the more difficult by the inability of authors to anticipate the varied reactions of their readership.

As learning increasingly shifts towards online platforms, learners are afforded more autonomy in managing their educational experience. This autonomy places a greater onus on the learners to regulate their own learning processes, which can result in both positive and negative outcomes. Research has shown that many learners, especially in self-directed or online environments, face significant challenges when trying to navigate through materials without sufficient support. This lack of guidance can lead to feelings of disorientation, frustration, and confusion, underscoring the need for instructors and course creators to play an active role in designing content that not only conveys information but also facilitates comprehension and cognitive engagement.

\textit{Learning analytics}, a field that employs data-driven methods to analyze learner behaviors and interactions with educational content, offers promising solutions to these challenges. By capturing detailed traces of learners' engagement with materials, learning analytics allows course creators to gain insights into how their content is being consumed, identify areas where learners struggle, and implement targeted improvements. Despite the potential of these methods, course creators often face a significant barrier: the technical complexity of analyzing large amounts of interaction data. Additionally, it is often difficult for non-expert authors to interpret these insights and translate them into actionable content revisions.

This research hypothesizes that by applying analytical methods to learners' reading traces, it is possible to uncover underlying comprehension difficulties and identify effective ways to improve instructional materials. One of the most pressing challenges in this area is the transformation of raw data into meaningful insights that can be readily understood and applied by course designers, who may not have technical expertise. The primary goal of this research is therefore to leverage learning analytics to identify comprehension issues based on learners' interactions with course materials and to assist course authors in making data-driven improvements to their content. 

In this context, this research will address the following research questions:

\begin{description}
    \item[\textbf{RQ1:}] \textit{What is the general conceptual framework to help authors enhance their courses and address learners' comprehension difficulties? }  The objective here is to define a clear methodology for analyzing reading patterns to identify comprehension issues. This will support authors in recognizing where learners might face difficulties and offer strategies for revising their content accordingly.
    \item[\textbf{RQ2:}] \textit{What are the comprehension issues learners face? } The aim is to identify the various document-inherent factors that influence how learners comprehend the material. This includes understanding which parts of the course materials are more challenging for learners, such as complex sentences, jargon, or multimedia integration, and the types of issues these factors cause during reading.
    \item[\textbf{RQ3:}] \textit{What corrective measures can be suggested to authors to address these issues?}  This question aims to provide actionable strategies for authors. Based on the identified comprehension issues, the research will propose specific ways that course materials can be improved, including simplifying complex text, adjusting the layout, or incorporating more effective multimedia elements.
    \item[\textbf{RQ4:}] \textit{How can comprehension issues be detected, and how can appropriate corrective measures be recommended?}  The goal is to develop an analytical approach that allows for the automatic detection of comprehension issues from learner behaviors. This will involve modeling reading activities and analyzing interaction traces to pinpoint areas where learners struggle. Using these insights, the research will propose a framework that offers tailored corrective actions for each identified issue.
    \item[\textbf{RQ5:}] \textit{What types of systems and tools should be developed to support authors in improving their content?} This question focuses on the development of tools that would assist authors in analyzing learner behavior and modifying their materials based on the insights gathered. The research will propose functional and design requirements for a system that effectively delivers these insights and allows for real-time content improvements. This could involve a user-friendly dashboard for course creators to interact with the data and receive actionable feedback.
\end{description}

In addressing these research questions, this study aims to provide a comprehensive framework for the reengineering of educational content based on learner interaction data. By linking the analysis of learners' reading behaviors with actionable content modifications, this work contributes to the development of a methodology that allows course authors to create more effective, learner-centered materials \citep{sadallah2019phd}.

The proposed methodology integrates learning analytics with a user-centered approach to content design, where the feedback loop between learners' interactions and content revisions is continuous and dynamic. The core of the framework lies in the development of an analytical tool, \textit{CoReaDa} (Course Reading Dashboard), that processes reading traces and suggests targeted revisions based on identified comprehension issues. Through this tool, authors are empowered to iteratively refine their materials, ensuring they align with learners' needs and cognitive processes.

Ultimately, this research offers a novel approach to educational content improvement, one that is data-driven, adaptable, and grounded in the real-world behaviors of learners. By equipping course authors with actionable insights into how their content is consumed, it is hoped that this work will lead to more engaging, comprehensible, and effective educational resources.

\section{Background and related research}

\subsection{Engineering and reengineering educational digital documents}
Modern technology is transforming education and is profoundly reshaping how teachers offer their courses and how students learn from them.
In this chapter, we begin by discussing some of the major paradigm shifts in educational practice that have followed the rise of the new digital age. Reading is one of the most important learning activities that have been significantly influenced by technology. We therefore review the impact of this transition to digital reading on learning behavior, reading performance and learning outcomes. One measure of learners' reading performance is their level of understanding, and an effective strategy to improve this level is to provide good quality contents. To fit learners' need, these contents need to be maintained through updates and revisions. This chapter thus concludes with a review of the requirements and challenges related to identifying and fulfilling learners' revision needs.

\subsubsection{Learning in the Digital Age}
\label{sec:doceng_learn}

Learning, at its core, involves the process of acquiring or modifying knowledge, skills, behaviors, values, or preferences \citep{Gross2015}. In today's interconnected world, learning happens in a wide array of contexts—formal, non-formal, and informal—which shape the methods and environments in which knowledge is transferred. Formal learning is structured, often occurring within the confines of educational institutions or workplaces where learning outcomes, schedules, and assessments are pre-defined. This form of learning typically leads to certification and credentials, affirming the learner's mastery of the subject.

Non-formal learning, while organized, occurs outside of formal educational settings. It includes organized programs or experiences that are not typically bound by rigid curricula or assessment methods, such as workshops, training sessions, or community-based educational initiatives. These learning activities are intentionally designed for educational purposes, but unlike formal learning, they may not always result in formal qualifications. While non-formal learning can be validated in certain contexts, it tends to be more flexible and adaptable, often tailored to specific needs or communities.

Informal learning, by far the most prevalent form of learning, constitutes the everyday, unplanned acquisition of knowledge and skills from daily experiences—such as those encountered at work, in the family, or through leisure activities. Over 80\% of human learning occurs in informal contexts, which \citep{Cross2011} emphasize as being crucial for personal development and adaptability in an ever-changing world. Informal learning is self-directed, typically unstructured, and does not involve formal assessment. It often occurs in non-educational contexts, yet it remains deeply impactful for lifelong learning.

In the digital age, the rapid expansion of digital technologies has profoundly transformed the landscape of education, particularly through the proliferation of online learning platforms, virtual classrooms, and e-learning technologies. E-learning, which emerged in its current form in the late 1990s, refers to the use of digital technologies to facilitate education, whether through online platforms, digital content, or interactive learning environments \citep{Gros2016}. This form of learning has not only expanded access to education but has also redefined traditional concepts of teaching and learning, making it more flexible, personalized, and interactive.

Distance education has a long-standing tradition, with its origins dating back to the 19th century through correspondence courses. In the digital era, distance learning has been redefined with the advent of the Internet and mobile technologies, allowing for more dynamic and engaging learning experiences. The core of distance education, as outlined by \citep{Kaplan2016}, involves providing learning opportunities to individuals who are geographically or temporally separated from instructors, often using mediums such as video, text, and real-time interactions. While distance education traditionally relied on one-way communication, contemporary e-learning technologies foster greater interaction and collaboration through synchronous and asynchronous methods.

The rapid evolution of e-learning technologies has accelerated the trend of self-directed learning, empowering learners to take greater control over their educational paths. According to \citet{Knowles1975}, self-directed learning involves learners actively participating in the planning, execution, and assessment of their learning experiences, which has become a hallmark of digital education. Today, platforms like MOOCs (Massive Open Online Courses) exemplify this shift, offering learners unprecedented access to high-quality educational content. MOOCs, first developed in 2008, exemplify the democratization of education by providing free or low-cost learning opportunities from prestigious institutions. These platforms have been embraced globally, breaking down barriers of cost, location, and institutional access to education \citep{Cormier2010}.

At the same time, the concept of “blended learning” has gained significant traction. Blended learning models combine traditional face-to-face education with digital elements, creating a hybrid model that takes advantage of both in-person interaction and the flexibility of online tools \citep{graham2009blended}. This approach aims to leverage the strengths of both delivery methods—while digital learning provides convenience and accessibility, face-to-face interactions foster engagement and direct feedback. Blended learning is increasingly seen as an effective strategy for higher education institutions to engage diverse learner populations.

A key development in the digital learning landscape has been the growth of Learning Management Systems (LMS) and Virtual Learning Environments (VLE). These platforms support both administrators and learners by organizing and tracking educational content, facilitating communication, and enabling formative assessments. The growth of LMS has been particularly important in corporate training, vocational education, and higher education, where they provide an integrated space for learners to access materials, track progress, and collaborate with peers \citep{Guy2009}. The growing sophistication of these platforms allows for highly personalized learning experiences, adapting the content and pace to the needs of the individual learner. Advances in artificial intelligence and machine learning are pushing LMS platforms to provide increasingly tailored experiences, leveraging learner data to offer personalized content and feedback in real-time.

The role of educational technology (EdTech) continues to expand, with digital tools facilitating everything from virtual classrooms to augmented reality (AR) and artificial intelligence (AI)-powered tutoring systems. AI is playing a transformative role in education by enabling adaptive learning technologies that adjust the difficulty and pace of instruction based on learner progress. This personalized approach has been shown to improve learner engagement and retention, particularly in STEM fields \citep{xu2022application}. Furthermore, the use of gamification and immersive technologies like AR and VR in education is reshaping how students interact with content, allowing for experiential learning that was previously not possible in traditional educational settings.

The rise of digital education also aligns with the growing trend of lifelong learning. As the global economy increasingly demands flexibility, innovation, and continuous skills development, lifelong learning has become a necessity for workers to stay relevant and adaptable in their careers. E-learning platforms and microcredentialing systems are enabling individuals to acquire new skills and knowledge at their own pace, empowering them to make career shifts and advance in their professions without the need to return to traditional institutions. In this context, learning is no longer confined to childhood or young adulthood, but is a continuous process that spans throughout a person’s life.

Overall, the digital revolution has irrevocably altered the education landscape, with technology enabling greater flexibility, accessibility, and personalization. As e-learning, MOOCs, LMS, and other digital platforms continue to evolve, the future of learning will increasingly be defined by the ability of learners to shape their own educational journeys, utilizing technology to support their diverse needs, preferences, and goals.

\subsubsection{Digital Documents and Their Usage in E-learning} 
\label{sec:doceng_doc}

The concept of a document extends beyond a mere piece of written evidence or a material object; historically, it has also been linked to the act of teaching and instruction. The Latin term \textit{doceo} means "teach," and \textit{documentum} is thus related to the act of teaching. The first serious reflection on the concept of a "document" came from \citep{Otlet1934}, who argued that objects such as sculptures and works of art could also be considered documents, as they serve as expressions of human thought. Extending this further, \citep{Briet1951} defined a document as "any physical or symbolic sign, preserved or recorded, intended to represent, reconstruct, or demonstrate a physical or conceptual phenomenon" \cite[see also][]{Briet1951}. She also noted that an antelope could be considered a document if it became an object of study or physical evidence of specific events, such as being captured and placed in a zoo.

In a broader sense, a document can be understood as any means capable of transmitting knowledge or information in a more or less durable form. Traditionally, this referred to records that carried written or graphical information. With the advent of digitization, this concept has evolved, with documents now encompassing a wide range of forms. For \citep{Buckland1997}, "whatever is displayed on the screen or printed out is a document," noting that even algorithms function as documents in a dynamic, non-physical sense, aligning with the trend to define documents in terms of function rather than physical format.

Digitization has transformed documents from simple text to dynamic representations of observed or expressed phenomena, with storage shifting from paper to electronic formats. This shift has enhanced the interactivity and richness of documents. According to \citet{Pedauque2006}, documents should be analyzed through three lenses: (1) as a \textit{form} (digital structure), (2) as a \textit{sign} (the content), and (3) as a \textit{medium} (a tool for communication). These dimensions help determine the "maturity" of a document from anthropological (legibility), intellectual (assimilation), and social (diffusion) perspectives \citep{Yahiaoui2011}.

A digital document can be understood as any digital composition created on a computer. Unlike physical documents, which have tangible, fixed characteristics, digital documents exist as strings of bits rendered through computer systems, lacking a physical reality \citep{Laha2010}. \citet{Levy2016} proposed a broad view, suggesting that a document is simply "a way to delegate the ability to speak to inanimate objects."

The emergence of digital documents also brought new forms of media and interaction. While physical documents are tangible and linear, digital documents are intangible, unlimited, and intertextual, offering interactive and flexible representations that can be modified in real-time. Paper-based documents are designed to be read from top to bottom, whereas digital documents can be represented differently across various platforms. Web-based documents, in particular, can change over time due to their ability to be updated and modified, providing new access means and a wider audience for content dissemination \citep{Thompson2005}.

The key difference between traditional paper documents and digital ones lies in the latter's potential for interactivity and the capacity for content updates. As \citep{Crystal2010} noted, web-based documents may not have a permanent form; each time a user accesses the document, it may appear differently due to these changes.

\paragraph{Hypertext, Hypermedia, and Multimedia}

The evolution of digital documents has introduced new ways of navigating and interacting with text, particularly through hypertext and hypermedia. The idea of hypertext emerged in 1945 when Vannevar Bush described a system called Memex in his article "As We May Think," which sought to organize human knowledge using microfilm technology \citep{Bush1945}. \citet{Nelson1965} later coined the term "hypertext" to describe a system that could interconnect written or pictorial material in complex ways, making it impossible to represent it conveniently on paper.

Hypertext enables users to navigate within and between documents in a non-linear fashion, using hyperlinks that establish connections between various pieces of content. This breaks the traditional sequential reading of texts, allowing readers more autonomy in their navigation. In hypertext, authors can only suggest a reading order through document structure and links, while readers are free to determine the order in which they read \citep{Nielsen1990}. This shifts the dynamic between authors and readers, blurring the distinction between the active author and the passive reader \citep{Van2001}.

Multimedia documents, which integrate various media types such as text, images, audio, and video, represent a more complex form of digital documents. These documents are interactive and organized both spatially and temporally. According to \citet{Jourdan1998}, multimedia documents differ from basic text-only documents due to their ability to organize information spatially and temporally and provide integrated navigation. \citet{Geurts2010} characterized multimedia documents by two main properties: (1) heterogeneous media types, including text, images, audio, and video, and (2) spatio-temporal dimensions, where media items are synchronized in meaningful ways.

Hypermedia extends hypertext by incorporating multimedia elements, linking not only text but also other media like images, audio, and video. This inclusion of temporal elements (e.g., synchronization of media content) enriches the hypertextual experience, offering a dynamic and immersive way of interacting with content. Hypermedia systems employ a node/link structure, with media types such as text, images, and videos acting as nodes that are connected through hyperlinks.

The integration of multimedia into hypermedia systems has enabled the creation of innovative tools for knowledge delivery. For example, hypervideos combine audiovisual content with data that is time-synchronized, providing additional interactive features and navigation options \citep{Aubert2008,Sadallah2011}. 

\paragraph{Document Structures in E-learning}

As the concept of a document has evolved, so too have the systems and tools used to manage and create them. One of the goals of document engineering is to design, develop, test, and maintain systems for producing electronic documents efficiently. The design of such documents requires understanding their composition, rendering, and storage, which necessitates the development of a suitable data model. This model defines the entities that make up a document, the relationships between them, and the rules for structuring the elements within the document's information space.

\citet{Volkel2007} proposed a generic data model where the document is seen as a knowledge artifact consisting of multiple layers, each layer representing a different aspect of the document's structure. This model helps to identify how documents are constructed from atomic objects and how these structures interact to form a cohesive whole. \citet{Christophides1998} further outlined a four-level description of digital documents, including semantic, logical, physical, and presentation layers. Each of these levels is characterized by a distinct data structure, which together define the document’s overall organization, meaning, and appearance.

The complexity of modern digital documents, combined with the increasing variety of media and interactivity they incorporate, presents new challenges in designing and managing documents effectively in e-learning environments. As a result, document engineering plays a critical role in the development of systems that can handle such complexity while maintaining usability and efficiency.
\subsubsection{Digital Reading and Learning}

Reading, whether in digital or paper format, involves a complex interaction between the reader and the medium. According to \citet{Grabe2009}, reading encompasses rapid, efficient, interactive, strategic, flexible, evaluative, and purposeful processes that engage both lower-level skills (such as recognizing and decoding words) and higher-level cognitive skills (such as making inferences and understanding the broader context) \citep{Alderson2000, Duran2013, McNamara2009}. 

Digital reading, which involves interacting with content in digital formats, became prevalent in the early nineties \citep{Bawden2008}, and it has grown significantly in popularity, especially among younger generations \citep{Liu2005}. This mode of reading offers several advantages, such as interactivity, non-linearity, instant access to information, and multimedia content (text, images, audio, and video) \citep{Liu2005}. With digital documents, the process of cross-referencing is significantly enhanced, allowing readers to navigate from one resource to another seamlessly \citep{Adler1998}. However, digital reading also introduces challenges, such as potential distractions and a tendency toward shallow reading behaviors, including skimming and selective reading \citep{Liu2005, Chou2012}. Some studies also note that the non-linear nature of digital reading may hinder deep comprehension, as it encourages more scanning and browsing rather than sustained, in-depth reading \citep{Liu2005, Mangen2013}.

Digital reading can also impact learning, particularly as educational materials increasingly shift from print to digital formats \citep{Coiro2012, Walsh2016}. While traditional print materials remain integral to learning, digital resources offer significant benefits, such as easy access to educational content anywhere and anytime \citep{Staiger2012}, and better facilitation of data-driven learning \citep{Stoop2013p2}. These digital documents, optimized for hypertext and multimedia, engage learners in more active and interactive reading behaviors, which have been shown to enhance learning outcomes \citep{Rockinson2013, Adler2014}. 

Active reading, a key feature of digital learning, involves engagement with the content through strategies like annotating, highlighting, summarizing, and cross-referencing. These strategies, which foster deeper understanding, are more easily implemented in digital formats where annotations and multimedia elements can be seamlessly integrated \citep{Rockinson2013, Mayer2002}. \citet{Sadallah2014} notes that digital annotations can significantly improve the learning process by allowing students to organize and contextualize their thoughts and reflections in relation to the material. This active engagement with the text leads to stronger memory traces and a more purposeful understanding of the material, making digital reading environments conducive to learning \citep{Ortlieb2014}.

The preference between paper and digital reading modes varies among learners, depending on context and purpose \citep{Liu2011}. While some students prefer print documents for in-depth study, digital formats are favored for their interactivity and convenience, allowing for easy access to a wide range of resources in one place \citep{Stoop2013p1, Levine2015}. Despite a growing preference for digital documents, particularly for tasks requiring integration of multimedia or interactive features, print remains favored for more complex or lengthy readings \citep{Mangen2013, Tuncer2014}. A study by \citep{Millar2015} revealed that 57.4\% of students preferred paper, citing factors like trustworthiness and the physical interaction with the material. In contrast, digital formats are seen as more convenient, offering greater portability and easier access to various forms of content \citep{Rockinson2013, Kurata2017}. 

Ultimately, while digital reading is becoming more widespread in educational contexts, traditional print-based media still hold strong appeal, particularly for tasks requiring deep, focused engagement. This dual preference reflects the unique advantages of both formats, depending on the learning or reading context.
\subsubsection{Comprehension in Reading for Learning}
\label{sec:doceng_comp}

Reading is the process of interpreting and giving meaning to written content, with comprehension often serving as a key measure of reading outcome \citep{Bulut2015}. Research typically evaluates reading success through efficiency (speed and accuracy) and effectiveness (level of comprehension) \citep{Oh2013}. While efficiency relates to the speed of reading and error detection, effectiveness focuses on comprehension, which is central to learning. Comprehension itself can be viewed at two levels: literal and inferential \citep{Mcnamara2012, Chen2014}. 

\textit{Literal comprehension} refers to the basic understanding derived from explicit knowledge in the text, often assessed with closed-ended questions. In contrast, \textit{inferential comprehension} requires integrating the text's explicit knowledge with the reader's prior knowledge, leading to deeper understanding, typically assessed using open-ended questions. While various methods have been proposed to measure comprehension, the number of correct answers on reading tests is commonly used as an indicator \citep{Dillon1992}.

\paragraph{Digital Reading and Its Impact on Comprehension}

The impact of digital reading on comprehension has been extensively studied, with researchers examining how the mode of reading—digital or paper—affects reading outcomes such as comprehension level, reading rate, and accuracy \citep{Margolin2013}. Early research focused more on the processes involved in digital reading (e.g., eye movements and navigation) rather than its outcomes. \citet{Dillon1992} found that while digital reading has some limitations, such as slower speed, these can be mitigated with appropriate reading strategies.

Subsequent studies (e.g., \cite{Farinosi2016, Porion2016}) found no significant difference in comprehension between digital and paper reading, both in educational and non-educational contexts. However, some researchers argue that the reading mode can indeed affect comprehension, particularly due to the unique characteristics of digital texts.

One advantage of digital texts is the ability to incorporate rich media, such as diagrams, animations, and hyperlinks, which can enhance understanding and retention \citep{Green2010, Duran2013}. Multimedia content provides multiple channels for information encoding, which can improve recall and engagement \citep{Ortlieb2014}. Additionally, multimedia can be especially helpful for struggling readers, who may benefit from visual aids to support word decoding and text comprehension \citep{Puchalski1992}. Overall, interactive, multimedia-rich digital texts are generally perceived to improve comprehension through active engagement with the material \citep{Ortlieb2014}.

However, despite these advantages, several studies \citep{Delgado2018, Kong2018} suggest that digital reading may be less effective than paper reading, particularly in terms of comprehension. This is attributed to the cognitive demands of navigating hypertext and non-linear digital content, which can cause disorientation and cognitive overload \citep{Salmeron2006, Conklin1987}.

\paragraph{Disorientation and Cognitive Overload}

Disorientation occurs due to the non-linear nature of digital texts, where readers can easily lose their sense of direction within the content. This can lead to confusion, especially for readers with lower meta-cognitive skills. Unlike traditional printed texts, where the reading order is fixed, hypertext requires readers to navigate through links and understand the overall structure of the text, often using graphical overviews or prior knowledge \citep{Britton1994, Baccino2008}.

Cognitive overload arises from the multiple decisions readers must make when navigating hypertexts, such as choosing which links to follow or how to return to previous topics. This mental effort can detract from comprehension and retention, as too much cognitive load can impair memory \citep{Mangen2013, Mizrachi2014}. \citet{Dundar2017} suggested that digital texts consume more cognitive resources than printed texts, potentially reducing retention and comprehension. Additionally, digital reading can cause eye fatigue, further diminishing concentration and comprehension \citep{Jabr2013, Lin2015}.

In conclusion, while digital texts offer advantages such as multimedia and interactivity, they also introduce challenges such as disorientation and cognitive overload, which can hinder comprehension compared to traditional paper reading.

\paragraph{Comprehension and Document readability}

Comprehension, the ultimate goal of reading, is influenced by both the reader's strategic reading skills and the content's processing ease \citep{McNamara2009}. Early research \citep{Gray1935} identified two key factors: intrinsic reader characteristics (e.g., intellectual capacity, reading skills, attitudes, and goals) and the document's readability.

The reader's individual characteristics significantly impact reading outcomes. Studies have shown that prior knowledge \citep{Calisir2003}, working memory \citep{Lee2003}, and age \citep{Lin2003} affect reading and navigation. Hyperlinks, which introduce non-linearity in digital content, can cause distractions and shallow reading \citep{Akyel2009, Liu2005}. Therefore, navigational skills are crucial for effective on-screen reading. 

Coiro \citep{Coiro2007} outlined a recursive cycle in online reading, including planning, predicting, monitoring, and evaluating. Skilled readers integrate prior knowledge effectively to improve comprehension \citep{Haenggi1992}, whereas low-skilled readers often struggle with strategy use and lack relevant background knowledge \citep{Leon1995}. Furthermore, digital reading allows for non-linear navigation, but many still apply linear reading strategies learned from paper-based reading \citep{Murphy2003}, which may be less effective.

Document readability, which impacts comprehension, involves content layout, organization, linguistic style, and the theme \citep{Francois2012, Nelson2012, McNamara2014}. According to \citet{Dale1949}, readability is ``\textit{the sum total (including all the interactions) of all those elements within a given piece of printed material that affect the success a group of readers have with it. The success is the extent to which they understand it, read it at an optimal speed, and find it interesting.}''

Readability assessments aim to predict the ease of understanding a text, distinguishing it from \textit{legibility}, which measures letter recognition. Early definitions by \citet{Dale1949} and  \citet{Laughlin1969} framed readability as a combination of factors that influence comprehension, reading speed, and interest. 

Readability is commonly measured using formulas that predict text difficulty based on lexical sophistication and syntactic complexity, typically related to word and sentence length \citep{Crossley2017}. Over 200 readability formulas have been developed, such as the \textit{Flesch Reading Ease formula} \citep{Flesch1943}, \textit{Flesch-Kincaid} \citep{Kincaid1975}, \textit{Fog Index} \citep{Gunning1969}, and \textit{SMOG} \citep{Laughlin1969}, which all compute a readability score based on these parameters. 
While these formulas are helpful, they have been criticized for focusing on surface structure and not accounting for deep syntactic and semantic features, which limits their validity from a psycholinguistic perspective \citep{Bruce1981}. They also perform poorly in predicting readers' comprehension judgments \citep{Crossley2017}.

Advancements in Natural Language Processing (NLP) have led to more sophisticated approaches to readability assessment, incorporating linguistic features beyond word and sentence length. These include measures of syntactic complexity, word frequency, and text cohesion \citep{Crossley2007, Pitler2008}. Modern methods, often using machine learning, aim to predict readability more accurately than traditional formulas. For instance, NLP tools like Coh-Metrix \citep{Graesser2004} have outperformed traditional methods in predicting readability for both first and second-language learners. 
New approaches use classification or ranking methods to assess readability. Classification methods assign texts to specific readability classes, while ranking methods position documents on a scale of ease to difficulty \citep{Tanaka2010, Ma2012}.

\subsubsection{Document Revision}
Document quality depends on factors related to its design and writing, which determine the ease of reading and comprehension. High-quality documents are essential for effective communication. One way to maintain or improve this quality is through regular document revisions, which help refine both the structure and content.

Revision is a crucial step in the writing process that has a direct impact on authorship success. Traditionally, revision was considered a mere copy-editing task \citep{Faigley1981}. However, as the focus shifted towards a more process-oriented view of writing \citep{Fitzgerald1987}, revision became recognized as a core component of the writing process. It involves evaluating the text by reading, comprehending, and criticizing it to detect problems, followed by selecting and applying strategies to resolve them.

According to \citet{Fitzgerald1987}, revision can occur at any point in the writing process. It involves identifying discrepancies between the intended and instantiated text, deciding what should be changed, and how to make those changes. These changes may or may not affect the meaning of the text and can range from minor corrections to significant alterations. Additionally, revision can be a mental process that occurs before, during, or after writing.

\paragraph{The Revision Process}
The process of revision involves several steps, starting with problem detection. This is the process by which the author identifies differences between the produced text and the intended one \citep{Hayes1987}. Detecting problems is a prerequisite for making revisions and improving the document \citep{Patchan2015}. Once a problem is identified, the author must diagnose it, which involves understanding why it is problematic and how to address it \citep{Flower1986}. This diagnosis may vary in clarity, from a well-defined problem that leads to a specific solution, to a more vague sense of something being wrong.
After diagnosing the problem, the author selects a revision strategy. This requires decision-making and problem-solving abilities. The author must choose which issues to prioritize and what strategies to apply, especially when the problem is ill-defined or the best approach is unclear. The effectiveness of these strategies depends on the writer’s ability and experience.

From a cognitive perspective, revision is a complex mental process. It involves problem detection, diagnosis, and the selection of appropriate strategies to address issues in the text. These cognitive aspects are central to understanding how revision works and how writers can improve their documents over time. As noted by \citep{Fitzgerald1987}, revision is not simply about making changes to the text, but also about thinking critically about the text and learning through the revision process.

\paragraph{Taxonomy of Revision}
A notable contribution to understanding revision is the taxonomy by \citet{Faigley1981}, which categorizes revisions into two broad classes based on their impact on the document's meaning:
\begin{itemize}
	\item \textit{Surface changes}: These changes do not alter the meaning of the text. They include:
		\begin{itemize}
			\item \textit{Formal changes}: Copy-edits like spelling corrections.
			\item \textit{Meaning-preserving changes}: Modifications that don’t alter overall meaning, such as adding words.
		\end{itemize}
	\item \textit{Text-based changes}: These changes affect the document's meaning and include:
		\begin{itemize}
			\item \textit{Microstructure changes}: Minor meaning shifts that wouldn't change the text's overall summary.
			\item \textit{Macrostructure changes}: Significant changes that would alter the text's summary.
		\end{itemize}
\end{itemize}
These changes can involve additions, deletions, substitutions, and rearrangements. The taxonomy has been adopted and expanded by various researchers \citep{Cho2010, Early2014}, with finer categorizations such as \textit{paraphrase}, \textit{markup}, and \textit{information} changes \citep{Liu2011, Daxenberger2012}.

Additionally, revision strategies can be classified as \textit{Editing}, which focuses on correcting errors without altering the meaning, and \textit{Re-writing}, which involves transforming content, rearranging organization, or changing meaning \citep{Allal2004}.

\paragraph{Challenges in Document Revision}
Revision is cognitively and procedurally challenging \citep{Flower1986, Hayes2006}. It requires authors to reconsider ideas, organization, wording, and identify problems \citep{Hayes2006, Olmanson2016}. Expert writers typically approach revision as an opportunity to refine their ideas and improve expression \citep{Hayes2006, Macarthur2016}, while novice writers often find the process difficult, focusing more on surface-level changes \citep{Fitzgerald1987}.

The revision process becomes more efficient as authors gain experience, with experts detecting and solving global meaning issues more effectively than novices \citep{Faigley1981}. Inexperienced writers often struggle with problem detection, diagnosis, and strategy selection, making their revisions less effective. Higher-ability authors detect more problems, particularly those related to global meaning \citep{Fitzgerald1987, Hayes1987}. However, detecting and diagnosing problems is challenging, as writers often fail to perceive errors in their own work. Some authors automatically correct perceived mistakes in their minds, which can hinder problem detection \citep{Flower1986}. After detecting and diagnosing problems, authors select appropriate strategies to resolve them. Expert writers have a broader repository of strategies and are better at selecting effective solutions \citep{Hayes2000}. However, when a problem is ill-defined, writers may resort to generic strategies such as deletion or rewriting without fully understanding the underlying issue. In such cases, revision may not effectively address the problem.

\subsection{Usage analytics and knowledge discovery in educational documents}

E-learning platforms typically include logging features to monitor learner behavior, driving the fields of learning analytics (LA) and educational data mining (EDM). These fields aim to optimize learning experiences and outcomes using data collected from these platforms. This chapter reviews trends in tracking and interpreting learner interactions, explores the methods and objectives of LA and EDM, and highlights the development of learning analytics dashboards—key tools for visualizing data and aiding pedagogical decision-making.

\subsubsection{Tracing Reading Usages in E-learning} 
Monitoring learning activities is common in education, whether in traditional classrooms or online settings, and is essential for evaluating teaching effectiveness. It involves the collection of data on specific indicators to track progress and outcomes. Learning monitoring can be defined as an ongoing process that systematically collects data to provide feedback on learner progress \citep{Marriott2009}.

Learners can monitor their own activities, promoting self-regulation \citep{Tabuenca2014}, or be monitored by teachers or administrators \citep{Rodriguez2017}, which helps improve instructional methods. This process fosters both self-awareness and state-awareness, enabling reflective practices crucial for learning.
Reflection builds on awareness and encourages critical thinking about one's experiences. Self-reflection allows learners to gain insights, while state-reflection involves external evaluations by instructors to improve understanding and guide future learning.

 Traditional monitoring methods, such as assessments, grade analysis, and attendance tracking, are often ineffective for online learning due to their intrusiveness and limited adaptability. They provide limited data, which leads to slow interventions. Automated approaches, supported by data from learner behaviors and performance, offer more timely insights, enabling the identification of learning patterns and fostering greater self- and state-awareness \citep{Gavsevic2015}.

 User behavior in digital environments is tracked through digital traces, which are marks left by activities and interactions within the system \citep{Mathern2012}. These traces can provide insights into user actions, though they may not be intentionally created. A digital trace is defined as a set of digital imprints left during the interaction process, either voluntarily or not \citep{Champin2012b}.

 Digital traces are crucial for understanding the interaction between users and the environment. Their analysis, through automatic or semi-automatic tools, can provide valuable information for reflecting on learners’ behaviors and progress. Interaction indicators help track these behaviors and can be used for various purposes, such as diagnosing problems, assessing retention, and monitoring engagement \citep{Gwizdka2007, Edwards2017}.

Indicators can be classified based on the user perspective and the data source. Common indicators track individual learner activity, such as the number of pages viewed or forum posts made, and can be used for both self-monitoring and instructor evaluation. Group learning indicators, such as the number of participants in discussions, help analyze collective learning behaviors. Content-related indicators track how students interact with course materials, like the number of unique users per resource or the frequency of resource revisits \citep{Zhang2007, Martin2007}. Data sources for these indicators include learner interactions, academic profiles, and performance evaluations.

\subsubsection{Analysis of learning traces}
\label{sec:edmla_analysis}

Digital learning environments are able to record very detailed information regarding learners' behavior, resulting in a huge amount of data that is getting more and more voluminous. Their analysis and interpretation therefore require advanced data analysis techniques to be able to deliver the appropriate information.

The interdisciplinary field of \textit{Knowledge discovery and data mining} focuses on designing suitable methodologies to extract useful knowledge from data. It leverages research in various fields, including statistics, databases, pattern recognition, machine learning and data visualization to provide advanced business intelligence and web discovery solutions. The term of \textit{data mining} refers to the ``step in the overall process of knowledge discovery that consists of pre-processing, data mining, and post-processing'' \citep{Witten2016}. It is the process of identifying valid, novel, potentially useful, and ultimately understandable patterns in data \citep{Frawley1992, Fayyad1996}. Rather than attempting to test prior hypotheses, it searches for new and generalizable relationships and findings from large amounts of data \citep{Slater2017}.

The use of analytics in education is relatively new, compared to other science disciplines such as physics and biology. According to \citet{Baker2014}, it has grown in recent years for four primary reasons: (1) a substantial increase in data quantity, (2) improved data formats, (3) advances in computing, and (4) increased sophistication of tools available for analytics.

The application of knowledge discovery and analytics methods on learning traces is attracting increasing interest. Current trends are characterized by an increased technological use of features related to optimizing learning. All this enables the emergence of tools that rely on a data-based infrastructure to collect a wide variety of data without user intervention. As the combination of ``big data'' and computational progress emerges, efforts are focusing increasingly on improving the overall learning process, both within and outside the formal framework. The objective is to take advantage of the increasing use of online courses and of databases containing assessment results and behavioral records for the creation of large repositories of educational data. In order to harness this vast amount of data, the fields of \textit{Learning Analytics} (LA) and \textit{Educational Data Mining} (EDM) have emerged as a middle ground between learning sciences and data analysis. Their objective is to give education actors the appropriate means to improve understanding of teaching and learning and, more specifically, to adapt education more effectively to learners.

The application of knowledge discovery from data in education is mainly addressed within the field of  \textit{Educational Data Mining} (EDM). EDM bridges between two disciplines: education and computing sciences (in particular \textit{Data Mining} and \textit{Machine Learning}) \citep{Bak2017}. Current research integrates the interdisciplinary research fields of \textit{Statistics and Visualization}, \textit{Psychological Education}, \textit{Knowledge Discovery and Database}, \textit{Machine Learning}, \textit{Information Science}, and \textit{Artificial Intelligence} to various educational data sets so as to resolve educational issues~\citep{Romero2010}. According to \citet{Baker2010}, 	``\textit{Educational data mining is the area of scientific inquiry centered around the development of methods for making discoveries within the unique kinds of data that come from educational settings, and using those methods to better understand students and the settings which they learn in}''.

The main reason for the rapid development of EDM research in recent years is due to the availability of huge amounts of educational data, mainly generated by online education systems, and the urgency of converting this data into useful information and knowledge.

The purpose of trace data analytics is to ``help us to evaluate past actions and to estimate the potential of future actions, so to make better decisions and adopt more effective strategies as organizations or individuals'' \cite[p. 3]{Cooper2012}. In the case of LA, this purpose is oriented towards education. Many definitions are associated with the learning analytics. One earlier definition discussed by the community suggested that ``Learning analytics is the use of intelligent data, learner-produced data, and analysis models to discover information and social connections for predicting and advising people's learning''.

The most cited definition emerged from an open online course on learning and knowledge analytics and was adopted by the ``Society for Learning Analytics Research'' (SoLAR)\footnote{SoLAR (\url{http://www.solaresearch.org}) was created in summer of 2011 to develop and advance a research agenda in learning analytics, and to educate in the use of analytics in learning.} that defines this field as follows: ``\textit{the measurement, collection, analysis, and reporting of data about learners and their contexts for purposes of understanding and optimizing learning and the environment in which it occurs }'' \citep{Siemens2012}

\subsubsection{Learning analytics dashboards} 
Learning data can quickly become overwhelming, leading to cognitive overload. One solution is the use of visual representations, which transform non-visual data into recognizable formats, aiding analysis and decision-making \citep{Kosara2007}. Information visualization supports users in exploring complex datasets by leveraging human cognitive abilities to find patterns and relationships \citep{Slingsby2011}. Visual analytics, an extension of this field, integrates interactive visualizations with automated analysis techniques, enabling effective reasoning and decision-making based on large datasets \citep{Cook2005, Keim2008}. These methods are especially useful in domains dealing with massive data, such as business and security \citep{Brouns2015}.

In learning analytics, information visualization techniques help translate data into actionable insights \citep{Charleer2017}, aiming to connect visualizations to decision-making \citep{Duval2011}. Educational dashboards, also known as Learning Analytics Dashboards (LAD), are interactive tools that present data on learners' progress, behaviors, and learning contexts \citep{Khalil2015}. These dashboards allow for intuitive and real-time monitoring of key performance indicators (KPIs) \citep{Podgorelec2011}. A dashboard aggregates various indicators into visualizations that facilitate informed decision-making and provide a snapshot of current and historical trends \citep{Brouns2015, Few2013}.

Learning analytics dashboards focus on visualizing learners’ data, such as engagement and performance metrics, to support educators and learners \citep{Brouns2015, Ramos2015}. They often use visualizations like tables, charts, and alerts to enhance understanding and prompt timely interventions \citep{Schwendimann2017}. These tools can be applied in diverse educational settings, including online, blended, and face-to-face environments \citep{Verbert2013}. Their use has been shown to improve learning outcomes and increase student motivation \citep{Arnold2012, Wise2016}.

Effective dashboard design is guided by principles from cognitive science, such as situational awareness and human perception. A well-designed dashboard should display relevant information in a format that is easy to interpret, fitting within a single screen and supporting quick decision-making \citep{Few2013, Yoo2015}. Dashboards must prioritize important information, use effective visualization techniques, and align with users’ goals to enhance their cognitive processes \citep{Endsley2016}.

\citet{Verbert2014} classified dashboards into three categories: 1) dashboards for traditional face-to-face lectures, 2) dashboards for group work, and 3) dashboards for awareness and behavior change. Dashboards for traditional face-to-face lectures help instructors track students’ understanding and engagement during lectures. For instance, \textit{Backstage} visualizes Twitter activity to foster peer comparison \citep{Pohl2012}, while \textit{Classroom Salon} supports collaborative document editing \citep{Barr2012}. Other tools like \textit{Slice 2.0} integrate student activity with lecture content \citep{Fagen2012}.

Dashboards for face-to-face group work target real-time monitoring of group dynamics. For example, \textit{TinkerBoard} visualizes group activity during collaborative work on tabletops \citep{Do2012}, while \textit{Collaid} helps teachers monitor collaborative learning data \citep{Martinez2012}.

Dashboards for awareness, reflection, sense-making, and behavior change aim to foster reflection and behavioral change in learners by visualizing their learning progress and predicting outcomes. \textit{Course Signals}, for example, predicts student success based on grades and activity \citep{Arnold2012}. Similarly, \textit{Student Activity Meter} visualizes progress and time spent on tasks to improve learner engagement \citep{Govaerts2012}.

Existing dashboards face several limitations that hinder their effectiveness. One key issue is the lack of theoretically informed design. Many dashboards fail to align with the objectives of the study, and their visualizations can be complex and difficult for users to interpret quickly \citep{Duval2011}. A survey by \citep{Reimers2015} found that many dashboards have poor interface design and inadequate usability testing, with data selection often not meeting stakeholders’ needs due to their limited involvement in the design process \citep{Holstein2017}. Designers need to carefully select visual representations and interactions to avoid promoting ineffective instructional practices \citep{Gavsevic2015}. The success of dashboards depends on the extent to which stakeholders are involved in their co-design \citep{Holstein2017}.

Another challenge is the selection of input data and computed indicators. While dashboards use a variety of data sources, including trace analysis, there is limited attention to other potential sources such as direct feedback or the quality of student work. Additionally, there is little research comparing the most suitable indicators and visualizations for users with different levels of data literacy \citep{Schwendimann2017}. Current visualizations often resemble those used in other domains, like web analytics, rather than being specifically tailored to teaching and learning \citep{Schwendimann2017}.

Finally, the impact of dashboards on learning remains largely unstudied. Most existing dashboards are exploratory or not deployed in real educational settings, and few have been rigorously evaluated \citep{Charleer2014, Leony2012, Schwendimann2017}. While some prototypes have been assessed in terms of user acceptance, usefulness, and ease of use, there is little research on their actual impact on learning outcomes \citep{Arnold2012, Brusilovsky2011, Kerly2008, Nakahara2005}. To better understand how dashboards affect teaching and learning, it is crucial to analyze actual behavior patterns of teachers and students and investigate the relationship between visualizations and user responses \citep{Kim2015, Park2015}.

\subsection{Summary and Discussion of the Related Research} 

This section reviews the key themes in the related research, emphasizing the importance of high-quality course content for learners’ comprehension and exploring the challenges faced by course authors in maintaining and revising educational materials. A critical factor for supporting learners is the delivery of well-designed courses that facilitate reading and comprehension. As learners' engagement with content evolves, particularly in the context of digital reading, it is essential for authors to continuously revise their courses based on detailed insights into learners' needs. However, revisions are not a simple process. Authors often face significant challenges in recognizing and addressing the barriers to comprehension that may exist within their materials. One approach to overcoming these difficulties is to utilize learners’ feedback, which, if properly collected and analyzed, can serve as a valuable source of insight. However, rather than relying solely on direct interactions with learners, the monitoring of learner behavior can offer an unobtrusive and effective means of detecting comprehension issues. This requires defining and analyzing behavioral indicators, which can be done through the application of learning analytics. As we will discuss, learning dashboards, which display the results of such analyses, could be a powerful tool for providing authors with the data they need to make informed decisions about course revisions.

Technological advancements have drastically transformed the educational landscape, particularly in the realm of digital reading. While the shift to digital media offers many advantages, it has also introduced new challenges, such as cognitive overload and disorientation, which can negatively impact learners' comprehension. Though traditional paper-based reading has been extensively studied, there has been comparatively little research on digital reading practices, despite their increasing prevalence in modern education \citep{Kong2018}. The central measure of reading performance remains comprehension, as it reflects the learner's ability to construct a mental representation of the text \citep{Al2016}. The quality of this comprehension is influenced by various factors, including learners' abilities, backgrounds, and reading strategies, as well as the inherent quality of the course itself. Factors such as content layout, linguistic properties, and structural organization play a pivotal role in shaping how easily learners can process and understand material \citep{McNamara2009, Dascalu2014}. While digital courses are widely offered, there remains a significant gap in efforts to optimize e-learning content \citep{Ma2003}. Providing accessible and high-quality content that aligns with learners' capacities is crucial for enhancing both comprehension and overall learning outcomes \citep{Crossley2017}. Thus, authors must continuously assess and refine their courses to ensure they are meeting the diverse needs of learners.

The revision process of course content is an intricate task that requires deep reflection and critical analysis on the part of the author \citep{Flower1986}. Identifying comprehension barriers within a course and diagnosing their causes are essential steps in making meaningful revisions \citep{Witte2013}. However, this process is often complicated by authors' difficulties in recognizing issues with their own work, especially when they lack clear strategies for identifying problems or understanding the needs of their audience \citep{Philippakos2017}. Research has identified several common challenges that authors face during revision: difficulties in detecting issues, diagnosing them accurately, and selecting appropriate strategies for addressing them \citep{Patchan2015}. For instance, errors may be overlooked because authors tend to unconsciously correct them in their minds, and when problems are detected, diagnosing their cause can be difficult, especially when the issue is ill-defined. Additionally, selecting a resolution strategy may not always be straightforward.

In order to initiate revisions, authors must identify the parts of the course that present comprehension barriers for learners. One traditional method for identifying these barriers is readability assessment, which involves applying formulas to measure the complexity of a text. However, such measures have been criticized for their inability to predict learners’ actual comprehension accurately \citep{Crossley2017}. A more reliable approach would involve collecting feedback directly from learners regarding their comprehension difficulties. However, apart from a few studies that have utilized explicit learner feedback \citep{Pattanasri2012}, little research has focused on understanding how learners actually comprehend content, especially from their perspective \citep{Dascalu2014}. This gap is partly due to the challenges involved in monitoring comprehension, as it requires careful observation and an active learning environment.

Although feedback is recognized as a valuable tool for identifying problems and suggesting improvements \citep{Cho2011}, gathering such feedback can be difficult due to the distance—either temporal or spatial—between the authors and their learners \citep{Couzijn2005}. Despite these challenges, feedback is often a critical factor in successful revisions. Confronting feedback from genuine readers allows authors to better understand the mental models of comprehension processes and learner needs \citep{Schriver1992}. Thus, feedback, whether direct or indirect, plays a key role in the revision process, providing authors with the insights necessary to refine their materials.

As learning platforms have become more sophisticated, automated methods for capturing and analyzing learners' behaviors have emerged. These methods offer an unobtrusive, objective, and reliable way of assessing learner interactions with educational content \citep{Cocea2011}. By analyzing the traces left by learners during their interactions with the course material, authors can identify specific sections or aspects that may be hindering comprehension. This data provides a foundation for informed decision-making, allowing authors to make targeted improvements to course content. Such behavioral data provides a more accurate and timely picture of learner needs than traditional feedback methods, making it a powerful tool for course revision.

Analyzing learner behavior through learning analytics has become an essential practice for understanding how learners engage with content and identifying potential barriers to comprehension. These methods allow for the tracking of various learner actions, such as time spent on particular sections, frequency of revisits, and overall progression through the course. By mining these traces, it is possible to uncover patterns of learner behavior that may indicate areas of difficulty. Such analyses enable the identification of content sections that require revision, as well as the development of personalized interventions that target specific learner needs. Learning analytics can also provide a broader picture of course effectiveness, offering insights into how learners as a whole engage with the material \citep{Dumais2014}. 

While the use of trace data to monitor learner behavior holds great promise, a key challenge remains in defining appropriate indicators to assess learning progress effectively. These indicators are typically derived from raw data and may be calculated at the course level (e.g., percentage of content read) or at the individual section level (e.g., popularity of specific parts based on visits or revisits). However, relying solely on request-based data has limitations, as requesting a page does not necessarily indicate that the learner has fully engaged with the content. A more insightful approach involves analyzing session-level data, which captures the learner's navigation behavior over time and provides a clearer picture of their understanding \citep{Hauger2011, Mobasher2007}.

One of the most promising applications of learning analytics is the development of learning dashboards, which provide visual representations of the insights derived from analyzing learner data. These dashboards offer course authors a means to make informed decisions based on real-time data, and can also suggest specific actions for course improvement \citep{Verbert2013, Gutierrez2012}. Dashboards are especially valuable in the context of digital reading, where they can reveal comprehension issues that may not be apparent from the content alone. However, the effectiveness of these tools depends on how well they are designed and tailored to the needs of the users. Co-designing dashboards with course authors is essential to ensure that the indicators, features, and user interface are aligned with their goals and expertise \citep{Holstein2017}. Furthermore, the design must draw on established theories from learning sciences and educational psychology to ensure that the data presented is actionable and meaningful for the revision process.

Despite their potential, many course authors face challenges when it comes to utilizing learning dashboards effectively. Studies have shown that instructors often lack the technical skills and training necessary to fully leverage these tools \citep{Peerani2013}. For dashboards to be truly effective, they must not only present insights into learner behavior but also guide authors in taking appropriate action. This means integrating features that help authors interpret the data and make informed decisions about how to revise their content. In this way, dashboards can motivate authors to engage in a continuous process of course improvement, ultimately enhancing the learning experience for students.

In conclusion, the application of learning analytics to digital reading data presents a significant opportunity to assess and improve comprehension within e-learning environments. By leveraging insights from learner behavior, course authors can make data-driven decisions to revise their content in ways that better align with learners' needs. Learning dashboards that present these insights in an accessible and actionable format can help authors identify comprehension issues and suggest solutions, ultimately fostering an environment of continuous improvement. However, the success of such tools depends on their design, the training provided to authors, and their ability to integrate seamlessly into the revision process. As we will explore in the next section, there is a need for further development of methods and tools to assist course authors in this regard, enabling them to create more effective and engaging learning experiences.

\section{Document Reengineering Framework}
\subsection{Framework overview}
We define ``\textit{document re-engineering}'' based on usage as the process of modifying the contents and structures of documents based on the analysis of the readers' usage, as recorded in their logs on the reading platform \citep{Sadallah2013}.
\begin{figure}[h]
	\centering
	\includegraphics[width=.75\linewidth]{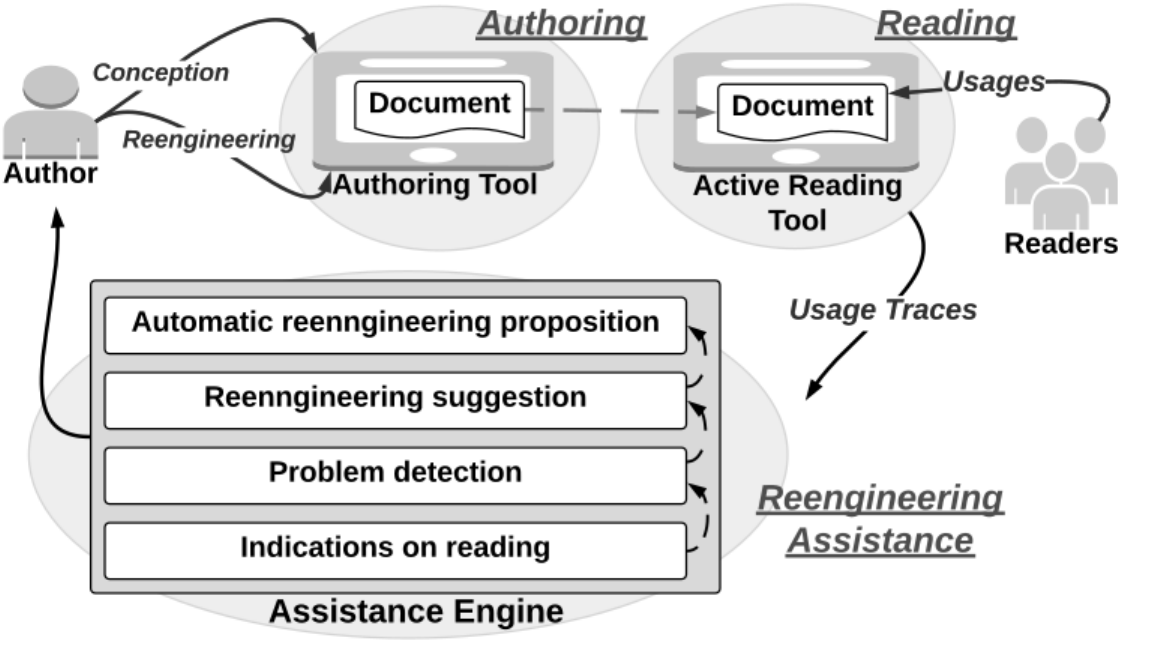}
	\caption{Overview of the Usage-Based Re-engineering Framework}
	\label{fig:overview}
\end{figure}

We propose the document re-engineering framework presented in Figure \ref{fig:overview}. By instrumenting an active reading tool, usage data is collected and analyzed to evaluate the modifications that need to be applied to the document. This framework defines three components: (1) the authoring tool, (2) the document reading tool (\textit{active reading tool}), which allows intercepting and recording actions (\textit{usage traces}) through a \textit{Source Collector}; and (3) the re-engineering assistance tool (\textit{assistance engine}), which receives the collected data for processing, analysis, and calculation of various \textit{reading indicators} to characterize the interaction of readers. Four main levels of assistance to authors are identified, each level building on the data from the previous one. 
\begin{itemize}
	\item \textit{Level 0: Reading Indicators.} The assistance engine can calculate and present the author with indications on how the document has been read. \textit{Ex.}: percentage of readers who followed a given link.
	\item \textit{Level 1: Problem Detection.} Based on the previous level, the assistance engine can detect problems in the reading process without suggesting ways to resolve them. \textit{Ex.}: if a video sequence is never watched past its first few seconds, the engine will signal this as an unexpected behavior to the author.
	\item \textit{Level 2: Re-engineering Suggestions.} At this level, the system not only detects problems but also provides suggestions. However, it is not capable of making the proposed modifications on its own. \textit{Ex.}: if a previous chapter is frequently revisited, the system might suggest including a reminder of the main concepts already covered.
	\item \textit{Level 3: Automatic Re-engineering Proposal.} At this level, the engine can detect problems and resolve them automatically. Consequently, a re-engineering proposal can be presented to the author for review and validation. \textit{Ex.}: if several zooms are performed on a specific part of the document, the system may automatically resize or adjust its font size.
\end{itemize}
The author can be assisted at all four levels of re-engineering. They can choose to consider the usage traces from a single reader, a given group of readers, or all readers. The final result is a new version of the document, which can, in turn, be subject to further improvements.

\subsection{Document Structures and Comprehension Issues}
To identify comprehension issues that may result from the design of a document, we first propose a model of the document and its structures. For each of these structures, we identify the factors that can influence the learner's level of comprehension. Each factor is examined to identify the problems it may be related to.
\subsubsection{Document Model.}
\begin{figure}[h]
	\centering
	\includegraphics[width=.7\linewidth]{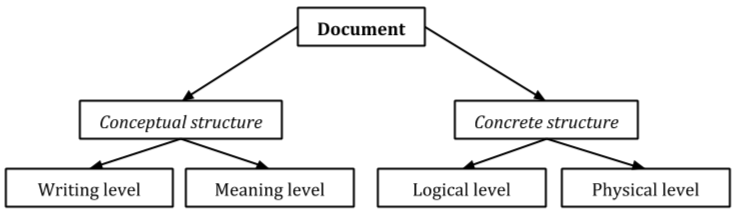}
	\caption{Document Structures}
	\label{fig:docstr}
\end{figure}

A digital document results from the translation of knowledge into a medium. It therefore has a concrete structure, which is rendered, and a conceptual structure that relates to knowledge, message, and meaning (see Figure \ref{fig:docstr}). The ease of understanding of the document depends on the conjunction of these two structures.
\paragraph{Surface (or Concrete) Structure.}
This describes the organization of the document and the relationships between its different elements. It is built on two complementary levels:
\begin{itemize}
	\item The \textit{logical level} defines the atomic units that compose the document, as well as the rules for composition. 
	\item The \textit{physical level} describes the spatial and temporal organization of the logical units on the rendering interface, as well as the navigation functions. 
\end{itemize}

A document results from the interleaving of composition units (content blocks), called \textit{document elements} (e.g., subsections, chapters), into other elements of the document that correspond to different levels of granularity (i.e., subsections into chapters, chapters into courses). Formally, we can express the organization of a document as follows:
\begin{verbatim}
			document = <doc_element+> 
			doc_element = <doc_element+> 
\end{verbatim}
\paragraph{Conceptual Structure.}
This reflects what is expressed by the author and how it is expressed, in terms of data, information, and/or knowledge, as well as their materialization.
This structure has a local level related to the expression of ideas, writing (the \textit{microstructure} in comprehension models), and a global level that represents the desired semantics (the \textit{macrostructure} in comprehension models): 
\begin{itemize}
	\item The \textit{writing level} is related to all the structures that are processed or described at the local level or short-term (graphics, words, phrases, clauses, sentences, and links between sentences). It represents the directly "expressed" structure of the document.
	\item The \textit{semantic level} (or \textit{signifier}) is a higher and more abstract level that organizes writing, interaction, and cognitive processing of different elements. It is related to the general meaning, the author's intent, and how they convey and translate their message.
\end{itemize}

Document composition corresponds to the author's encoding of the conceptual structure into a concrete structure. Through a decoding process, reading allows the reverse path (inferring meaning from writing). The gap between the decoded conceptual structure and the original one reflects the reader's level of comprehension. Therefore, comprehension support aims to minimize this gap as much as possible. This support can be provided upfront by examining the elements likely to cause decoding gaps.

\subsection{Taxonomy of Document Reengineering Actions}

To associate editing actions with various types of reading issues, we first developed a taxonomy of reengineering primitives (atomic actions) that reflect the most common editing operations used in the production of digital content.

\subsubsection{Modeling Reengineering}
Document reengineering involves applying a set of actions to one or more elements of a document (referred to as \textit{targets}) in order to produce a modified version. The targets of these actions depend on the document units that make up the document model. While most existing editing taxonomies focus on the sentence or paragraph level, our model operates at the \textit{document element} level. This allows for greater precision and provides more modification opportunities.

A reengineering action can be broken down into a series of elementary actions, which we call editing \textit{primitives}. The impact of a primitive is a specific aspect of its target: \emph{style}, \emph{structure}, \emph{content}, or \emph{links} of the document element. Each primitive has one of three possible effects on the target: \textit{addition}, \textit{modification}, or \textit{deletion}. The following formalism expresses this definition of reengineering:

\begin{verbatim}
reengineering = <action+> 
action = <primitive, target, dimension> 
dimension = <(style | structure | content | link)+> 
\end{verbatim}

 \begin{table}[h]
 	\centering
 	\footnotesize
 	\begin{tabular}{@{}lp{3cm}p{6cm}l@{}}
 		\toprule
 		\textbf{} & \textit{Addition}                               & \textit{Modification}                                                                  & \textit{Deletion}   \\ \midrule
 		Style   & \textit{Add style}                               & \textit{Alter style}                                                               & \textit{Delete style}   \\ \midrule
 		Structure & \textit{Add element}                               & \begin{tabular}[c]{@{}l@{}}\textit{Retitle} -- \textit{Move} \textit{Merge} -- \textit{Split}\end{tabular}                                   & \textit{Delete element}  \\ \midrule
 		Content  & \begin{tabular}[c]{@{}l@{}}\textit{Insert}\\ \textit{Explain}\\ \textit{Illustrate}\\ \textit{Remind}\end{tabular} & \begin{tabular}[c]{@{}l@{}}\textit{Organize} -- \textit{Summarize} --  \textit{Extend}\\  \textit{Deepen} -- \textit{Reformulate} -- \textit{Simplify}\\ \textit{Correct} -- \textit{Update} -- \textit{Translate} \end{tabular} & \textit{Delete content}  \\ \midrule
 		Links   & \textit{Add ref./link} & \textit{Modify ref./link} & \textit{Delete ref./link} \\ \bottomrule
 	\end{tabular}
 	\caption{Reengineering primitives taxonomy}
 	\label{tab:actions}
 \end{table}
\subsubsection{Reengineering Primitives}
By specializing the effects of reengineering according to different dimensions, we define four classes of primitives: \emph{restyling}, \emph{restructuring}, \emph{rewriting}, and \emph{linking} (see Table \ref{tab:actions}). Formally, a primitive can be expressed as follows:

\begin{verbatim}
primitive = <type , effect> 
type = <restyling | restructuring | rewriting | linking> 
effect = <addition | modification | deletion> 
\end{verbatim}

\paragraph{Restyling Primitives}
This class alters the presentation of the target element on the user interface (e.g., for personalization or accessibility purposes). Since this research focuses on content rather than presentation, this class is not discussed in detail.

\paragraph{Restructuring Primitives}
This class primarily targets the \textit{logical structure}, which may have implications for both content and other concrete or conceptual structures.
\begin{itemize}
	\item The \textit{addition} primitive introduces a new level in the document structure by adding a new element with a title and content.
	\item \textit{Modification} involves changing an entry point by either relocating the element, merging it with another, splitting it into several parts, or simply rearranging it.
	\item The \textit{deletion} primitive removes an element, along with its content, sub-elements, and entry point, from the document.
\end{itemize}

\paragraph{Rewriting Primitives}
This class focuses on modifying the content itself. The actions within this class are inspired by Bloom's Taxonomy, which defines levels of cognitive objectives, each associated with specific action verbs. From the "comprehension" level onward, we have selected verbs that are relevant to our context.
\begin{itemize}
	\item \textit{Content addition} can involve inserting new content or updating existing content.
	\item \textit{Content modification} includes tasks such as organizing, summarizing, expanding, deepening, rephrasing, simplifying, correcting, updating, or translating the content.
	\item \textit{Content deletion} involves removing content or parts of content from the element.
\end{itemize}

\paragraph{Navigation Primitives}
These primitives modify the document's navigation structure by adding, modifying, or deleting links (external links) or references (internal links within the document).

\subsection{Issues Related to Document Structures and Corresponding Reengineering Actions}
A number of comprehension issues encountered by learners arise from the structure of the document itself, which is influenced by the various characteristics of its organizational components. The following section examines these structures and identifies the problems associated with them. Each problem is assigned a \texttt{code} to facilitate reference.

\subsubsection{Comprehension at the Surface Structure Level}
At the surface level, comprehension is influenced by factors related to both the logical and physical aspects of the document.

\paragraph{Logical Structure.}
The logical structure pertains to the document's organization and layout. Factors affecting this level include the definition of the document's elements and the order in which they are presented. Table \ref{tab:logissues} outlines significant problems that can arise from these factors and suggests reengineering actions that can be employed to address them.

\begin{table}[h]
	\centering
	\footnotesize
	\begin{tabular}{|l|l|l|l|}
		\hline
		\multicolumn{2}{|c|}{\em{Issue}}             & \multicolumn{2}{c|}{\em{Reengineering action}}                  \\ \hline
		\multicolumn{1}{|c|}{\textit{Code}} & \multicolumn{1}{c|}{\textit{title}} & \multicolumn{1}{c|}{\textit{Type}} & \multicolumn{1}{c|}{\textit{Primitives}}         \\ \hline
		
		\multicolumn{4}{|c|}{\textit{Selection of elements}}     \\ \hline
		$LL1$ & Unnecessary/bulky element   & \multirow{4}{*}{Restructuring} & \texttt{\textsc{Remove}}                 \\ \cline{1-2} \cline{4-4} 
		$LL2$ & Non suitable title      &           & \texttt{\textsc{Retitle}}                \\ \cline{1-2} \cline{4-4} 
		$LL3$ & Element to decompose   &      & \texttt{\textsc{Split}}\\ \cline{1-2} \cline{4-4} 
		$LL4$ & Element to combine with others  &          & \texttt{\textsc{Combine}} (with) \\ \hline
		\multicolumn{4}{|c|}{\textit{Document outline and elements sequence }} \\ \hline
		$LL5$ & Element not in its best position & \multirow{3}{*} {Restructuring}      & \texttt{\textsc{Move}} (to)        \\ \cline{1-2} \cline{4-4} 
		$LL6$ & Late position of the element   &       & \texttt{\textsc{Move}} (backward)             \\\cline{1-2} \cline{4-4} 
		$LL7$    & Early position of the element  &           & \texttt{\textsc{Move}} (forward)              \\ \hline
	\end{tabular}
	\caption{Problems and reengineering primitives associated with the logic level}
	\label{tab:logissues}
\end{table}

\paragraph{Physical Structure.}
The physical structure governs how the document is visually rendered. Any imbalance in the spatial definition (e.g., the size, placement, or arrangement of elements) and/or temporal definition (e.g., timing and synchronization) can negatively impact reading, thus hindering comprehension and the overall perception of the content. Additionally, since digital reading often involves interaction through hyperlinks, the navigational structure is equally critical. Therefore, comprehension challenges may stem from issues such as the positioning and synchronization of elements or the proper definition of navigational links. Table \ref{tab:physissues} enumerates key problems associated with these factors and suggests reengineering actions that could resolve them.

\begin{table}[h]
	\centering
	\small
	\begin{tabular}{|l|l|l|l|}
		\hline
		\multicolumn{2}{|c|}{\em{Issue}}             & \multicolumn{2}{c|}{\em{Reengineering action}}                  \\ \hline
		\multicolumn{1}{|c|}{\textit{Code}} & \multicolumn{1}{c|}{\textit{title}} & \multicolumn{1}{c|}{\textit{Type}} & \multicolumn{1}{c|}{\textit{Primitives}}         \\ \hline
		\multicolumn{4}{|c|}{\textit{Placement on the layout}}                                      \\ \hline
		PL1           & Bad location        & \multirow{2}{*}{Restyling}   & \texttt{\textsc{Modify}} (location) \\ \cline{1-2} \cline{4-4} 
		PL2           & Inadequate size       &            & \texttt{\textsc{Modify}} (size) \\ \hline
		\multicolumn{4}{|c|}{\textit{Timing and synchronization}}     \\ \hline
		PL3           & Inadequate temporal information   & \multirow{2}{*}{Restyling}   & \texttt{\textsc{Modify}} (timing)\\ \cline{1-2} \cline{4-4} 
		PL4           & Bad synchronization      &            & \texttt{\textsc{Modify}} (synchronization) \\ \hline
		\multicolumn{4}{|c|}{\textit{Linking and navigation}}                                                          \\ \hline
		PL5           & Inappropriate/useless link    & \multirow{3}{*} {Linking}        & \texttt{\textsc{Delete}} (link)\\ \cline{1-2} \cline{4-4}
		PL6           & Needed link missing      &         & \texttt{\textsc{Add}}(link) \\ \cline{1-2} \cline{4-4}
		PL7           & Broken link        &         & \texttt{\textsc{Modify}} or \textsc{\texttt{Delete}} (link) \\ \hline
	\end{tabular}
	\caption{Issues and reengineering primitives associated with the physical level}
	\label{tab:physissues}
\end{table}

\subsubsection{Comprehension at the Conceptual Structure Level}
At the conceptual level, comprehension involves extracting semantic meaning, transforming words into understanding, and creating a cognitive representation of the text both locally and globally. The efficiency of this process is influenced by the properties of two major levels: \textit{writing} (microstructure) and \textit{semantics} (macrostructure).

\paragraph{Writing Level (Microstructure).}
Microstructural analysis focuses on how meaning is conveyed by the author. Two primary factors influencing comprehension at this level are \textit{productivity}, which relates to the syntactical structure (e.g., the number of words or ideas), and \textit{complexity}, which refers to the sophistication of the writing style (e.g., average sentence length or clause density). These factors help identify the key issues at this level (see Table \ref{tab:wrissues}), along with the corresponding reengineering measures that may be used to resolve them.

\begin{table}[h]
	\centering
	\small
	\begin{tabular}{|p{5.5mm}|l|p{25mm}|p{6.5cm}|}
		\hline
		\multicolumn{2}{|c|}{\em{Issue}}                       & \multicolumn{2}{c|}{\em{Reengineering action}}                           \\ \hline
		\multicolumn{1}{|c|}{\textit{Code}} & \multicolumn{1}{c|}{\textit{title}}            & \multicolumn{1}{c|}{\textit{Type}} & \multicolumn{1}{c|}{\textit{Primitives}}                  \\ \hline
		\multicolumn{4}{|c|}{\textit{Productivity and readability}}                                                        \\ \hline
		WL1           & Language and lexical weakness              & \multirow{2}{*}{Rewriting}   & \texttt{\textsc{Reformulate}} and \textsc{\texttt{Correct}}           \\ \cline{1-2} \cline{4-4} 
		WL2           & Bad syntactic construction               &            & \texttt{\textsc{Reformulate}} and \textsc{\texttt{Correct}}           \\ \hline
		\multicolumn{4}{|c|}{\textit{Complexity}}                                                              \\ \hline
		\multirow{2}{*}{WL3}     & \multicolumn{1}{c|}{\multirow{2}{*}{Many new complex information}} & Rewriting         & \texttt{\textsc{Reformulate}}, \textsc{\texttt{Summarize}} and \textsc{\texttt{Clarify}} \\ \cline{3-4} 
		& \multicolumn{1}{c|}{}                & Restructuring        & \textsc{\texttt{Split}}                     \\ \hline
		WL4           & Complex construction                 & Rewriting         & \texttt{\textsc{Reformulate}} and \textsc{\texttt{Correct}}           \\ \hline
		\multirow{2}{*}{WL5}     & \multirow{2}{*}{Recall problems}              & Rewriting         & \texttt{\textsc{Add}} (reminders)                   \\ \cline{3-4} 
		&                       & Linking         & \textsc{\texttt{Add}} (links)                     \\ \hline
	\end{tabular}
	\caption{Issues and reengineering primitives associated with the writing level}
	\label{tab:wrissues}  
\end{table}

\paragraph{Semantic Level (Macrostructure).}
The semantic level corresponds to the macrostructure of the document, which relates directly to the meaning conveyed by the text. Quality at this level is evaluated according to textuality criteria, which include \textit{cohesion}, \textit{coherence}, \textit{intentionality}, \textit{acceptability}, \textit{informativity}, \textit{situationality}, and \textit{intertextuality}. Table \ref{tab:meanissues} outlines the problems associated with these semantic factors and presents the potential reengineering actions to address them.

\begin{table}[h]
	\centering
	\small
	\begin{tabular}{|p{7.5mm}|p{5cm}|l|p{7cm}|}
		\hline
		\multicolumn{2}{|c|}{\em{Issue}}                  & \multicolumn{2}{c|}{\em{Reengineering action}}                                                    \\ \hline
		\multicolumn{1}{|c|}{\textit{}} & \multicolumn{1}{c|}{\textit{title}}        & \multicolumn{1}{c|}{\textit{Type}} & \multicolumn{1}{c|}{\textit{Primitives}}                                           \\ \hline
		\multicolumn{4}{|c|}{\textit{Consistency}}                                                                                 \\ \hline
		\multirow{2}{*}{ML1}     & \multirow{2}{*}{Lack or loss of thematic unit}    & Rewriting         & \texttt{\textsc{Update}}, \texttt{\textsc{Correct}}  \\ \cline{3-4} & & Restructuring & \texttt{Move} or \textsc{\texttt{Delete}} \\ \hline
		ML2          & Contradictions              & Rewriting         & \texttt{\textsc{Update}} and \texttt{\textsc{Correct}}                                     \\ \hline
		\multirow{2}{*}{ML3}     & \multirow{2}{*}{Unclear semantic relationship}    & Rewriting         & \texttt{\textsc{Reformulate}} and \textsc{\texttt{Correct}}                                    \\ \cline{3-4} 
		&                  & Restructuring        & \textsc{\texttt{Delete}}                                              \\ \hline
		\multicolumn{4}{|c|}{\textit{Cohesion}}                                                                                  \\ \hline
		ML4          & Unclear connection between ideas        & Rewriting         & \texttt{\textsc{Reformulate}}, \textsc{\texttt{Organize}} , \textsc{\texttt{Explain}} and \texttt{\textsc{Extend}}                  \\ \hline
		\multirow{2}{*}{ML5}     & \multirow{2}{*}{Incoherent ideas}         & Rewriting         & \texttt{\textsc{Reformulate}}, \textsc{\texttt{Correct}}, \textsc{\texttt{Explain}} and \textsc{\texttt{Clarify}}                  \\ \cline{3-4} 
		&                  & Restructuring        & \textsc{\texttt{Move}} or \textsc{\texttt{Delete}}                                      \\ \hline
		\multicolumn{4}{|c|}{\textit{Intentionality and acceptability}} \\ \hline
		ML6 & Misunderstanding & Rewriting  & \textsc{\texttt{Reformulate}}, \textsc{\texttt{Explain}}, \textsc{\texttt{Correct}}, \texttt{\textsc{Clarify}}, \texttt{\textsc{illustrate}} and \textsc{\texttt{Deepen}} \\ \hline
		\multicolumn{4}{|c|}{\textit{Informativity}}                                                                                 \\ \hline
		\multirow{2}{*}{ML7}     & \multirow{2}{4cm}{Marginal or uninformative} & Rewriting         & \texttt{\textsc{Deepen}}, \texttt{\textsc{Add}}   \\ \cline{3-4} 
		& & Restructuring  & \texttt{\textsc{Merge}} or \texttt{\textsc{Delete}} \\ \hline
		\multirow{2}{*}{ML8}     & \multirow{2}{*}{Overwhelming}         & Restructuring        & \textsc{\texttt{Split}}  \\ \cline{3-4} 
		& & Rewriting         & \textsc{\texttt{Clarify}}, \textsc{\texttt{Explain}}, \textsc{\texttt{simplify}} and \textsc{\texttt{Summarize}}                   \\ \hline
		\multicolumn{4}{|c|}{\textit{Situationality}}                                                                                 \\ \hline
		ML9          & Inadequacy                & Restructuring        & \texttt{\textsc{Move}} or \texttt{\textsc{Delete}}                                      \\ \hline
		\multicolumn{4}{|c|}{\textit{Intertextuality}}                                                                                \\ \hline
		\multirow{2}{*}{ML10}    & \multirow{2}{*}{Prerequisites needed}       & Rewriting         & \texttt{\textsc{Add}}                                                \\ \cline{3-4} 
		&                  & Restructuring        & \texttt{\textsc{Move}} the element or \textsc{\texttt{Add}} links                                 \\ \hline
	\end{tabular}
	\caption{Issues and reengineering primitives associated with the semantic level}
	\label{tab:meanissues}
\end{table}

\subsection{Summary of the Reengineering Approach}
\begin{figure}[!h]
	\centering
	\includegraphics[width=\linewidth]{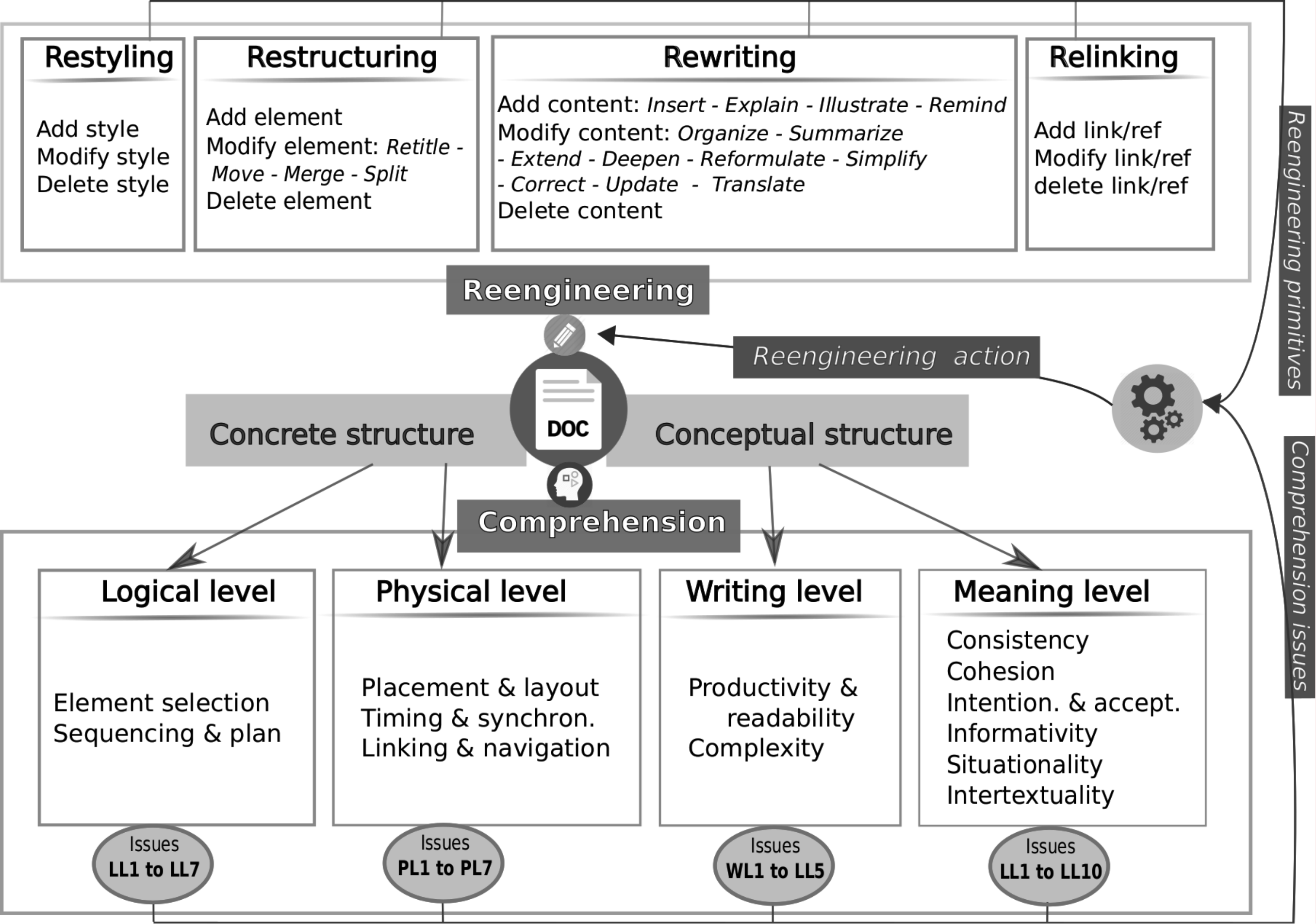}
	\caption{The Document Reengineering Approach \citep{sadallah2019phd}}
	\label{fig:chap5}
\end{figure}

Figure \ref{fig:chap5} provides a schematic overview of the usage-based reengineering approach. Initially, we defined the core structures of a document and identified the factors associated with them that affect comprehension. This analysis allowed us to outline the range of comprehension issues that readers may encounter, which are linked to the document's structural components. Furthermore, we cataloged the different editing actions an author can take on a document, which in turn led to the identification of various reengineering actions aimed at modifying the document’s structure or content. By linking each comprehension issue with a set of possible reengineering measures, we offer practical strategies for authors to enhance the effectiveness and comprehensibility of their documents. The framework we presented in this chapter provides a theoretical foundation for the subsequent chapter, where we propose a methodology for course revision based on learners' online reading traces. This approach aims to foster better comprehension and improve learning outcomes.

\section{An Analytical Approach to Reading for Course Revision}

Building upon the reengineering framework, we propose an approach to course revision based on the analysis of learners' reading traces. We refer to the tracking and analysis of learners' reading behavior as \textit{reading analytics}, which we define as a subfield of \textit{learning analytics} \citep{sadallah2020leveraging}. It involves the tracking, collection, analysis, and communication of data related to learners' reading of educational content, as well as the context in which the reading activity occurs.

\begin{figure}[h]
	\centering
	\includegraphics[width=\linewidth]{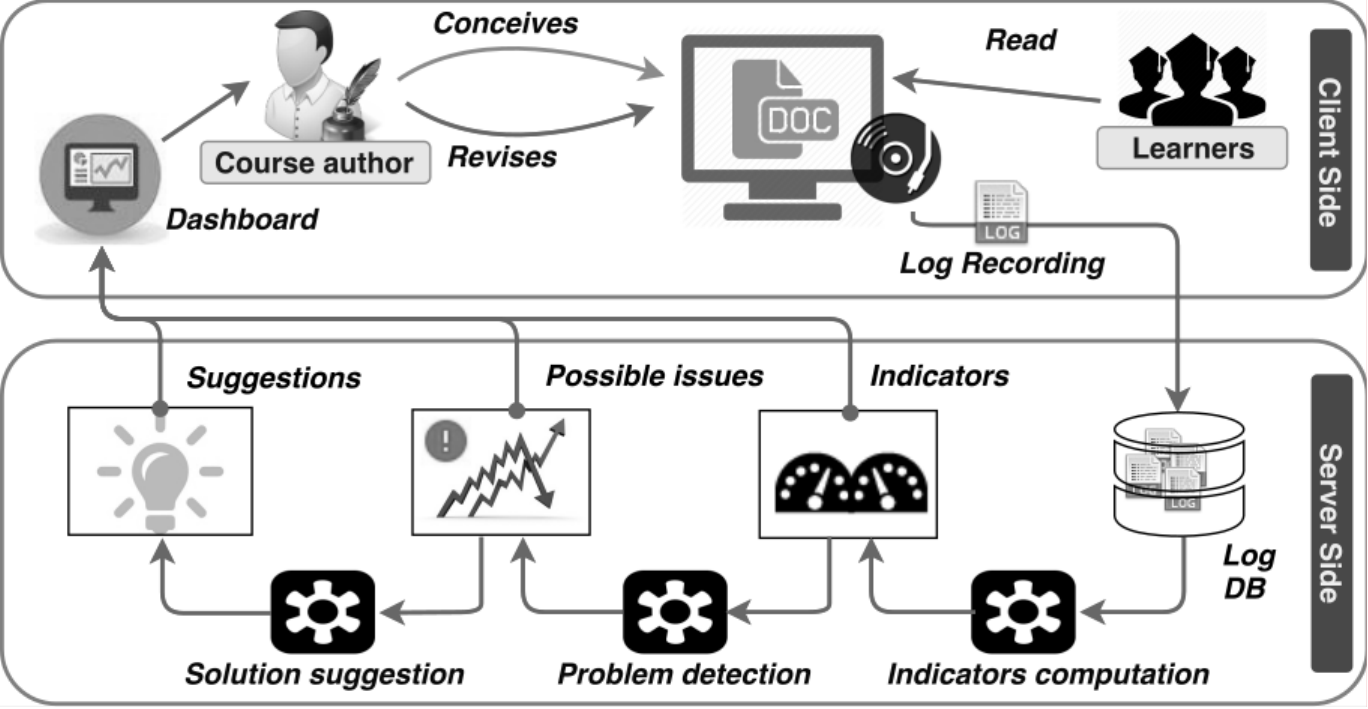}
	\caption{Author assistance approach}
	\label{fig:model}
\end{figure}

The proposed approach (Figure \ref{fig:model}) is non-intrusive and is based on the general document reengineering framework. It focuses on the first three levels of support: calculation of \textit{reading indicators}; detection of \textit{reading problems}; and provision of \textit{revision proposals}. Given the complexity and sensitivity of the educational context, we do not address the fourth level of support, which pertains to the automatic generation of revised courses.

\subsection{The Concept of "Reading Session"} 
A session represents the actions performed by a user over a period of time or in relation to the completion of a specific task. In the context of online learning, we use the concept of a "reading session" to refer to the active period during which a reading activity takes place. It consists of a sequence of consecutive actions by a learner, which can be considered continuous (with the exception of brief interruptions, such as reading emails). This means that a learner who spends one hour on a course would have a reading session lasting one hour.

\subsection{Session Identification Approach}
We propose a new approach for identifying sessions, which is more efficient by: (1) considering only reading activity; (2) using actual learner data that represents their interactions within the learning system; and (3) calculating the reading time for each course element \citep{Sadallah2015}. This approach defines a dynamic process that allows for the updating of detected sessions with the arrival of new traces (each page has its fixed values until new data is available).

\begin{figure}[h]
	\centering
	\includegraphics[width=\linewidth]{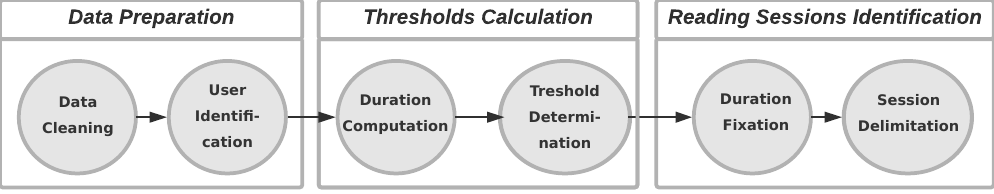}
	\caption{Steps to Calculate Reading Sessions from Learner Logs}
	\label{fig:rsalgo}
\end{figure}

Technically, the approach consists of five consecutive steps, with the first two steps (\textit{user identification} and \textit{action duration estimation}) serving as preprocessing steps. The phases of the session identification algorithm are depicted in Figure~\ref{fig:rsalgo}.

\subsubsection{Data Preparation}
This phase involves preparing the data for processing through various cleaning tasks applied to the raw collected data. This helps identify and eliminate any errors or inconsistencies, thereby improving the data quality.

\subsubsection{Threshold Calculation}
The traces (logs) result from learner interactions, represented as an ordered set of timestamped queries. Since online systems lack a means to explicitly mark the end of an action, the duration of these actions is not directly available. Therefore, the chronological order of event occurrences is used. For each sequence of actions by a given user, the start time of each action is considered to be the end time of the previous action.

The recorded data may have durations that are either too long or too short. For example, a learner may access a part of the course and then temporarily change activities without disconnecting, or the learner may modify their activity for a long or indefinite period. On the other hand, some events may be very short and therefore do not represent actual reading actions. To minimize the impact of such actions on threshold calculations, we only use "normal actions" by excluding those with excessively long or insignificant durations. For this purpose, we apply the \textit{Peirce Criterion}, a method for eliminating suspicious or outlying values. For each course element, the maximum value of the subset of data obtained after removing outliers is taken as the reading threshold for that element. This threshold is used to delimit the reading sessions.

\subsubsection{Identification of Reading Sessions}
We calculate the duration of each action as the time interval between its occurrence and the occurrence of the following action. Unknown durations arise for the last action, as no subsequent request can be used to define its end time. In order to avoid affecting the data corpus, and instead of ignoring these actions, we assign them the threshold values of the course elements read during these actions.

We use the reading thresholds of course elements to organize the learner logs into sessions. A reading session is considered complete when the time spent on reading an element exceeds its time threshold (\textit{threshold}). The following element is then assumed to belong to a different session.

\subsection{Taxonomy of Indicators Based on Reading Sessions}
Several indicators derived from learners' reading sessions are proposed, based on widely used metrics in navigation analysis. Organized into four classes (stickiness, rereading, navigation, and stop \& resume), these indicators aim to characterize reading behavior from the following perspectives: (1) learners' interest and reading pace; (2) learners' rereading habits; (3) learners' navigation through the course; and (4) interruptions, pauses, and resumption of reading \citep{sadallah2020towards}.

\subsubsection{Stickiness}
This class reflects the ability of each course element to attract and retain learners' interest. The indicators in this class include: 
\begin{itemize}
	\item \texttt{Visits}: the rate of visits to the element compared to all course visits.
	\item \texttt{Readers}: the rate of unique learners who read the element among all course readers.
	\item \texttt{Reading sessions}: the rate of reading sessions containing the element among all learners' reading sessions.
	\item \texttt{Reading speed}: the average reading speed for the element, expressed in words per minute.
	\item \texttt{Interest}: a global measure calculated as the average of the various indicators in this class.
\end{itemize}

\subsubsection{Rereading}
Rereading is one of the most common strategies used in reading for learning. The indicators in this class include: 
\begin{itemize}
	\item \texttt{Rereads}: the rate of revisits by the same learners to the course element.
	\item \texttt{Within-session rereads}: the rate of rereading that took place within the same reading session.
	\item \texttt{Between-session rereads}: the rate of rereading that occurred across different reading sessions.
\end{itemize}

\subsubsection{Navigation}
A learner's navigation refers to the reading paths formed by transitions between visited elements (arrivals and departures). A transition is linear when the arriving element is located immediately after the departing element in the course plan. The indicators in this class include: 
\begin{itemize}
	\item \texttt{Navigation Linearity}: the rate of navigation \textit{to} the element immediately following or \textit{from} the element directly preceding it in the course plan.
	\item \texttt{Arrival Linearity}: the rate of arrivals from the element immediately preceding in the course plan.
	\item \texttt{Departure Linearity}: the rate of departures to the element immediately following in the course plan.
	\item \texttt{Future Arrivals}: the rate of arrivals from elements located after the current element in the course plan.
	\item \texttt{Past Arrivals}: the rate of arrivals from the element located before the current element in the course plan.
	\item \texttt{Future Departures}: the rate of departures to elements located after the current element in the course plan.
	\item \texttt{Past Departures}: the rate of departures to elements located significantly before the current element, excluding the next element.
\end{itemize}

\subsubsection{Stop and Resume}
This class identifies how learners interrupt or stop their reading activity, and how they resume it. The indicators in this class include: 
\begin{itemize}
	\item \texttt{Reading Halt}: the rate of reading session endings on the element.
	\item \texttt{Reading Stop}: the rate of definitive reading stoppages that occurred on the element.
	\item \texttt{Resume Linearity}: the rate of resumptions that take place on the same element where reading stopped, or on the next element (in the course plan).
	\item \texttt{Past Resume}: the rate of reading resumptions that occur on previous elements.
	\item \texttt{Future Resume}: the rate of reading resumptions that occur on future elements.
\end{itemize}

\subsection{Problem Detection}
Since the indicators are univariate numerical variables, problem detection involves identifying potential outliers in their values. To this end, we use the MAD (\textit{Median Absolute Deviation}) method, which is a robust technique that is insensitive to the presence of outliers.

\subsection{Comprehension Issues and Associated Revisions} 
We use the taxonomy of revision actions to formulate appropriate revision actions for the different types of reading problems we have identified. These actions are eventually rewritten as clear and actionable statements for course authors.

\subsubsection{Problems and Revision Suggestions Related to Stickiness}
\paragraph{Low Interest.}
The lack of appeal of an element can be reflected by a low number of visits and/or learners engaging in reading the element, as well as a reduced number of sessions that include it.
\begin{itemize}
    \item \textit{Related factors:} logical level ($LL2$) and semantic level ($ML7$, $ML1$, $ML9$).
    \item \textit{Associated suggestion:} ``If the element needs to be presented: (1) move it to a more appropriate location; (2) give it a more meaningful and engaging title; and (3) update, correct, and deepen its content, enriching it with new material. If not, merge it with another relevant element or remove it.''
\end{itemize}

\paragraph{Fast Reading Speed.}
A high reading speed may indicate that readers find little interest in the element.
\begin{itemize}
    \item \textit{Related factors:} logical level ($LL1$) and semantic level ($ML3$, $ML9$, $ML7$).
    \item \textit{Associated suggestion:} ``If the element needs to be presented: (1) move it to a more appropriate location; (2) give it a more meaningful and engaging title; and (3) update, correct, and deepen its content, enriching it with new material. If not, merge it with another relevant element or remove it.''
\end{itemize}

\paragraph{Slow Reading Speed.}
An unintentional slow reading speed of an element often indicates difficulties in understanding and interpreting the content. When the reading speed of an element is generally slower than that of other course elements, this can suggest that its content is complex and difficult to comprehend.
\begin{itemize}
    \item \textit{Related factors:} logical level ($LL3$), readability level ($WL2$, $WL3$, $WL4$), and semantic level ($ML10$, $ML8$).
    \item \textit{Associated suggestion:} ``Rewrite the content of the element to reduce its complexity. Rephrase, summarize, and clarify the complicated or lengthy parts, and simplify the writing. It might be useful to break the element into sub-elements to allow for a progressive reading of the information, or move it to a position that would facilitate its reading and comprehension.''
\end{itemize}

\subsubsection{Problems and Revision Suggestions Related to Proofreading}
\paragraph{Frequent Proofreading.}
Proofreading is a common strategy used by readers who have difficulty assimilating the information conveyed. In this case, the author is called to clarify the discourse further to make it more accessible.
\begin{itemize}
    \item \textit{Related factors:} physical level ($PL5$), readability level ($WL4$ and $WL5$), and semantic level ($ML8$).
    \item \textit{Associated suggestion:} ``To minimize the need for proofreading, facilitate memorization by rephrasing the content, synthesizing and clarifying complicated or lengthy sections, and simplifying the writing. Some proofreading may be due to the presence of many links to this element: remove some of them and replace them with brief reminders.''
\end{itemize}

\paragraph{Frequent Intra-session Proofreading.}
Successive proofreading of the same content may indicate difficulties in assimilation by the readers. Sometimes, a rewrite of the content may be necessary to support the readers.
\begin{itemize}
    \item \textit{Related factors:} readability level ($WL1$, $WL2$, $WL3$, and $WL4$), and semantic level ($ML4$ and $ML8$).
    \item \textit{Associated suggestion:} ``To minimize this type of proofreading, improve readability and facilitate comprehension by rephrasing, synthesizing, and clarifying complicated or lengthy sections, and simplifying the writing.''
\end{itemize}

\paragraph{Frequent Proofreading Across Sessions.}
Proofreading occurring across different sessions can be seen as an indication of readers needing reminders of previously visited content. This can be reduced by using reminders of the read content and simplifying it to make it easier to remember.
\begin{itemize}
    \item \textit{Related factors:} logical level ($LL5$), physical level ($PL5$), readability level ($WL5$), and semantic level ($ML2$, $ML4$, and $ML10$).
    \item \textit{Associated suggestion:} ``To minimize this type of proofreading, try moving the element to a more appropriate position that would enhance memorization, or distribute it across different positions. Rewrite thoroughly by rephrasing, synthesizing, and clarifying complicated or lengthy sections, and simplifying the writing. Some proofreading may be due to numerous references to this element: remove those links or replace them with brief reminders.''
\end{itemize}

\subsubsection{Problems and Revision Suggestions Related to Navigation}
The linearity of navigation reflects the degree to which the reading order conforms to that predefined by the author (through the outline). This degree, closely linked to understanding, characterizes the deviation of the reading paths from the expected one. A significant deviation often indicates potential disorientation of the reader and cognitive overload. In such cases, the author of the course can reduce structural disorientation caused by the document's structure to better guide learners in constructing an effective reading path and a coherent mental model for it.

\paragraph{Excessive Non-linear Navigation}
\begin{itemize}
    \item \textit{Related factors:} logical level ($LL5$), physical level ($PL5$), readability level ($WL5$), and semantic level ($ML3$, $ML4$, and $ML5$).
    \item \textit{Associated suggestion:} ``To encourage linear reading of the element, move it to a more appropriate position. To facilitate memorization: rephrase its content, synthesize and clarify complicated or lengthy sections, and simplify the writing. Consider also \textit{removing certain links} to/from distant elements and replacing them with quick \textit{reminders} of relevant content where and when necessary.''
\end{itemize}

\paragraph{Excessive Arrival of Future Elements}
\begin{itemize}
    \item \textit{Related factors:} logical level ($LL5$ and $LL7$), physical level ($PL5$), readability level ($WL5$).
    \item \textit{Associated suggestion:} ``Consider moving the element to a more appropriate location. Otherwise, remove links to future elements and replace them as needed with reminders of relevant content. It is important to facilitate memorization: rephrase its content, synthesize and clarify complicated or lengthy sections, and simplify the writing.''
\end{itemize}

\paragraph{Excessive Arrival of Past Elements}
\begin{itemize}
    \item \textit{Related factors:} logical level ($LL5$ and $LL6$), physical level ($PL5$).
    \item \textit{Associated suggestion:} ``Consider \textit{moving} the element backward to a more appropriate location or \textit{deleting links} to it from earlier elements.''
\end{itemize}

\paragraph{Excessive Departure to Future Elements}
\begin{itemize}
    \item \textit{Related factors:} logical level ($LL1$, $LL2$, $LL5$, and $LL7$), and semantic level ($ML9$).
    \item \textit{Associated suggestion:} ``If the following element is worth presenting, give it a more meaningful and engaging title, enrich it with new content, use graphics and rich media if possible, and update, correct, and deepen the existing content. Otherwise, move it to a more appropriate position or simply \textit{delete} it.''
\end{itemize}

\paragraph{Excessive Departure to Past Elements}
\begin{itemize}
    \item \textit{Related factors:} logical level ($LL5$ and $LL6$), readability level ($WL5$), and semantic level ($ML3$, $ML4$, $ML5$).
    \item \textit{Associated suggestion:} ``Review this element and the related preceding elements to simplify their understanding, facilitate memorization, and help learners establish meaningful connections between the ideas presented: rephrase, update, and correct the content of these elements, synthesize and clarify complex or lengthy sections, and simplify the writing. Also consider moving this element to a more appropriate position or adding reminders of the content presented.''
\end{itemize}

\subsubsection{Problems and Revision Suggestions Related to Pauses and Regressions}

\paragraph{Multiple Pauses, With or Without Resumption.}
While some pauses in reading are trivial (e.g., the last chapters of the course), certain cases may indicate that learners have lost motivation and interest in the course. The author may need to review the elements where reading stops.
\begin{itemize}
    \item \textit{Related factors:} readability level ($WL1$, $WL2$, $WL3$, and $WL4$), and semantic level ($ML1$, $ML2$, $ML3$, $ML4$, and $ML10$).
    \item \textit{Associated suggestion:} ``If the element is worth presenting, move it to a more appropriate location. Otherwise, merge it with a suitable element or remove it. Also, rewrite the content to improve comprehension by rephrasing, simplifying, explaining, and illustrating ideas. Check for any possible errors, correct them, and update the content as needed.''
\end{itemize}

\paragraph{Non-linear Resumption of Reading.}
Many abnormal resumptions suggest that learners need to navigate elsewhere to understand the information presented. Therefore, the author may need to improve the writing of the document and provide the necessary explanations.
\begin{itemize}
    \item \textit{Related factors:} logical level ($LL5$), physical level ($PL5$), readability level ($WL5$), and semantic level ($ML3$, $ML4$, and $ML5$).
    \item \textit{Associated suggestion:} ``To ensure a linear curriculum vitae after pausing on the element, move it to a more appropriate position. Facilitate its memorization by rephrasing the content, synthesizing and clarifying complicated or lengthy sections, and simplifying the writing. Also, consider \textit{removing some links} from/to distant elements and replacing them with quick \textit{reminders} wherever and whenever needed.''
\end{itemize}

\paragraph{Resumption on Future Elements.}
Such resumption may reflect the lack of appeal or relevance of the current element and the one that follows. Therefore, the author should review the skipped elements and possibly merge them with others or remove them.
\begin{itemize}
    \item \textit{Related factors:} logical level ($LL1$, $LL2$, $LL5$, and $LL7$), and semantic level ($ML5$ and $ML9$).
    \item \textit{Associated suggestion:} ``If the next element needs to be presented: (1) move it to a more appropriate location; (2) give it a more meaningful and attractive title; and (3) update, correct, deepen its content, and enrich it with new material. Otherwise, merge it with another suitable element or remove it.''
\end{itemize}

\paragraph{Resumption on Past Elements.}
A regression can indicate that learners need to recall content already read. Therefore, it is suggested that the author either review this knowledge to facilitate its memorization or add reminders as needed.
\begin{itemize}
    \item \textit{Related factors:} logical level ($LL5$ and $LL6$), readability level ($WL5$), and semantic level ($ML3$, $ML4$, and $ML5$).
    \item \textit{Associated suggestion:} ``Review this element and related previous elements to simplify their understanding, facilitate memorization, and help learners establish a meaningful connection between expressed ideas: rephrase, update, and correct their content, synthesize and clarify complicated or lengthy sections, and simplify the writing. Also, consider moving this element to a more appropriate position or adding reminders of the concepts presented.''
\end{itemize}

\section{CoReaDa -- the Course Reading Analytics Dashboard}

The analytical approach to course revision is implemented using taxonomies and features introduced by its instantiation. CoReaDa - the Course Reading Analytics Dashboard - is an analysis and visualization tool co-designed with online course authors and HCI researchers \citep{sadallah2020coreada}. Its interface is developed using modern technologies and has been instantiated for courses delivered by OpenClassrooms, one of the largest e-learning platforms in Europe.

\subsection{Design Methodology}

The co-design process, involving online course authors and HCI researchers, led to multiple preliminary and intermediate versions of the prototype. CoReaDa was designed to closely follow the requirements identified in the fields of visual analysis and dashboard design:
\begin{itemize}
    \item The dashboard is designed as a single-page web application (\textit{SPA}), displaying only the relevant information in a scattered manner.
    \item The single-page design prevents fragmentation of information across different screens or pages, thus avoiding loss of context and ensuring that users can maintain a continuous flow of thought while analyzing data.
    \item Appropriate visualizations are used, avoiding purely decorative components.
    \item Graphical representations are employed for complex data, condensed to reveal trends or comparisons in a visually digestible format.
    \item Graphical components are combined with textual components to provide explanations and further details for the data presented.
\end{itemize}
We have integrated three approaches: \textit{Overview+Detail}, \textit{Focus+Context}, and \textit{Contextual Cues} in the design:
\begin{itemize}
    \item The \textit{Overview+Detail} interface allows simultaneous display of an overview and a detailed view of an information space, each in a distinct presentation area.
    \item The \textit{Focus+Context} approach enables users to view detailed information in the context of the current task, while also allowing interaction with other information within the system. It seamlessly integrates detailed and contextual information into a single view.
    \item Finally, \textit{Contextual Cues} enhance the detailed view with abstract shapes (e.g., arrows, arcs) that visually refer to content outside the screen's immediate view, providing a contextual reference.
\end{itemize}

\subsection{System Architecture and Technological Choices}

\begin{figure}[h]
    \centering
    \includegraphics[width=\linewidth]{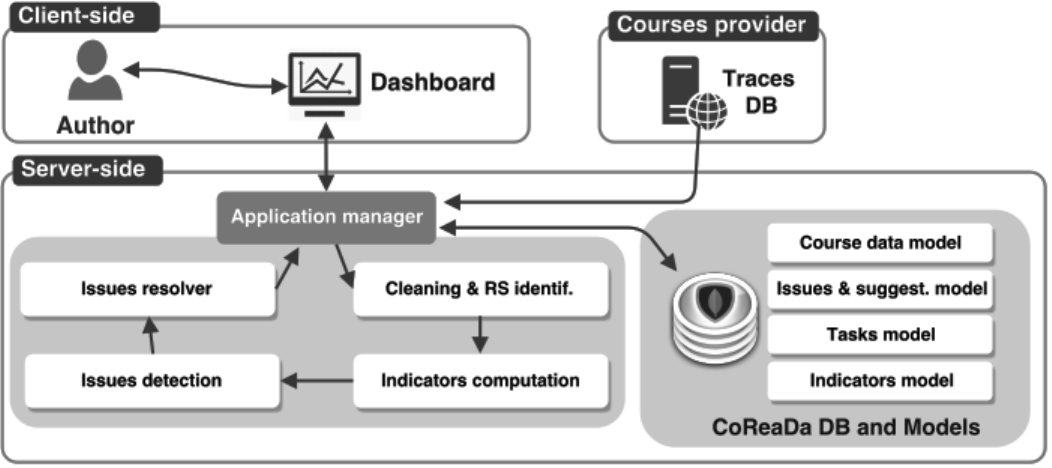}
    \caption{CoReaDa Architecture}
    \label{fig:arch}
\end{figure}

CoReaDa is built upon an \textit{MVC} (Model-View-Controller) architecture, where the presentation, processing, and management of the application are logically separated. The architecture is depicted in Figure \ref{fig:arch}. The application structure consists of a database, server-side logic, and client-side logic with a user interface. The client-side code is responsible for coordinating the interaction with the author, while the server-side code implements the analysis and business logic, determining the flow of control in the application. The application's persistent data is stored in a backend database, which is accessed and modified by the server-side code based on the author's interactions.

CoReaDa is implemented using the \textit{MEAN} stack, which has gained popularity due to its combination of highly efficient open-source technologies: \textit{MongoDB}, \textit{Express.js}, \textit{AngularJS}, and \textit{Node.js}. 
\textit{Node.js} provides efficient server-side execution, while \textit{Express.js} assists in website design. 
The flexibility of \textit{MongoDB} ensures efficient data storage and retrieval. On the client side, \textit{AngularJS} is an ideal framework for enhancing cooperative functions and Ajax-driven rich components. 
Communication between the client and server is streamlined as \textit{JavaScript} is fully supported on both the browser and server sides. 
Data analysis functions are implemented in \textit{R}, a platform-independent open-source analytics environment.

\subsection{CoReaDa Analytics (server-side)}

The analysis process begins once the learner logs are collected from the course provider's server. It consists of four consecutive tasks:
\begin{enumerate}
    \item Preparation of log files and identification of reading sessions for each learner.
    \item Calculation of indicator values for each course element.
    \item Detection of potential issues based on indicator values.
    \item Generation of revision suggestions for each identified issue using different revision actions.
\end{enumerate}

\subsection{CoReaDa Interface}

\begin{figure}[]
    \centering
    \includegraphics[width=\linewidth]{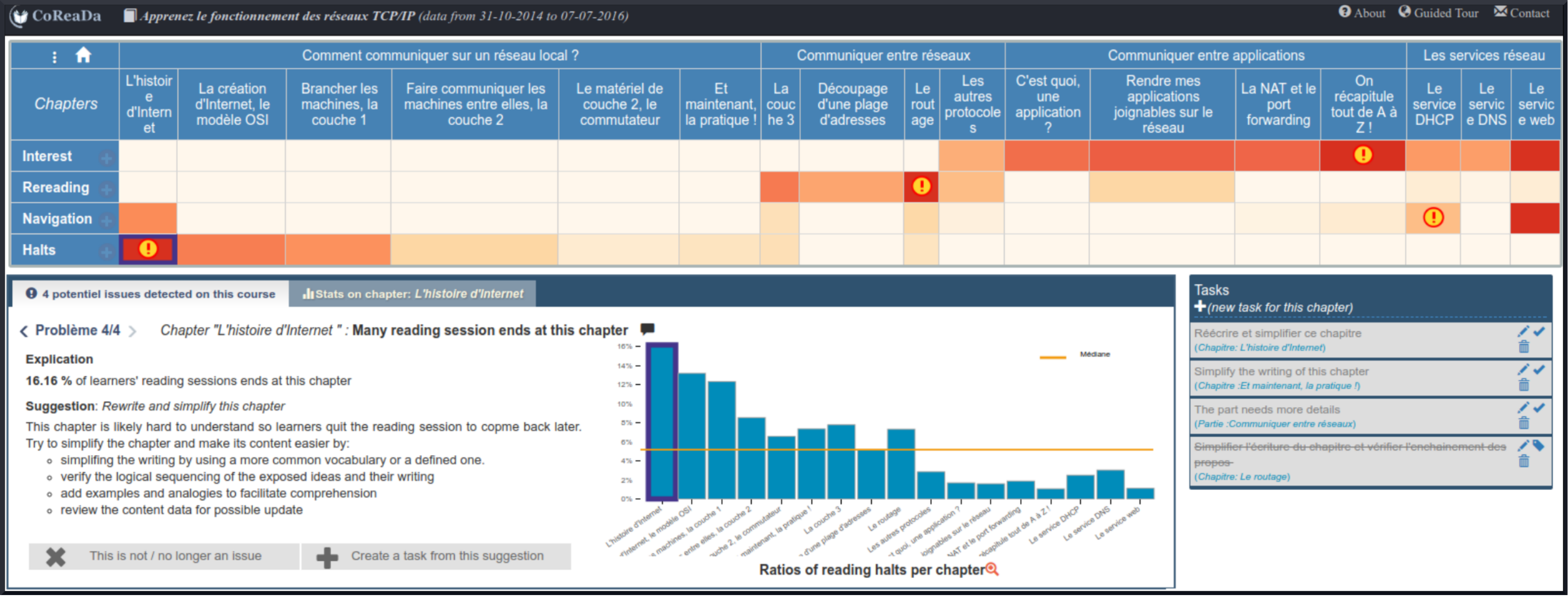}
    \caption{CoReaDa Instance Screenshot}
    \label{fig:shot}
\end{figure}

Figure \ref{fig:shot} illustrates the dashboard of a course. The user interface consists of three main areas. The top area (\textit{Data Grid Area}) displays the various indicator values, color-coded in shades (forming a heatmap). A cell's color tends to turn red to indicate an abnormal value. Detected issues are highlighted with a yellow exclamation icon. This allows the author to quickly gain an overview of their course data. The author can also focus on a particular cell to get more information about the selected indicator for the corresponding chapter. Several exploration features are available to allow the author to drill deeper into the analysis or customize the view.

The bottom-left area (\textit{Inspector Area}) is intended for the author to inspect contextual information about the selected item (course, chapter, cell, indicator, etc.) in text mode or through graphical visualization. Depending on their needs, the author can obtain either basic statistics or a detailed report that examines a specific reading issue or revision task.

The bottom-right area (\textit{Tasks Area}) allows the author to plan and manage revision tasks. A task can target a specific issue or its context (elements involved in the issue). The task specification can either come from the generated suggestions or be manually entered by the author.

\section{Evaluation and Validation Studies}
\subsection{Objectives and Context}
\label{sec:eval_context}

The studies were conducted using data from \emph{OpenClassrooms}\footnote{\url{http://fr.OpenClassrooms.com}}, a leading French e-learning platform offering MOOCs and vocational training programs. OpenClassrooms hosts over 1000 courses in English, French, and Spanish, covering diverse fields like entrepreneurship, digital marketing, and web development. The platform, launched in 1999, serves 2.5 million users worldwide and aims to make education accessible through engaging, community-driven learning experiences.

The evaluation aimed to address the following objectives:
\begin{description}
	\item[Study 1] Evaluate the session identification algorithm's accuracy in detecting learners' actual reading sessions.
	\item[Study 2] Assess course authors' perceptions of the relevance of the proposed indicators for reading analysis and course revision.
	\item[Study 3] Examine the problem detection and resolution mechanisms with course authors.
	\item[Study 4] Validate with learners whether identified issues align with their actual difficulties.
	\item[Study 5] Evaluate the dashboard's usability and authors’ willingness to adopt it for course revisions.
\end{description}

The studies employed online questionnaires and task-based experiments. Questionnaires for \textit{Studies 2} and \textit{3}  were validated by three independent researchers and two platform-affiliated authors. Final adjustments were made based on their feedback.

\subsection{Participants and Data}
\label{sec:dataused}
A total of 403 OpenClassrooms authors were invited via email to participate in the evaluation. Of these, 125 authors joined the first phase, while a subset of twelve courses was selected for \textit{Studies 3} and \textit{5}. Selection criteria included representativeness (based on the number of chapters) and popularity (measured by visits and unique readers). Table \ref{tab:coursestat} provides key statistics about the selected courses.

The learner data consisted of server logs from \textit{31 October 2014} to \textit{07 July 2016}, capturing timestamps, request identifiers, users (if not anonymous), session IDs, course IDs, and course element IDs. A typical record structure is as follows:

{\small
\begin{verbatim}
<request_id, user_id, course_id, element_id, server_session_id, timestamp>
\end{verbatim}
}

For \textit{Study 4}, 26 master's students (10 female, 16 male, aged 23-26) in Computer Science and Information Systems at CERIST participated, focusing on advanced training in Information Systems.

\begin{table}[!h]
	\centering
	\small
	\begin{tabular}{@{}lrrrrrrr@{}}
		\toprule
		Course & \# Chapters & \# Logs & \# Learners & \# Sessions \\ 
		\midrule
		\textit{Bootstrap} & 7 & 229,362 & 13,045 & 94,654 \\
		\textit{Web}       & 18 & 240,978 & 11,793 & 53,695 \\
		\textit{Twitter}   & 9 & 17,576  & 1,560  & 5,223 \\
		\textit{Arduino}   & 14 & 66,911  & 4,864  & 26,797 \\
		\textit{JavaScript}& 13 & 289,153 & 12,829 & 101,614 \\
		\textit{Ionic}     & 19 & 27,283  & 2,020  & 8,663 \\
		\textit{Ruby}      & 18 & 4,895   & 706    & 2,794 \\
		\textit{Project}   & 14 & 49,255  & 3,156  & 14,607 \\
		\textit{TCP}       & 17 & 111,026 & 7,239  & 43,392 \\
		\textit{Symfony}   & 27 & 402,039 & 9,357  & 236,635 \\
		\textit{Startups}  & 21 & 11,772  & 1,223  & 3,574 \\
		\textit{GitHub}    & 19 & 109,092 & 5,826  & 29,452 \\
		\midrule
		Median & 17.5 & 88,002 & 5,345 & 28,125 \\
		Mean   & 16.33 & 129,945 & 6,135 & 51,758 \\
		SD     & 5.38 & 129,909 & 4,654 & 67,300 \\
		\bottomrule
	\end{tabular}
	\caption{Basic statistics about the selected courses.}
	\label{tab:coursestat}
\end{table}

\subsection{Study 1 -- capabilities of the session identification approach} 
\subsubsection{Methodology}
Reconstructing learners' sessions from navigation traces without precise knowledge of actual sessions poses challenges in validating their accuracy. Prior studies reveal that session sizes, measured as the number of pages visited, follow a \textit{Power Law} distribution, where a small subset of pages receives most visits, while others are less frequented \citep{Berendt2001, Vazquez2006}. We evaluated our session reconstruction method by examining its adherence to this distribution, a common practice in similar studies \citep{Arce2014, Roman2014, Dell2008}.

Additionally, we assessed the alignment between estimated action durations and the complexity of course elements, where complexity was approximated by element size, a key factor in website complexity \citep{Butkiewicz2011}. Using data from the twelve selected courses (\ref{tab:coursestat}), we applied our session detection approach, evaluated reconstruction quality, and compared results with established web usage mining methods. 

\subsubsection{Results}
\paragraph{Quality of the reconstruction}
\begin{figure}[h]
	\centering
	\includegraphics[width=\linewidth]{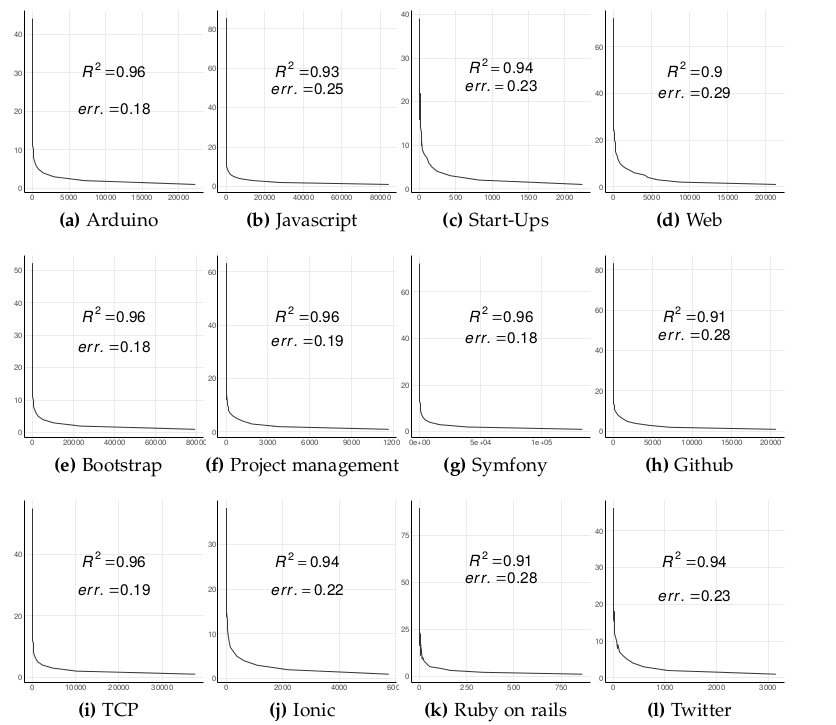}
	\caption{Taille des sessions trouvée par la distribution de la loi de puissance sur les 12 cours}
	\label{fig:rs_distr}
\end{figure}

\begin{table}
	\footnotesize
	\begin{tabular}{p{2 cm}|p{1 cm}p{1 cm}|p{1 cm}p{1 cm}|p{1 cm}p{1 cm}}
		\toprule
		& \multicolumn{2}{c|}{\textit{Reading Session}} & \multicolumn{2}{c|}{\textit{10-min Page Threshold}} & \multicolumn{2}{c}{\textit{30-min Session Threshold}}\\ 
		\textit{} & \textit{$R^{2}$}& \textit{Err} & \textit{$R^{2}$} & \textit{Err} & \textit{$R^{2}$} & \textit{Err}\rule[0.33cm]{0cm}{0cm}\\ \midrule
		\multicolumn{1}{c|}{Bootstrap} & \textbf{0.96} & \textbf{0.18} & 0.95 & 0.22 & \textbf{0.96} & 0.21 \\
		\multicolumn{1}{c|}{Web} & \textbf{0.90} & \textbf{0.29} & 0.88 & 0.31 & 0.86 & 0.31 \\
		\multicolumn{1}{c|}{Twitter} & \textbf{0.94} & \textbf{0.23} & \textbf{0.94} & 0.24 & 0.91 & 0.25 \\ 
		\multicolumn{1}{c|}{Adruino} & \textbf{0.96} & \textbf{0.18} & 0.94 & 0.20 & 0.93 & 0.23 \\ 
		\multicolumn{1}{c|}{JavaScript} & \textbf{0.93} & \textbf{0.25} & 0.92 & 0.26 & 0.90 & 0.28 \\ 
		\multicolumn{1}{c|}{Ionic} & \textbf{0.94} & 0.22 & 0.93 & \textbf{0.21} & 0.92 & 0.24 \\ 
		\multicolumn{1}{c|}{Ruby} & 0.91 & 0.28 & 0.91 & 0.29 & \textbf{0.93} & \textbf{0.26} \\ 
		\multicolumn{1}{c|}{Project} & \textbf{0.96} & \textbf{0.18} & 0.92 & 0.24 & 0.92 & 0.23 \\ 
		\multicolumn{1}{c|}{TCP} & \textbf{0.96} & \textbf{0.19} & 0.95 & 0.20 & 0.95 & 0.20 \\ 
		\multicolumn{1}{c|}{Symfony} & \textbf{0.96} & \textbf{0.29} & 0.88 & 0.31 & 0.86 & 0.35 \\ 
		\multicolumn{1}{c|}{Startups} & \textbf{0.94} & 0.23 & 0.93 & \textbf{0.22} & 0.93 & 0.23 \\ 
		\multicolumn{1}{c|}{Github} & 0.91 & 0.28 & \textbf{0.94} & \textbf{0.23} & 0.93 & 0.25 \\ \bottomrule
	\end{tabular}
	\caption{Constructed sessions using three methods : our proposal, fixed page threshold (10-min) and fixed session threshold (30-min).}
	\label{table:rs_s_fsa}    
\end{table}

Evaluating reconstruction quality using the Power Law involves linear regression on the logarithm of distinct read elements versus total reading sessions. The quality is assessed by the regression correlation coefficient $R^{2}$ (measuring fit, with $R^{2} \to 1$ indicating a perfect fit) and the standard error \textit{err} (indicating fit accuracy, with \textit{err} $\to 0$ being optimal). Results (Figure \ref{fig:rs_distr}) demonstrate excellent session identification capabilities, with $R^{2} \geq 0.90$ (mean $0.94$) and acceptable \textit{err} values (mean $0.22$). 

To reinforce conclusions, we recalculated sessions using two common methods:
\begin{itemize}
    \item A fixed \textit{page stay time} threshold of 10 minutes.
    \item A fixed \textit{total session time} threshold of 30 minutes.
\end{itemize}
Reconstruction quality was evaluated for each, and comparative results (Table \ref{table:rs_s_fsa}) show our method performed best for $75\%$ of courses (9/12) and achieved optimal fitness and accuracy in $58\%$ of cases (7/12). Notably, thresholds on page stay time, whether fixed or dynamic, were more effective for session delimitation than fixed total session duration in this context.

\paragraph{Compliance with elements size and complexity}
\begin{figure}[!h]
	\centering
    \includegraphics[width=.7\linewidth, height = 8cm]{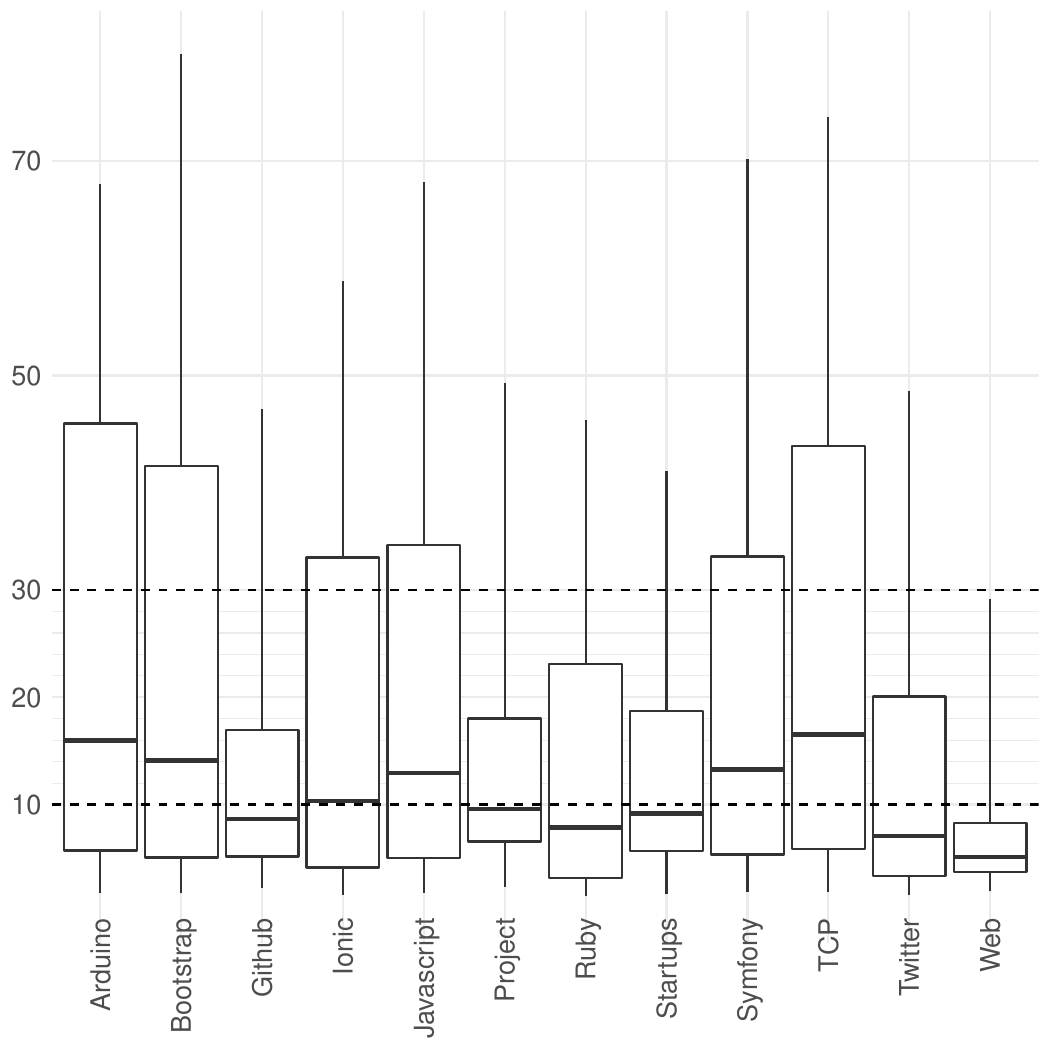} 
	\caption{Course chapters reading duration statistics}
	\label{fig:time}
\end{figure}
We used duration as an indicator of the element complexity. We estimated the size of each element of the courses by counting its significant words and in-line images (with each image considered as a short paragraph of 30 words). \ref{fig:time} It indicates for each course the distribution of reading time for each of its elements (chapters). As shown by the evolution of the distribution, reflected by the size and shape of the boxplots, it is not possible to determine a single fixed threshold value that is appropriate either for all chapters of all the courses or even for chapters of the same course. Whatever the values selected, there will always be some elements that can be read in less time and others that may require much more time.

Defining dynamic thresholds per course elements can better reflect element complexity. Indeed, Pearson correlation coefficient between element size (computed as the words and figure count) and time threshold for that element is $r = 0.82$ ($p < 0.001$). This positive and significant correlation means that the method succeeds in assigning important durations for pages with significant content and, conversely, in associating reduced durations for relatively short pages. It is therefore fair to conclude that the approach is sufficiently generic and robust to take into account the size of the elements without having any further knowledge about them (and therefore without the need to calculate them).
We can make the hypothesis that it is also the case for element complexity, even if element size does not directly indicate the complexity level of the content. This also confirms the need to take into account the characteristics of the elements for more accurate threshold values.

\subsection{Study 2 -- relevance of the indicators} 
\label{sec:indicators_eval}
\subsubsection{Protocol}
This study was carried out during one month (April 2016). A document presenting the project, its motivation, and its objectives was distributed to the 125 OpenClassrooms authors.  Thereafter, the author received an online questionnaire titled \textit{Indicators Relevance Survey}  which is composed of four sections, one for each class of indicators. Each section explains the associated class and its indicators. The authors were asked to evaluate the relevance of each of the twenty indicators for course revision. The rating used a 5-point Likert scale, ranging from 1 (\textit{strongly disagree}) to 5 (\textit{strongly agree}). In average, the questionnaire needed 18 minutes to complete.

\subsubsection{Results}
To evaluate the perceived relevance of the indicators, we used the results from the \textit{Indicators' relevance questionnaire}. 
There are two schools of thoughts on analyzing Likert-scale data: \textit{ordinal vs. interval} \citep{Carifio2007}.
A significant amount of empirical evidence exists supporting that Likert scales can be used as interval data  \citep{Carifio2007} or aggregated to create a new composite metric scale. Accordingly, we analyze the results both by individual indicators and by aggregating them into their corresponding classes.

The  $Cronbach's$ $\alpha$ coefficient obtained for the ratings is $0.82$. The reliability coefficient for internal consistency if an individual indicator is removed from the scale gave values ranging from $0.78$ to $0.91$. This shows that the reliability of the results had appropriate internal consistency. No strong correlation between the ratings of the indicators was found (\textit{Pearson correlation} coefficients ranging from $-0.16$ to $0.62$). This suggests that the authors considered these indicators to cover relatively distinct aspects of reading behavior.
\begin{table}[htp]
	\centering
	\small
	\begin{tabular}{@{}llll@{}}
		\toprule
		\textit{Variable}                            & \textit{Category} & \textit{Frequency} & \textit{Percentage} \\ \midrule
		\multirow{2}{*}{\textit{Gender}}             & Female            & 57                 & 45.6\%               \\
		& Male              & 68                 & 54.4\%               \\\midrule
		\multirow{3}{*}{\textit{Age}}                & 19-25             & 48                 & 38.4\%               \\
		& 26-40             & 66                 & 52.8\%               \\
		& 41-58             & 11                 & 8.8\%                \\\midrule
		\multirow{3}{*}{\textit{Level of education}} & Bachelor          & 11                 & 8.8\%                \\
		& Master            & degree             & 49\%                 \\
		& Doctorate         & degree             & 65\%                 \\
		\bottomrule
	\end{tabular}
	\caption{Demographic description of the participating authors}\label{tab:demographic}
\end{table}

We conducted a series of $ANOVA$ (analysis of variance) to study the influence of each independent demographic variable (gender, age, and level of education) on the author ratings (the significance level was set at $0.05$). The demographic data (gender, age, and education level) of the 125 course authors are provided in \ref{tab:demographic}. The results showed no significant effect of the gender and level of education on the ratings ($p>0.05$). However a tangible effect of the age on the results was found for two classes: stickiness ($F(2,125)=3.83$, $p=0.024$) and navigation ($F(2,125)=3.41$, $p=0.036$).  The variable age can have three values ($N=125, Min=19, Max=58, Mean=29.48, Median=27, SD=8.12$), as indicated in \ref{tab:demographic}, allowing to split the participants into three groups. A $Tukey$ post hoc test revealed that the ratings of the stickiness and navigation indicators were statistically different ($p<0.05$) between the three groups of participants. In addition, they showed that the ratings of young participants were higher, while those of older participants were the lowest.

\begin{table}[htp]
	\centering
	\small	
	\begin{tabular}{@{}LLLL|LLL|LLL@{}}
		\toprule
		& \multicolumn{3}{p{3cm}}{Descriptive statistics}   & \multicolumn{3}{p{2cm}}{Student t-test$^{*}$} &\multicolumn{3}{p{4cm}}{Inter-correlations (Spearman)}\\ 
		\textit{} & \textit{Mean} & \textit{SD} & \textit{Median} & \textit{t}  & \textit{df}  & \textit{p}& Navigation & Rereading & Stops \\
		\midrule
		Stickiness  & 3.66& 0.53 &3.71&0.41  &105   & 0.67 & 0.01 & 0.16 & 0.13 \\
		Navigation  & 3.44&0.54&3.33& 0.47  &105   & 0.64 & & -0.01 & -0.04 \\
		Rereading  &3.67&0.59&3.67& -0.04  & 105  &0.96  & & & 0.01 \\
		Stops &3.55&0.60&3.67&-0.89  &105   &0.33 & & &  \\
		\bottomrule
		\multicolumn{10}{L}{^{*}group~1: female~participants (n=42); group~2: male~participants (n=63)}\\
	\end{tabular}
	\caption{Statistics about authors' ratings}\label{tabl:indictorsrating}
\end{table}

In order to investigate for possible differences between the genders, we ran an independent \textit{t-test} for each class of indicators (the significance level for the mean variation was set at $p<0.05$). 
The results, shown on~\ref{tabl:indictorsrating} (columns 5, 6 and 7), revealed no significant difference for all the classes of indicators. Associations between the ratings of the different classes of indicators were examined using \textit{Pearson's correlation coefficients}. As reported in the table, no significant correlation was found.


\begin{figure}
	\centering
	\includegraphics[width=.7\linewidth]{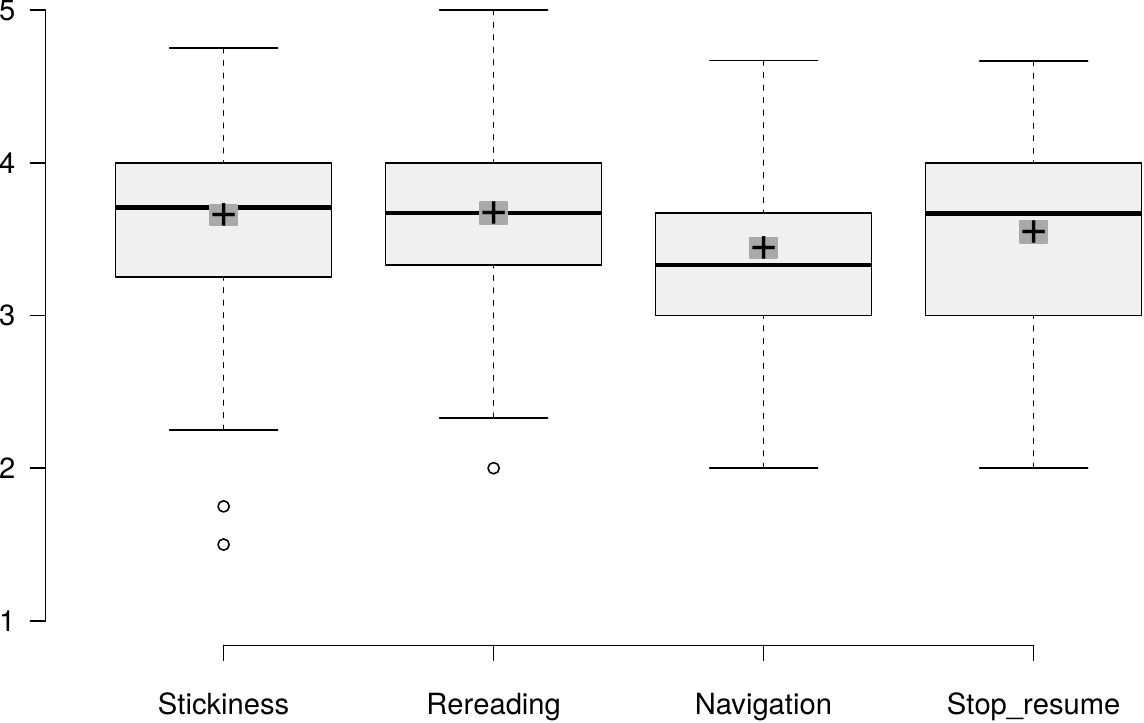}
	\caption{Authors' rating of the indicators, aggregated by classes}
	\label{fig:resp}
\end{figure}

The descriptive statistics of the aggregated results are presented on the first three columns of~\ref{tabl:indictorsrating} and the boxplots represented on~\ref{fig:resp}. The boxplots are relatively short, suggesting that the participants had a high-level agreement with each other. The results show that participants mostly agreed that the classes of indicators are relevant for analyzing course reading.
All the proposed classes of indicators were highly rated; indeed, the median and mean points were all above the neutral point of 3 (corresponding to the mention ``neither agree nor disagree''). In other words, globally the four classes of indicators were deemed useful. 

\begin{figure}
	
	\subfloat[Stickiness indicators]{%
		\label{fig:stick}
		\includegraphics[clip,width=.85\columnwidth]{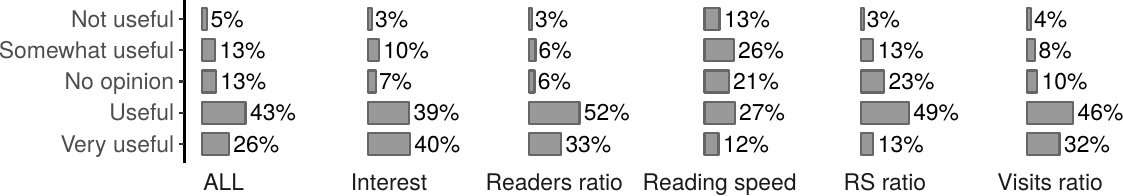}%
	}
	
	\subfloat[Rereading indicators]{%
		\label{fig:reread}
		\includegraphics[clip,width=.85\columnwidth]{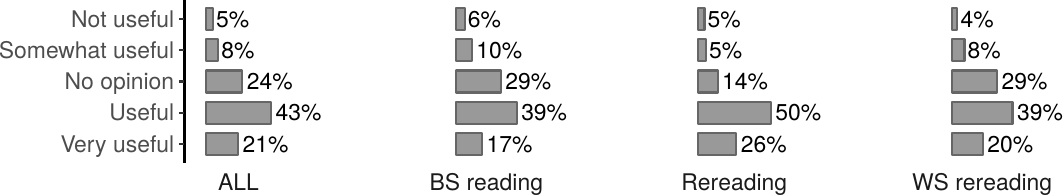}%
	}
	
	\subfloat[Navigation indicators]{%
		\label{fig:nav}
		\includegraphics[clip,width=.85\columnwidth]{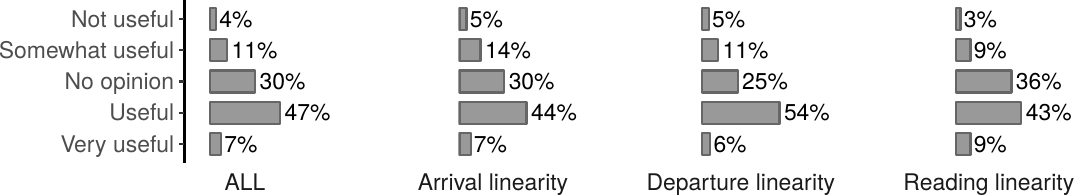}%
	}
	
	\subfloat[Stops and resume indicators]{%
		\label{fig:stop}
		\includegraphics[clip,width=.85\columnwidth]{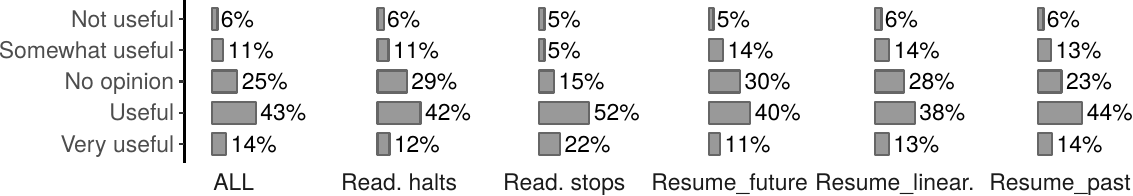}%
	}
	
	\caption{The relevance of the reading indicators, rated by 125 authors}
	\label{fig:indrat}    
\end{figure}

The authors' rating of the reading indicators is given on~\ref{fig:indrat}. We define as \textit{positive rating} any rating that is either \textit{useful} or \textit{very useful}, while a \textit{negative rating} corresponds to either \textit{somewhat useful} or \textit{not useful}. 
According to the results, the indicators were found globally suitable for analyzing reading and performing revisions, as reflected by the generally positive rating of each of them, the aggregated results for all the classes being 61\% positive, 23\% no opinion and 16\% negative. The stickiness class was perceived as the most relevant (average of 69\% of positive rating). Indeed, authors were very interested in the popularity of their courses given that the most highly rated indicators were \textit{readers ratio}, \textit{interest} and \textit{visits ratio} (resp. 85\%, 79\% and 78\% of positive ratings). The authors recognized the importance of the indicators related to rereading (average of 64\% of positive rating for this class). Yet, while the rereading ratio indicator is highly rated (76\% of positive rating), the indicators related to within and between sessions rereading seem too technical and complicated for some authors, as reported in their comments and ratings (less than 60\% of positive rating and 30\% of no opinion).

Indicators related to navigation are those rated the less useful (54\% of positive rating). About 30\% of authors responses on these indicators have no opinion, while 15\% have a negative rating. Two authors argued that they do not expect their courses to be read in a totally linear way, and another pointed out that his course could be read in any order.
The relatively low-rated indicators are related to reading speed (about. 39\% is negative and 21\% no opinion). According to their feedbacks, many authors did not pay much attention to the speed of reading since learners differ in background and learning styles. 
One common suggestion of three authors is to combine some indicators into higher level ones, to lower the number of indicators and to provide aggregated results.

\paragraph{Comments and opinions}
Many authors think that these indicators are numerous enough and judicious to give a good idea of the way learners read courses. The following sample quotes illustrate some authors' opinion:
\begin{itemize}
	\item ``What is interesting is this opportunity given to the authors of online courses and their learning to communicate, in a direct and indirect way.''
	\item ``Why not include direct exchanges between authors and learners through comments and forums?''
	\item ``These are important metrics about course consumption, they could help me understand how to rethink my course material.''
	\item ``While they seem interesting, I think you would have to select the more important indicators to present to authors. The other ones can serve for deeper analysis.''
	\item ``Be careful not to abuse the personal data of users. The reader should actually be informed that his reading is logged and analyzed.''
\end{itemize}

The authors recognize that the interaction with readers is essential to create interesting and productive courses. While more than 60 authors found the list of indicators to be comprehensive enough to analyze reading, five of them considered that there were too many indicators: without a meaningful presentation to the authors, this would be counterproductive.
The fact that readers' usages logs allow considering the end-user perspective on consuming the course is deemed interesting. 

All participating authors valued the usefulness of using reading traces to detect parts or aspects of the course that necessitate review. Many among them have appreciated the definition of indicators that result from aggregating data, since this may better reflect recurrent reading problems.
For one authors, the approach seemed complicated to implement technically and therefore may generate some unreliable results. Similarly, another author felt that we would need a good level of abstraction so that authors would not have to consult many tables and endless statistics. Several authors proposed to consider the supplementation of computed indicators with explicit readers feedback (courses ratings, comments and annotations, etc.) that would help them to better understand readers needs.
A last aspect reported by two authors was related to privacy: they suggested to ask learners before logging them. 

\subsection{Study 3 -- capabilities of the issue detection and resolution mechanisms} 
\label{sec:issues_detect_resolve_eval}
\subsubsection{Protocol}

This study was conducted from January 13th to January 26th, 2017. Each author received a pre-filled version of an online questionnaire with data specific to his course. Titled \textit{Issues and Suggestions Survey}, the questionnaire contained two sequential parts. The first part was a blank list that the author had to complete with his predictions on the possible problems encountered by learners related to each of the indicator classes. The second part consists of a listing of the issues detected and the revision suggestions generated. Using five-point Likert scales, ranging from 1 (\textit{strongly disagree}) to 5 (\textit{strongly agree}), the author was asked to estimate the plausibility of each issue and then to evaluate the usefulness of the related revision suggestion. The questionnaire took an average of 34 minutes to complete.

\subsubsection{Results}
\paragraph{Effectiveness of the detected issues for enhancing authors' awareness}
To evaluate the quality of the issue detection, we used the gathered data about authors' expectations of possible issues that may exist on their courses.
In order to quantify and qualify the gain in awareness provided by the detection tool, we classified the set of issues (provided by the author and/or by the tool) based on three criteria: whether they were expected or not, whether they were detected or not, and whether they were rated useful for revision or not. We transposed this classification into classes of knowledge as follows:
\begin{itemize}
	\item In terms of detection of the expected issues: the \textit{undetected knowledge} reflects the set of issues expected by the author but not detected by the tool; and the \textit{confirmed knowledge} results from the set of issues expected by the author and detected by the tool.
	\item In terms of expectation of the detected issues: the \textit{confirmed knowledge} (the same above-defined class) reflects the set of issues expected by the author and detected by the tool; and the \textit{new knowledge} is the set of issues that were not expected by the author but detected by the tool. 
	\item In terms of usefulness for revision of the expected and detected issues: the \textit{useless confirmed knowledge} results from the set of issues that were expected by the author and detected by the tool, but were nor rated useful for revision; and the \textit{useful confirmed knowledge} reflects the set of issues that were expected by the author, detected by the tool and rated useful for revision. 
	\item In terms of usefulness for revision of the detected but not expected issues: the \textit{useless new knowledge} reflects the issues that were detected by the tool but neither expected nor rated useful for revision; and the \textit{useful new knowledge} reflects the set of issues that were not expected by the author but were detected by the tool and rated useful for revision.     
\end{itemize}

\begin{figure}[htp]
	\subfloat[Distribution of the knowledge expected by the authors with regards to its detection by the tool]{%
		\label{fig:exp}
		\includegraphics[clip,width=\columnwidth]{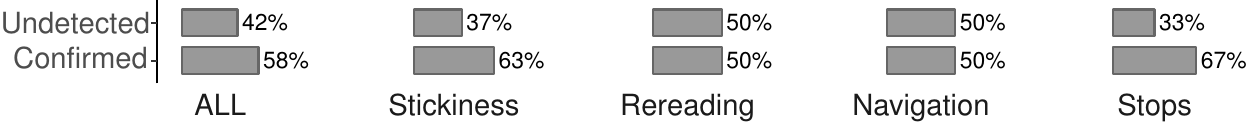}%
	}
	
	\subfloat[Authors' expectation and relevance for revision rating, of the knowledge detected]{%
		\label{fig:comp}
		\includegraphics[clip,width=\columnwidth]{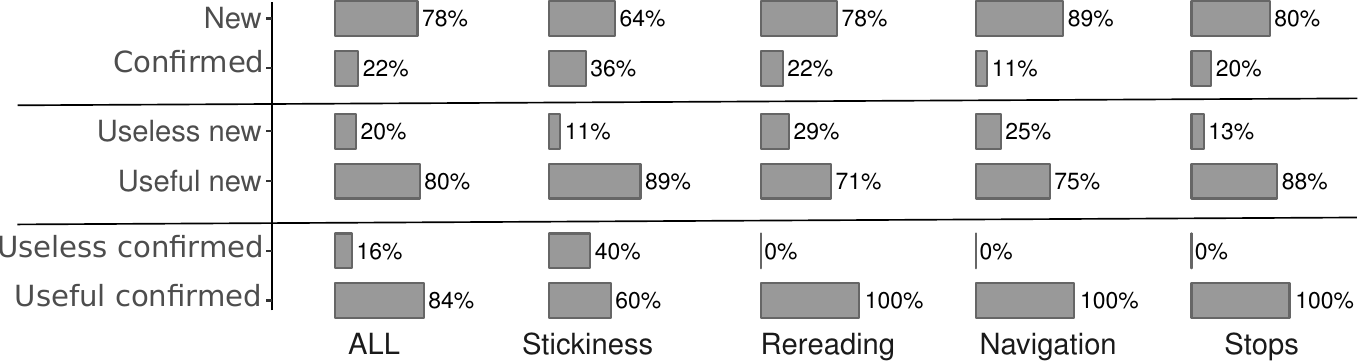}%
	}
	
	\caption{Distribution of the knowledge expected and the knowledge provided}
	\label{fig:expcomp}    
\end{figure}
\ref{fig:expcomp} represents (a) the distribution of the expected knowledge in terms of confirmed/undetected classes; (b) the distribution of the detected knowledge in terms of authors expectation and relevance for revision.
Slightly more than half of the knowledge expected by the authors was discovered by the tool. The authors had particularly good guesses about the issues related to reading stops and reading stickiness and interest. 
The undetected knowledge likely corresponds to false expectations and beliefs and is more related to rereading and navigation. 
About 78\% of the detected knowledge was new for authors, of which 80\% was found relevant for triggering revision actions. 22\% of the knowledge provided by the tool was expected, of which 84\% was deemed useful. On the stickiness class, some issues were expected and detected but were not considered relevant for revision (\textit{useless expected knowledge} class). According to some authors, these problems were predictable; for example, chapters containing only additional information do not attract much interest.

\paragraph{Usefulness of the suggestions for guiding authors in course revision}
\begin{figure}[ht]
	\centering
	\includegraphics[width=\linewidth]{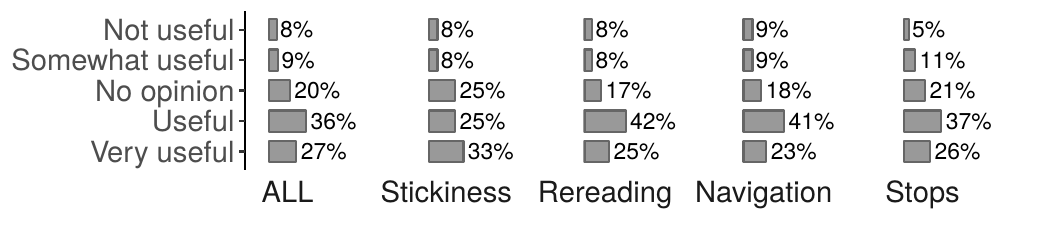}
	\caption{Relevance of the suggestions}
	\label{fig:issug}
\end{figure}
\ref{fig:issug} presents the results of the \textit{Issues and suggestions questionnaire} related to the authors' ratings of the usefulness of the suggestions. 63\% of these suggestions were rated either useful or very useful, i.e. found to provide hints and guidance for revising the course so as to resolve the associated issues. In one-fifth of cases, the authors were not sure whether the proposed solutions could effectively be the best revision to undertake; they found the suggestions too broad and they noted that deeper analysis was first needed. 
The suggestions marked as not relevant (17\%) were mostly related to issues not highly rated as needing revisions, in more than 90\% of cases. For the remaining cases, authors suggested reformulations that took more into account the context of the provided suggestions without calling into question the primitives used for generating the suggestions.

\subsection{Study 4 -- conformance of the detected issues with learners' problems}
\label{sec:study4}
\subsubsection{Description}
Among the twelve courses used is \textit{Study 3} (\ref{tab:coursestat}), the students selected four courses that they indicated having already followed on the platform during the first and/or the second semester of the year 2017 (\textit{TCP}, \textit{JavaScript},\textit{Bootstrap} and \textit{Symfony}). 
The reading logs of these courses, provided by \textit{OpenClassrooms}, were used to compute the values of the different indicators. We examined the statistical distribution of the values of these indicators for each course, to assess possible reading issues. In order to not overwhelm the learners, we considered some major problems that can be encountered by learners and detected by our tool. The rules we followed for marking a given value of a specific indicator as reflecting a potential reading issue are provided on~\ref{table:issueswording}. The numerical results of the issue detection on the four courses are shown on \ref{table:issues4courses}.

\begin{table}[!h]
	\centering
	\small
	\begin{tabular}{p{2cm}|p{5.5cm}|p{6cm}}
		\toprule
		Class & Issue triggering& Issue description   \\ \hline
		\multirow{3}{1.25cm}{Stickiness}  & \multirow{2}{5cm}{Low values of: visits, readers, reading session; High reading speed}   & ($SI_1$) Low popularity due to low attractiveness of the chapter and/or its low readability \\ 
		\cline{2-3} & Low reading speed & ($SI_2$) Low stickiness due to the complexity of the content \\ \hline
		\multirow{3}{1.25cm}{Navigation}  & \multirow{3}{5cm}{Low linearity of reading (arrivals and/or departures); High ratios of navigation to distant chapters} & ($NI_1$) Disorientation due to bad structuring   \\ 
		\cline{3-3} & & ($NI_2$) Non linear reading due to low memorability   \\ 
		\cline{3-3} & & ($NI_3$) Non linear reading due to low content complexity  \\ \hline
		\multirow{2}{1.25cm}{Rereading}  & High values of rereads and/or within-session rereads    & 
		($RRI_1$) Many consecutive rereading due to content complexity  \\ \cline{2-3} 
		& High values of between-session rereads & 
		($RRI_2$) Many distant rereading,due to low memorability   \\ \hline
		\multirow{8}{1.25cm}{Stop \& resume} & \multirow{3}{5cm}{High values of final reading stops}    & 
		($SRI_1$) Permanently stop reading the course because of loss of interest, poor readability and/or high complexity \\ 
		\cline{2-3} & High values of reading halts (non final stops) & ($SRI_2$) Reading halts due to content complexity \\ 
		\cline{2-3} & High values of nonlinear resume; High values of resume on distant chapters & ($SRI_3$) Resuming on previous on future distant chapters due to content complexity, low memorability and/or bad structuring \\ 
		\bottomrule
	\end{tabular}
	\caption{Issue detection using indicator value}
	\label{table:issueswording} 
\end{table}

\begin{table}[!h]
	\centering
	\small
	\begin{tabular}{@{}l|LL|LLL|LL|LLL|l@{}}
		\toprule
		& \multicolumn{2}{L}{Stickiness} & \multicolumn{3}{L}{Navigation} & \multicolumn{2}{L}{Rereading} & \multicolumn{3}{L}{Stop ~\&~resume} & \textit{ALL} \\
		Cours  & SI_1   & SI_2  & NI_1  & NI_2  & NI_3  & RRI_1  & RRI_2  & SRI_1 & SRI_2 & SRI_3 &  \\ \midrule
		TCP & 3   & 1   & 2 & 3 & 0 & 1   & 2   & 1  & 0  & 1   & 14 \\
		Javascript & 1   & 1   & 1 & 1 & 0 & 1   & 0   & 1  & 0  & 2  & 8  \\
		Symfony & 3   & 1   & 2 & 1 & 1 & 2   & 1   & 2  & 1  & 1  & 15 \\
		Bootstrap  & 1   & 1   & 1 & 2 & 1 & 2   & 2   & 0  & 1  & 2   & 13 \\
		\bottomrule
	\end{tabular}
	\caption{Main reading issues detected on four courses}
	\label{table:issues4courses}
\end{table}

After gathering basic demographic characteristics of the students, we presented them a paper-based questionnaire that listed for each class of indicators the issues we assessed, supplemented with summary explanations. We asked the students to carefully examine the marked issues and to rate their effectiveness using five-points \textit{Likert scales} (1=absolutely disagree, 5=absolutely agree).

\subsubsection{Results}
\begin{table}[!h]
	\centering
	\small
	\begin{tabular}{@{}LLLLL|LLL@{}}
		\toprule
		& & \multicolumn{3}{p{5cm}}{Descriptive statistics$^{*}$}   & \multicolumn{3}{p{4cm}}{Student t-test$^{**}$ results} \\ 
		\textit{} & \textit{} & \textit{Mean} & \textit{SD} & \textit{Median} & \textit{t}  & \textit{df}  & \textit{p}  \\
		\midrule
		\multicolumn{2}{l}{\textit{Stickiness issues}} & & & & & \\ 
		&SI_1  & 4.04 & 0.96   & 4.00  & -0.57  & 24  & 0.57  \\
		&SI_2  & 3.15 & 1.26   & 3.00  & -0.81  & 24  & 0.43  \\
		\multicolumn{2}{l}{\textit{Navigation issues}} &  & & & & \\ 
		&NI_1  & 3.58 & 0.86   & 4.00  & -1.32  & 24  & 0.20  \\
		&NI_2  & 3.50 & 1.03   & 4.00  & -1.18  & 24  & 0.25  \\
		&NI_3  & 3.42 & 1.03   & 4.00  & -1.28  & 24  & 0.21  \\
		\multicolumn{2}{l}{\textit{Rereading issues}} & & & & & \\ 
		&RRI_1  & 3.96 & 1.04   & 4.00  & -1.88  & 24  & 0.07  \\
		&RRI_2  & 3.39 & 1.02   & 3.50  & -0.33  & 24  & 0.75  \\
		\multicolumn{2}{l}{\textit{Stop \& resume issues}} &  & & & & \\ 
		&SRI_1  & 3.73 & 1.00   & 4.00  & -1.35  & 24  & 0.19  \\
		&SRI_2  & 3.46 & 0.95   & 4.00  & -1.54  & 24  & 0.13  \\
		&SRI_3  & 3.31 & 1.12   & 3.00  & -0.74  & 24  & 0.47  \\ 
		
		\bottomrule
		\multicolumn{8}{L}{^{*}Scales: 1 = very~low~ \textit{to}~ 5 = very~ high}\\
		\multicolumn{8}{L}{^{**}group~1: female~participants~(n=10); group~2: male~participants (n=16)}\\
	\end{tabular}
	\caption{Statistics about learners' rating of the effectiveness of the issues, and \textit{t-test} results based on gender difference}
	\label{table:issuesrating}
\end{table}

\begin{figure}
	\centering
	\includegraphics[width=.8\linewidth]{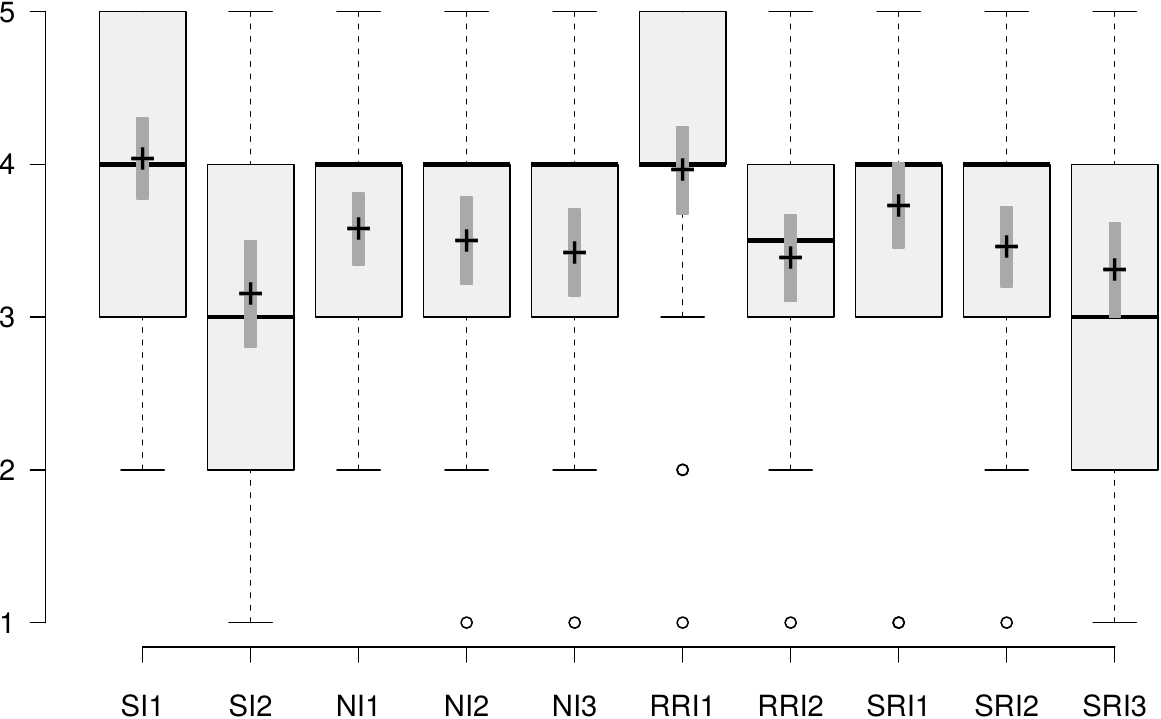}
	\caption{Learners' rating of the effectiveness of the issues (1 = very low, 5 = very high)}
	\label{fig:learners_bp}
\end{figure}

The descriptive statistics of the results are shown on \ref{table:issuesrating} and \ref{fig:learners_bp}. To examine the difference in ratings between male and female participants, we conducted an independent-samples \textit{t-test} analysis (with the significance level for the mean variation set at $p<0.05$). The results (the last three columns of \ref{table:issuesrating}) show no statistically significant difference between the two groups (female vs. male). There is a clear difference in the rating distribution for the different indicators in terms of skewness. 

All the issues had good rating values, with a median that is superior to the neutral point of 3 except for $SRI_3$ which median is equal to the neutral point.
This suggests that the participating students acknowledged that most of the detected issues correspond to real problems within the course that may hamper easily reading it and understanding its ideas. 
Issues related to course element popularity ($SI1$) and content complexity ($RRI_1$, $SRI_1$ and $NI_1$) were the most highly rated.

\begin{table}[!htp]
	\centering
	\small
	\begin{tabular}{@{}LLLLLLLLLL@{}}
		\toprule
		& SI_2 & NI_1 & NI_2  & NI_3 & RRI_1 & RRI_2  & SRI_1  & SRI_2  & SRI_3 \\ \midrule
		SI_1 & -0.30 & -0.13 & -0.02 & 0.23 & 0.01 & -0.18 & -0.19 & 0.16 & -0.46^{*} \\
		SI_2 &  & 0.04 & 0.28 & 0.08 & -0.18 & -0.17 & -0.09 & -0.50^{**} & 0.08 \\
		NI_1 &  &  & 0.52^{**} & 0.21 & 0.30 & 0.24 & 0.37 & 0.15 & 0.39^{*} \\
		NI_2 &  &  &   & 0.48^{*} & 0.35 & 0.38 & 0.36 & 0.20 & 0.31 \\
		NI_3 &  &  &   &  & 0.20 & 0.22 & -0.01 & 0.24 & 0.02 \\
		RRI_1 &  &  &   &  &  & 0.70^{***} & 0.64^{***} & 0.55^{**} & 0.42^{*} \\
		RRI_2 &  &  &   &  &  &   & 0.57^{**} & 0.51 & 0.24 \\
		SRI_1 &  &  &   &  &  &   &   & 0.64^{***} & 0.54^{**} \\
		SRI_2 &  &  &   &  &  &   &   &   & 0.35 \\ \midrule
		\multicolumn{10}{L}{Note. ^{*}p<.05, ^{**}p<.01, ^{***}p<.001}  \\
		\bottomrule
	\end{tabular}
	\caption{Inter-correlations (Spearman) among learners' ratings of the detected issues}
	\label{table:learnerscorr}
\end{table}

To further investigate the results, we used Pearson correlations between the issue ratings to determine if any association existed between them. The results that are shown on \ref{table:learnerscorr} reveal many significant relations. There are moderate negative correlations between score of stickiness issues and issues related to reading halts and nonlinear resumes ($SI_1$ and $SRI_3$ with $r=-0.46, p<.05$; $SI_2$ and $SRI_2$ with $r=-0.50, p<.01$). This suggests that the failure to attract learners' interest is also reflected by the tendency of learners to stop or to momentously interrupt reading. There are also moderate to strong positive correlations between rereading and reading stops and nonlinear resume issues ($RRI_1$ and $SRI_2$ with $r=0.64, p<.001$; $RRI_1$ and $SRI_2$ with $r=0.55, p<.01$; $RRI_1 $ and $SRI_3$ with $r=0.64, p<.05$; and $RRI_2$ and $SRI_1$ with $r=0.57, p<.001$). The majority of these issues reflect the difficulty for learners to grasp the meaning. Positive moderate inter-correlations within class issues exist: issues related to stops and resume ($SRI_1$ and $SRI_2$ with $r=0.64, p<.001$; $SRI_1$ and $SRI_3$ with $r=0.54, p<.01$), navigation issues ($NI_1$ and $NI_2$ with $r=0.52, p<.01$; $NI_2$ and $NI_3$ with $r=0.47, p<.05$), and rereading issues ($RRI_1$ and $RRI_2$ with $r=0.69, p<.001$). Finally, we found a weak correlation between two issues ($NI_1$ and $SRI_3$ with $r=0.39, p<.05$), both of them reflect possible disorientation due to the course structuring.

\subsection{Study 5 -- evaluation of the dashboard}
\label{sec:dashboard_eval}
\subsubsection{Protocol}
This study, conducted from April 5th to April 11th, 2017, aimed to evaluate the dashboard interface in terms of usability and acceptance. The authors first received their personal credentials for accessing the tool running on their courses. They were then instructed to access the interface, to complete the usability experiment and then to fill an acceptance questionnaire.
\paragraph{Usability experiment}
\label{sec:study6}
Usability assessment is a means of ensuring that an interactive system is adapted to users and their tasks and that there are no negative consequences of its use.
Evaluating interactive system usability is a fundamental step in the user-centered design process. Its goal is to assess the degree to which the system is effective (i.e., how well the system's performances meet the tasks for which it was designed), efficient (i.e., how much resources such as time or effort is required to use the system in order to achieve tasks for which the system was design), and favors positive attitudes and responses from the intended users \citep{Bevan2001}.

\begin{table}[hp]
	\centering
	\small
	\begin{tabular}{ll}
		\toprule
		\#&Task \\ \midrule
		T1&Follow the guided tour \\ 
		T2&Find a specific indicator value for a given chapter   \\
		T3&Find a specific issue, review it and mark it as not an actual problem. \\
		T4&Select an issue, add the suggestion as a task, modify the task and then mark it as done. \\ 
		T5&Display all the available indicators and issues to find chapters with the most issues.\\ 
		\bottomrule
	\end{tabular}
\caption{Authors' tasks}
\label{tab:tasks}
\end{table}

In this study, we aimed at evaluating the usability of the dashboard using a task-based experiment. 
The authors were asked to accomplish the set of tasks, described on Table~\ref{tab:tasks}, on their course dashboard. 
The task $T1$ consisted in obtaining assistance with the use of the tool. The tasks $T2$, $T3$ and $T4$ were related to performing diverse instructional design activities, by using features such as visualizing data, interpreting the analysis results, and taking relevant decisions. To perform the task $T2$, the author must scan the available data looking for a specific information. In the task $T3$, the author had to examine the source of a detected problem and then decide whether an intervention is appropriate. During the task $T4$, the author had to consider the suggestions provided before using them for designing and implementing appropriate corrective actions. The last task $T5$ involved some of the tool's advanced features to plan and execute complex pedagogical decisions.

The task list was integrated into the dashboard as a non-modal floating window that displays the tasks one-by-one in sequence and that collects the authors' answers. All the authors' actions were recorded. The experiment took an average time of 11 minutes.

\paragraph{Acceptance evaluation}
\label{sec:study7}
At the end of their task-based sessions, authors were invited to describe their willingness to adopt the dashboard in their revision work by answering an online questionnaire. We relied on the \textit{Technology Acceptance Model} (TAM) (Davis 1989), \citep{Davis1989}, a theoretical model that helps to predict user adoption of information technology.
Two measures of acceptance are posited by TAM: \textit{Perceived Usefulness} (\textit{PU}), and \textit{Perceived Ease of use} (PE). 
Perceived usefulness is ``the prospective user's subjective probability that using a specific application system will increase his or her job performance within an organizational context'', and perceived ease of use reflects ``the degree to which the prospective user expects the target system to be free of effort'' \cite[p.~985]{Davis1989}.
This model is among the most widely used in investigating technology acceptance, and has been validated by many empirical studies in the context of e-learning,  and in educational research (e.g., \citep{Cheung2013}). 
A statistical meta-analysis of TAM applied to 88 published studies showed it to be valid and robust \citep{King2006}. 

\begin{table}[!h]
	\centering
	\small	
	\begin{tabular}{ll}
		\toprule     
		\multicolumn{2}{c}{ \textit{Perceived Ease of Use (PE)}} \\ \midrule
		Q1& Learning to use CoReaDa would be easy for me \\ 
		Q2& I would find it easy to get CoReaDa to revise my course\\ 
		Q3& My interaction with CoReaDa would be clear and understandable\\ 
		Q4& I would find CoReaDa to be flexible to interact with\\ 
		Q5& It would be easy for me to become skillful at using CoReaDa\\ 
		Q6& I would find CoReaDa easy to use\\ 
		\toprule
		\multicolumn{2}{c}{\textit{Perceived Usefulness (PU)}} \\ \midrule
		Q7& Using CoReaDa to revise my course would enable me to accomplish tasks more quick\\ 
		Q8& Using CoReaDa would improve my revision performance\\ 
		Q9& Using CoReaDa to revise my courses would increase my productivity\\ 
		Q10& Using CoReaDa would enhance my effectiveness on course revision\\ 
		Q11& Using CoReaDa would make it easier to revise my courses\\ 
		Q12& I would find CoReaDa useful in revising my courses\\ 
		\bottomrule
	\end{tabular}
\caption{TAM questionnaire items}
\label{tab:tam}
\end{table}

Being correlated to predicted future usage, these two measures can reflect the authors' attitudes towards adopting the dashboard in their work. 
Consequently, based on TAM, we designed the Acceptance Questionnaire (Table~\ref{tab:tam}), the online version of which was provided to the authors for completion. They were asked to assess their level of agreement with each of the statements, using a 7-point Likert scale, ranging from 1 (\textit{strongly disagree}) to 7 (\textit{strongly agree}).  The questionnaire necessitated an average time of 6 minutes to complete.

\subsubsection{Results}
\paragraph{Dashboard usability}
Using the logs collected during the tasks-based experiment, we computed four performance metrics (results on Table~\ref{tab:learnresults}): 
\begin{enumerate}
	\item the \textit{success ratio} gives the ratio of tasks that were achieved successfully; 
	\item the \textit{average clicks} metric gives the average number of clicks that were performed to accomplish the task; 
	\item the \textit{average erroneous clicks} is the number of clicks that cannot help the author successfully do the task; and
	\item the \textit{average time in seconds} is the mean time spent by authors doing the task. 
\end{enumerate}

Different successful paths (in terms of click sequences and associated times) can be followed to achieve a given task.
Instead of using reference values, we analyzed the results in absolute terms regardless the underlying paths, since our objective is to evaluate whether or not authors were able to quickly and effectively use the dashboard from the first attempt.
\begin{table}[!h]
	\centering
	\footnotesize
	\begin{tabular}{rrrrr}
		\toprule
		& {Success ratio} & {Average \#clicks} & {Average \#erroneous clicks} & {Average time (sec.)} \\
		
		\midrule
		T1     & 100\%         & 20   & 0    & 171   \\
		T2     & 100\%         & 6   & 0.7   & 36    \\
		T3     & 100\%         & 4.3  & 1.1   & 27   \\
		T4     & 87\%          & 7   & 1.6   & 43   \\
		T5     & 75\%          & 13   & 3.1   & 89   \\
		\bottomrule    
		\multicolumn{5}{p{14cm}}{\textit{Used metrics}: Success ratio, Average number of clicks (\textit{avg. \#clicks}), Average number of erroneous clicks (\textit{avg. \#err.clicks}) and Average time spent in seconds (\textit{avg. time (sec.)})}             \\  
	\end{tabular}
	\smallskip
	\caption{Performance metrics computed from the tasks results}
	\label{tab:learnresults}
\end{table}

The results show that the tasks that involve options available by default on the interface ($T1$, $T2$, and $T3$) are performed easily, quickly and successfully. 
The guided visit ($T1$) contains $18$ mini-pages organized in sequence, and thus requires a significant amount of time with an average of $8.5$ seconds per mini-page. The authors pointed out the capital gain of this stage for rapidly learning to use the dashboard, which comforts our choice to prompt the guided visit automatically at the dashboard load. 
Tasks $T2$ and $T3$ are related to the use of the main features of the tool and require an average time of about half a minute to be accomplished, with an average of one erroneous manipulation click. The task $T4$ implies the use of the task manager and takes less than one minute to completion, with one failure (an author deleted a task instead of marking it as \textit{done}).
The task $T5$ required the use of advanced/hidden features of the tool (activating an advanced view) since the authors needed to figure out and locate the corresponding options. Two authors were not able to correctly find the chapter with more issues, they both provided chapters with fewer issues than the expected one. This task, despite its complexity, took an average of less than one minute and a half to be accomplished.

\paragraph{Dashboard acceptance}
The TAM scale ranges from 1 (strongly disagree) to 7 (strongly agree), with 4 (neither agree nor disagree) as the neutral midpoint. A score above 4 indicates that the respondent agrees to some extent with the corresponding statement.  The descriptive statistics of the results on Figure \ref{tab:tamresults} show that the mean scores for $PE$ were between $4.38$ and $5.25$, suggesting that a significant number of respondents had no major technical concerns when using the tool.  They also reveal that the respondents were not very dispersed around their mean scores on individual statements (standard deviations between $1.49$ and $1.92$).  The mean scores of the statements used to measure PU were between $4.75$ and $5.50$ with a standard deviation ranging from $1.31$ to $1.83$. This shows that most respondents tend to perceive the dashboard as having a rather positive impact in terms of effort, time and performance in conducting course reading analysis and revision tasks.

\begin{table}[]
	\centering\small
	\begin{tabular}{@{}p{1cm}p{1cm}p{1cm}p{2mm}|p{1cm}p{1cm}p{1cm}p{2mm}@{}}
		\toprule
		\multicolumn{4}{l|}{Perceived Ease of Use}      & \multicolumn{4}{l}{Perceived Usefulness} \\ \midrule
		\textit{Item}        & \textit{Mean*} & \textit{SD} &                    &\textit{Item} & \textit{Mean*} & \textit{SD}   &    \\ \midrule
		\textit{Q1} & 4.38 & {1.92}& & \textit{Q7}       & 4.75      & 1.83 &    \\
		\textit{Q2} & 5.00 & {1.60}& & \textit{Q8}       & 5.00      & 1.69 &    \\
		\textit{Q3} & 5.00 & {1.51}& & \textit{Q9}       & 4.75      & 1.67 &    \\
		\textit{Q4} & 4.88 & {1.55}& & \textit{Q10}      & 5.13      & 1.81 &    \\
		\textit{Q5} & 5.00 & {1.51}& & \textit{Q11}      & 5.25      & 1.49 &    \\
		\textit{Q6} & 5.25 & {1.49}& & \textit{Q12}      & 5.50      & 1.31 &    \\ \hline
		\textit{PE} & 4.91 & {1.62}& & \textit{PU}       & 5.06      & 1.49 &    \\\bottomrule
		\multicolumn{8}{l}{\textit{*Scale: 1=Strongly disagree to 7=Strongly agree}}             \\ 
	\end{tabular}
\caption{Results of the TAM questionnaire}
\label{tab:tamresults}
\end{table}

Items related to the perceived usefulness were combined into a composite variable PU ($mean = 5.06, std = 1.62$) and the items related to the perceived ease of use were combined into a composite variable PE ($mean = 4.91, std = 1.49$). 
A one-sample t-test (with the midpoint 4 as test value) for each of these variables indicated that the mean was significantly higher than the neutral midpoint (PU: $t = 1.736, df = 7, p=.125; PE: t = 1.736, df = 7, p=.125$). 

These results reflect a good authors' opinion about the studied aspects. Indeed, 77\% of the responses on the perceived usefulness of the dashboard were positive. This indicates that the dashboard is found convenient by authors for easily, quickly and effectively planning the revision of their courses. Moreover, 72\% of the responses expressed a positive level of agreement of the perceived ease of use.
This indicates that: (1) they found the tool easy to learn, to master and to use in a concise and convenient way; (2) using the tool could contribute to improving their performances since they have to deploy little effort to use it.

Within the comment section, five authors expressed their willingness to see such functionalities within their private space on the platform. An author said that this would help authors integrate course revision to their agenda as a routine. Another author, although having done successfully the experiments, suggested simplifying the interface even more, for a better user experience.

\section{Discussion}
\label{sec:eval_discuss}

The proposed method for delimiting learners' reading sessions involves computing dynamic page-specific thresholds based on learners' interactions with course elements and the characteristics of each page. These thresholds are recalculated each time new reading actions are logged, ensuring they adjust automatically to incoming reading data and reflect any changes in the course, such as page restructuring or content updates. Evaluation results show that this approach better represents learners' reading behavior and aligns with the expected statistical patterns of real reading sessions. The results also support our assertion that fixed-value methods (e.g., for session duration or page stay time) are unsuitable for educational websites, as unique threshold values fail to account for the specificities of courses and the diverse content on their pages. Nonetheless, further studies could refine this method by defining more accurate metrics for element complexity and by comparing the detected reading sessions with actual learner behavior.

Modeling reading activity using sessions enabled us to define indicators that describe the underlying behavioral processes of learners. These indicators not only help characterize reading behavior but also serve as cues to identify potential comprehension issues. The reading indicators evaluated with course authors were found to be highly relevant, with more than 60\% receiving positive ratings. These indicators were chosen for their robustness and popularity in online behavior analysis, although they represent only a subset of possible metrics. Future work should focus on identifying a comprehensive methodology that captures all relevant behavioral factors, which can provide insights into learners' understanding and the quality of the content they engage with.

These reading indicators are instrumental in detecting comprehension issues, which are then linked to appropriate revision actions. This process provides authors with valuable insights into how they can improve their courses. In many cases, these insights align with authors' expectations (more than 50\% of expected issues were detected), but in other cases, they challenge existing assumptions (about 50\% of authors' expectations were not confirmed). This often leads to new realizations, as nearly 80\% of detected issues were unexpected. Furthermore, 63\% of the suggested remediation actions were deemed useful and capable of resolving the reported issues. In contrast, the suggestions that were found unhelpful typically corresponded to issues that authors did not consider relevant. A study with learners demonstrated that this approach effectively identifies reading difficulties related to course structure and content, raising authors' awareness of the challenges learners face. This encourages authors to rethink their course design to enhance its readability and overall comprehension.

The dashboard was found to be intuitive and easy to use by the participating authors, as evidenced by the task-based study where authors succeeded in using it from their first attempt. This was further supported by the acceptance study, where authors expressed positive attitudes toward the dashboard and indicated a willingness to continue using it. According to the Technology Acceptance Model (TAM), both perceived usefulness and ease of use are strong predictors of user attitude and willingness to adopt a system. The study results suggest that the dashboard is both useful and easy to use, which contributes to its desirability among authors. However, as noted by \citep{Park2015}, a visualization tool is only effective if it drives meaningful changes in user behavior. Although learning analytics dashboards have gained popularity, studies have pointed out their limitations and potential pitfalls, especially when addressing the complex skill sets involved in learning \citep{Corrin2015, Teasley2017}. Therefore, more research is needed to assess the long-term impact of such dashboards on educational practices. Additionally, to design these tools in line with principles of educational psychology, large-scale empirical studies are required to establish more precise design guidelines \citep{Klerkx2014, Verbert2014}.

Despite the positive feedback received, this study has several limitations. The experiment was conducted with courses in informal learning settings through a self-directed learning platform. Moreover, while 125 authors participated in the first phase of the study, only 8 authors took part in the subsequent stages. Although \citet{Nielsen2000} suggests that five users are sufficient for reliable usability testing, broader studies involving a larger and more diverse group of authors, spanning different learning environments—such as formal, informal, and blended settings—are necessary to generalize our findings.

\section{Conclusion and Outlook}
This research aimed to provide means and tools for analyzing reading traces to better understand learners’ reading behavior and uncover knowledge that could improve authors' awareness of their course consumption. The central research objective was:
\textit{To investigate the use of reading analysis on learners' traces to identify comprehension issues and assist authors in improving their content accordingly.}
To address this goal, we proposed various solutions to answer our research questions.

The first research question (\textbf{RQ1}) asked: \textit{What is the general conceptual framework for supporting authors in improving their courses and addressing learners' understanding issues?} In response, we introduced a generic usage-based document reengineering model. This model conceptualizes how digital usage data can be leveraged for reengineering tasks on documents, providing authors with a framework for using learner feedback to continuously evolve their materials. This model lays the groundwork for systems that empower authors to respond effectively to learners' needs and enhance their courses.

The second research question (\textbf{RQ2}) was: \textit{What are the understanding issues?} We examined the structural factors affecting comprehension and identified document properties that significantly impact understanding. We introduced a document model analyzing both surface and conceptual structures, exploring factors such as readability and meaning. This analysis highlights where comprehension issues arise and the specific challenges faced by learners when engaging with course materials.

The third research question (\textbf{RQ3}) asked: \textit{What remediation can be proposed to authors based on understanding issues?} To answer this, we developed a model of revision activity and a taxonomy of revision primitives. These actions, ranging from enrichment to restructuring, help authors address comprehension issues identified in their materials. We proposed a set of revision suggestions aimed at guiding authors to improve problem areas and enhance learners' understanding of the course content.

The fourth research question (\textbf{RQ4}) was: \textit{How can comprehension issues be detected and linked to appropriate remediation actions?} To address this, we introduced a course revision approach that applies the reengineering model in an educational context. By analyzing learners' reading data and identifying reading indicators, this approach provides authors with insights into how their documents are consumed and suggests actions for improvement. While this data-driven approach offers valuable insights, it relies on indirect feedback, and incorporating learners' opinions could further improve accuracy. We also proposed a method for delimiting reading sessions based on dynamically computed thresholds, which automatically adjust as new data is collected. The evaluation confirmed that this method more accurately simulates reading behavior and highlights the limitations of fixed-value methods in educational contexts.

Additionally, we developed a taxonomy of session-based reading indicators that describe how learners engage with content. These indicators were validated by course authors, who found them useful for identifying areas where documents could be improved. A strategy was introduced to detect reading issues and propose remediation actions by analyzing indicator values and identifying outliers, signaling comprehension challenges. A learner study confirmed that this approach effectively identifies and addresses reading difficulties related to course structure and content.

The fifth research question (\textbf{RQ5}) was: \textit{What systems and tools can effectively support authors in improving their courses?} In response, we developed \textit{CoReaDa}, the "Course Reading Dashboard," which integrates the proposals for analyzing and revising online course reading. CoReaDa was co-designed with course authors, ensuring it met both functional and design requirements. A task-based experiment showed that the dashboard was intuitive and easy to use, with authors successfully adopting it from the first attempt. Positive feedback from an acceptance study further demonstrated authors' interest in using the dashboard for ongoing course improvement.

These results confirm the effectiveness of using learning analytics dashboards to support authors in analyzing reading behavior, identifying improvement opportunities, and performing relevant revisions. Despite the study's limitations, the findings indicate a strong interest among authors in understanding learners' comprehension and its impact on course quality. Overall, the studies demonstrate the usefulness of this approach in helping authors improve course content and enhance comprehension. Furthermore, these findings suggest that systems like CoReaDa can empower authors to make data-driven decisions, ultimately leading to courses that better support learner understanding and engagement.

Assuming that high-quality, relevant content plays a crucial role in the success of learning, we developed a learning analytics approach and tools that exploit learners' reading logs to assist course authors in improving the quality of their delivered content. This approach was instantiated on a major European e-learning platform.

Our proposals were grounded in theoretical frameworks from document engineering, reading comprehension, and content revision, which we applied to the learning analytics field. To the best of our knowledge, the findings from these domains have not been explicitly integrated into the e-learning context until now.

For the field of learning analytics, it is essential to draw from educational research and theories when developing applications \citep{Wise2014a}. Within the learning community, aligning learning design with learning analysis tools is a critical challenge that requires collective effort \citep{Bakharia2016, Lockyer2013, Echeverria2018}. This research serves as a starting point for further investigation into how reading analytics and learning dashboards can support course authors in enhancing the quality of course content over time.

There are numerous opportunities for future research stemming from the results of this dissertation. Our first direction is to conduct large-scale studies across diverse educational contexts to refine and validate our proposals. The concept of a session could also be further developed by considering learner-specific factors, such as background and reading pace, ensuring that reading traces more accurately reflect the behavior of different learner categories.

We also plan to integrate traces of other learning activities, such as video lectures and exercises, along with learner profiles and assessment data. This would create a comprehensive framework that supports not only the enhancement of course content quality but also the improvement of the entire pedagogical experience.

In terms of reading indicators, we aim to define more sophisticated metrics, or even multi-level indicators, which would be derived from combinations of existing ones. This would enable a more precise characterization of learners' reading behaviors and provide deeper insights into their needs. To achieve this, further experiments with both course authors and learners, including pre-tests and post-tests, will be necessary to identify which indicators are most predictive of learners' comprehension levels and overall performance.

Regarding the visual aspect of reading analytics with CoReaDa, a potential next step would be to design and implement a variety of customizable visualizations for more granular analysis of reading activity. We would also like to integrate editing tools that would allow authors to redesign their courses directly within the platform. This would represent the fourth level of assistance: automatically generating revised course versions based on analysis results. A particularly exciting initiative would be to create a modular framework of indicators and visualizations that can be easily adapted to different courses and platforms, thus enhancing interoperability across e-learning environments.

Technology is fundamentally reshaping the educational landscape. Ultimately, the goal of education and training is to equip learners with the skills and knowledge they can apply in relevant real-world situations. One of the key questions raised by the use of technology is its impact on the quality and effectiveness of teaching and learning outcomes.

We firmly believe that when technology fails to enhance the learning experience and outcomes, it may represent a change, but not a pedagogical innovation. For technology to be effective, investing in the right tools is not enough; educators, instructors, and course authors must also be supported to improve their practices and teaching methodologies.

\small
\bibliographystyle{apalike}
\bibliography{references}

\begin{thebibliography}{}

\bibitem[Adler et~al., 1998]{Adler1998}
Adler, A., Gujar, A., Harrison, B.~L., O'hara, K., and Sellen, A. (1998).
\newblock A diary study of work-related reading: design implications for digital reading devices.
\newblock In {\em Proceedings of the SIGCHI conference on Human factors in computing systems}, pages 241--248. ACM Press/Addison-Wesley Publishing Co.

\bibitem[Adler and Van~Doren, 2014]{Adler2014}
Adler, M.~J. and Van~Doren, C. (2014).
\newblock {\em How to read a book: The classic guide to intelligent reading}.
\newblock Simon and Schuster.

\bibitem[Akyel and Er{\c{c}}etin, 2009]{Akyel2009}
Akyel, A. and Er{\c{c}}etin, G. (2009).
\newblock Hypermedia reading strategies employed by advanced learners of english.
\newblock {\em System}, 37(1):136--152.

\bibitem[Al~Madi and Khan, 2016]{Al2016}
Al~Madi, N.~S. and Khan, J.~I. (2016).
\newblock Measuring learning performance and cognitive activity during multimodal comprehension.
\newblock In {\em 2016 7th International Conference on Information and Communication Systems (ICICS)}, pages 50--55. IEEE.

\bibitem[Alderson, 2000]{Alderson2000}
Alderson, J.~C. (2000).
\newblock Technology in testing: The present and the future.
\newblock {\em System}, 28(4):593--603.

\bibitem[Allal et~al., 2004]{Allal2004}
Allal, L., Chanquoy, L., and Largy, P. (2004).
\newblock {\em Revision cognitive and instructional processes}, volume~8.
\newblock Springer.

\bibitem[Arce et~al., 2014]{Arce2014}
Arce, T., Rom{\'a}n, P.~E., Vel{\'a}squez, J., and Parada, V. (2014).
\newblock Identifying web sessions with simulated annealing.
\newblock {\em Expert Systems with Applications}, 41(4):1593--1600.

\bibitem[Arnold and Pistilli, 2012]{Arnold2012}
Arnold, K.~E. and Pistilli, M.~D. (2012).
\newblock Course signals at purdue: Using learning analytics to increase student success.
\newblock In {\em Proceedings of the 2nd international conference on learning analytics and knowledge}, pages 267--270. ACM.

\bibitem[Aubert et~al., 2008]{Aubert2008}
Aubert, O., Champin, P.-A., Pri{\'e}, Y., and Richard, B. (2008).
\newblock Canonical processes in active reading and hypervideo production.
\newblock {\em Multimedia Systems}, 14(6):427--433.

\bibitem[Baccino et~al., 2008]{Baccino2008}
Baccino, T., Salmer{\'o}n, L., and Ca{\~n}as, J. (2008).
\newblock La lecture des hypertextes.
\newblock In {\em Ergonomie des documents {\'e}lectroniques}, pages 9--34. Presses Universitaires de France.

\bibitem[Baker et~al., 2010]{Baker2010}
Baker, R. et~al. (2010).
\newblock Data mining for education.
\newblock {\em International encyclopedia of education}, 7(3):112--118.

\bibitem[Baker and Inventado, 2014]{Baker2014}
Baker, R.~S. and Inventado, P.~S. (2014).
\newblock Educational data mining and learning analytics.
\newblock In {\em Learning analytics}, pages 61--75. Springer.

\bibitem[Bakharia et~al., 2016]{Bakharia2016}
Bakharia, A., Corrin, L., de~Barba, P., Kennedy, G., Ga\v{s}evi\'{c}, D., Mulder, R., Williams, D., Dawson, S., and Lockyer, L. (2016).
\newblock A conceptual framework linking learning design with learning analytics.
\newblock In {\em Proceedings of the Sixth International Conference on Learning Analytics \& Knowledge}, LAK '16, pages 329--338, New York, NY, USA. ACM.

\bibitem[Bakhshinategh et~al., 2017]{Bak2017}
Bakhshinategh, B., Zaiane, O.~R., ElAtia, S., and Ipperciel, D. (2017).
\newblock Educational data mining applications and tasks: A survey of the last 10 years.
\newblock {\em Education and Information Technologies}, pages 1--17.

\bibitem[Barr and Gunawardena, 2012]{Barr2012}
Barr, J. and Gunawardena, A. (2012).
\newblock Classroom salon: a tool for social collaboration.
\newblock In {\em Proceedings of the 43rd ACM technical symposium on Computer Science Education}, pages 197--202. ACM.

\bibitem[Bawden et~al., 2008]{Bawden2008}
Bawden, D. et~al. (2008).
\newblock Origins and concepts of digital literacy.
\newblock {\em Digital literacies: Concepts, policies and practices}, 30:17--32.

\bibitem[Berendt et~al., 2001]{Berendt2001}
Berendt, B., Mobasher, B., Spiliopoulou, M., and Wiltshire, J. (2001).
\newblock Measuring the accuracy of sessionizers for web usage analysis.
\newblock In {\em Workshop on Web Mining}, pages 7--14.

\bibitem[Bevan, 2001]{Bevan2001}
Bevan, N. (2001).
\newblock International standards for hci and usability.
\newblock {\em International journal of human-computer studies}, 55(4):533--552.

\bibitem[Briet, 1951]{Briet1951}
Briet, S. (1951).
\newblock {\em Qu'est-ce que la documentation?}
\newblock \'Editions documentaires et industrielles, Paris.

\bibitem[Britton, 1994]{Britton1994}
Britton, B.~K. (1994).
\newblock {\em Understanding expository text: Building mental structures to induce insights}.
\newblock Academic Press.

\bibitem[Brouns et~al., 2015]{Brouns2015}
Brouns, F., Zorrilla~Pantale{\'o}n, M.~E., Solana~Gonz{\'a}lez, P., Cobo~Ortega, {\'A}., Collantes~Via{\~n}a, M., Rodr{\'\i}guez~Hoyo, C., De~Lima~Silva, M., Marta-Lazo, C., Gabelas~Barroso, J.~A., Arranz, P., et~al. (2015).
\newblock Eco d2. 5 learning analytics requirements and metrics report.

\bibitem[Bruce et~al., 1981]{Bruce1981}
Bruce, B., Rubin, A., and Starr, K. (1981).
\newblock Why readability formulas fail.
\newblock {\em IEEE Transactions on Professional Communication}, PC-24(1):50--52.

\bibitem[Brusilovsky et~al., 2011]{Brusilovsky2011}
Brusilovsky, P., Hsiao, I.-H., and Folajimi, Y. (2011).
\newblock Quizmap: open social student modeling and adaptive navigation support with treemaps.
\newblock In {\em Towards Ubiquitous Learning}, pages 71--82. Springer.

\bibitem[Buckland, 1997]{Buckland1997}
Buckland, M.~K. (1997).
\newblock {What is a "document"}.
\newblock {\em Journal of the American Society for Information Science}, 48(9):804--809.

\bibitem[Bulut, 2015]{Bulut2015}
Bulut, M. (2015).
\newblock The impact of functional reading instruction on individual and social life.
\newblock {\em Educational Research and Reviews}, 10(4):462--470.

\bibitem[Bush et~al., 1945]{Bush1945}
Bush, V. et~al. (1945).
\newblock As we may think.
\newblock {\em The atlantic monthly}, 176(1):101--108.

\bibitem[Butkiewicz et~al., 2011]{Butkiewicz2011}
Butkiewicz, M., Madhyastha, H.~V., and Sekar, V. (2011).
\newblock Understanding website complexity: measurements, metrics, and implications.
\newblock In {\em Proceedings of the 2011 ACM SIGCOMM conference on Internet measurement conference}, pages 313--328. ACM.

\bibitem[Calisir and Gurel, 2003]{Calisir2003}
Calisir, F. and Gurel, Z. (2003).
\newblock Influence of text structure and prior knowledge of the learner on reading comprehension, browsing and perceived control.
\newblock {\em Computers in Human Behavior}, 19(2):135--145.

\bibitem[Carifio and Perla, 2007]{Carifio2007}
Carifio, J. and Perla, R.~J. (2007).
\newblock Ten common misunderstandings, misconceptions, persistent myths and urban legends about likert scales and likert response formats and their antidotes.
\newblock {\em Journal of Social Sciences}, 3(3):106--116.

\bibitem[Champin et~al., 2012]{Champin2012b}
Champin, P.-A., Cordier, A., Lavou{\'e}, {\'E}., Lefevre, M., and Skaf-Molli, H. (2012).
\newblock User assistance for collaborative knowledge construction.
\newblock In {\em Proceedings of the 21st International Conference on World Wide Web}, pages 1065--1074. ACM.

\bibitem[Charleer et~al., 2017]{Charleer2017}
Charleer, S., Moere, A.~V., Klerkx, J., Verbert, K., and De~Laet, T. (2017).
\newblock Learning analytics dashboards to support adviser-student dialogue.
\newblock {\em IEEE Transactions on Learning Technologies}.

\bibitem[Charleer et~al., 2014]{Charleer2014}
Charleer, S., Odriozola, S., Luis, J., Klerkx, J., and Duval, E. (2014).
\newblock Larae: Learning analytics reflection \& awareness environment.
\newblock In {\em CEUR Workshop Proceedings}, volume 1238, pages 85--87. CEUR-WS.

\bibitem[Chen et~al., 2014]{Chen2014}
Chen, G., Cheng, W., Chang, T.-W., Zheng, X., and Huang, R. (2014).
\newblock A comparison of reading comprehension across paper, computer screens, and tablets: Does tablet familiarity matter?
\newblock {\em Journal of Computers in Education}, 1(2):213--225.

\bibitem[Cheung and Vogel, 2013]{Cheung2013}
Cheung, R. and Vogel, D. (2013).
\newblock Predicting user acceptance of collaborative technologies: An extension of the technology acceptance model for e-learning.
\newblock {\em Computers \& Education}, 63:160--175.

\bibitem[Cho and MacArthur, 2010]{Cho2010}
Cho, K. and MacArthur, C. (2010).
\newblock Student revision with peer and expert reviewing.
\newblock {\em Learning and Instruction}, 20(4):328--338.

\bibitem[Cho and MacArthur, 2011]{Cho2011}
Cho, K. and MacArthur, C. (2011).
\newblock Learning by reviewing.
\newblock {\em Journal of Educational Psychology}, 103(1):73.

\bibitem[Chou, 2012]{Chou2012}
Chou, I.-C. (2012).
\newblock Understanding on-screen reading behaviors in academic contexts: a case study of five graduate english-as-a-second-language students.
\newblock {\em Computer Assisted Language Learning}, 25(5):411--433.

\bibitem[Christophides, 1998]{Christophides1998}
Christophides, V. (1998).
\newblock Electronic document management systems.

\bibitem[Cocea and Weibelzahl, 2011]{Cocea2011}
Cocea, M. and Weibelzahl, S. (2011).
\newblock Disengagement detection in online learning: validation studies and perspectives.
\newblock {\em IEEE transactions on learning technologies}, 4(2):114--124.

\bibitem[Coiro, 2007]{Coiro2007}
Coiro, J. (2007).
\newblock {\em Exploring changes to reading comprehension on the Internet: Paradoxes and possibilities for diverse adolescent readers}.
\newblock University of Connecticut.

\bibitem[Coiro, 2012]{Coiro2012}
Coiro, J. (2012).
\newblock Understanding dispositions toward reading on the internet.
\newblock {\em Journal of adolescent \& adult Literacy}, 55(7):645--648.

\bibitem[Conklin, 1987]{Conklin1987}
Conklin, J. (1987).
\newblock Hypertext: An introduction and survey in ieee computer, 20 (9), 17--41.

\bibitem[Cook and Thomas, 2005]{Cook2005}
Cook, K.~A. and Thomas, J.~J. (2005).
\newblock Illuminating the path: The research and development agenda for visual analytics.
\newblock Technical report, Pacific Northwest National Lab.(PNNL), Richland, WA (United States).

\bibitem[Cooper, 2012]{Cooper2012}
Cooper, A. (2012).
\newblock A brief history of analytics, series 1 (9).
\newblock {\em JISC CETIS Analytics}, 1(9).

\bibitem[Cormier and Siemens, 2010]{Cormier2010}
Cormier, D. and Siemens, G. (2010).
\newblock Through the open door: Open courses as research, learning, and engagement.
\newblock {\em EDUCAUSE Review}, 45(4):30--39.

\bibitem[Corrin and de~Barba, 2015]{Corrin2015}
Corrin, L. and de~Barba, P. (2015).
\newblock How do students interpret feedback delivered via dashboards?
\newblock In {\em Proceedings of the Fifth International Conference on Learning Analytics And Knowledge}, pages 430--431. ACM.

\bibitem[Couzijn and Rijlaarsdam, 2005]{Couzijn2005}
Couzijn, M. and Rijlaarsdam, G. (2005).
\newblock Learning to write instructive texts by reader observation and written feedback.
\newblock In Rijlaarsdam, G., van~den Bergh, H., and Couzijn, M., editors, {\em Effective Learning and Teaching of Writing: A Handbook of Writing in Education}, pages 209--240. Springer Netherlands, Dordrecht.

\bibitem[Cross, 2011]{Cross2011}
Cross, J. (2011).
\newblock {\em Informal learning: Rediscovering the natural pathways that inspire innovation and performance}.
\newblock John Wiley \& Sons.

\bibitem[Crossley et~al., 2007]{Crossley2007}
Crossley, S.~A., Dufty, D.~F., McCarthy, P.~M., and McNamara, D.~S. (2007).
\newblock Toward a new readability: A mixed model approach.
\newblock In {\em Proceedings of the Annual Meeting of the Cognitive Science Society}.

\bibitem[Crossley et~al., 2017]{Crossley2017}
Crossley, S.~A., Skalicky, S., Dascalu, M., McNamara, D.~S., and Kyle, K. (2017).
\newblock Predicting text comprehension, processing, and familiarity in adult readers: New approaches to readability formulas.
\newblock {\em Discourse Processes}, pages 1--20.

\bibitem[Crystal, 2010]{Crystal2010}
Crystal, D. (2010).
\newblock The changing nature of text: a linguistic perspective.
\newblock In {\em Text comparison and digital creativity}, pages 227--252. Brill.

\bibitem[Dale and Chall, 1949]{Dale1949}
Dale, E. and Chall, J.~S. (1949).
\newblock The concept of readability.
\newblock {\em Elementary English}, 26(1):19--26.

\bibitem[Dascalu et~al., 2014]{Dascalu2014}
Dascalu, M., Dessus, P., Bianco, M., Trausan-Matu, S., and Nardy, A. (2014).
\newblock Mining texts, learner productions and strategies with readerbench.
\newblock In {\em Educational Data Mining: Applications and Trends}, pages 345--377. Springer International Publishing, Cham.

\bibitem[Davis, 1989]{Davis1989}
Davis, F.~D. (1989).
\newblock Perceived usefulness, perceived ease of use, and user acceptance of information technology.
\newblock {\em MIS quarterly}, pages 319--340.

\bibitem[Daxenberger and Gurevych, 2012]{Daxenberger2012}
Daxenberger, J. and Gurevych, I. (2012).
\newblock A corpus-based study of edit categories in featured and non-featured wikipedia articles.
\newblock In {\em COLING}, pages 711--726.

\bibitem[Delgado et~al., 2018]{Delgado2018}
Delgado, P., Vargas, C., Ackerman, R., and Salmer{\'o}n, L. (2018).
\newblock Don't throw away your printed books: A meta-analysis on the effects of reading media on reading comprehension.
\newblock {\em Educational Research Review}.

\bibitem[Dell et~al., 2008]{Dell2008}
Dell, R.~F., Roman, P.~E., and Velasquez, J.~D. (2008).
\newblock Web user session reconstruction using integer programming.
\newblock In {\em Proceedings of the 2008 IEEE/WIC/ACM International Conference on Web Intelligence and Intelligent Agent Technology-Volume 01}, pages 385--388. IEEE Computer Society.

\bibitem[Dillon, 1992]{Dillon1992}
Dillon, A. (1992).
\newblock Reading from paper versus screens: A critical review of the empirical literature.
\newblock {\em Ergonomics}, 35(10):1297--1326.

\bibitem[Do-Lenh, 2012]{Do2012}
Do-Lenh, S. (2012).
\newblock {\em Supporting reflection and classroom orchestration with tangible tabletops}.
\newblock PhD thesis, {\'E}cole Polytechnique Federale De Laussane.

\bibitem[Dumais et~al., 2014]{Dumais2014}
Dumais, S., Jeffries, R., Russell, D.~M., Tang, D., and Teevan, J. (2014).
\newblock Understanding user behavior through log data and analysis.
\newblock In {\em Ways of Knowing in HCI}, pages 349--372. Springer.

\bibitem[D{\"u}ndar and Ak{\c{c}}ay{\i}r, 2017]{Dundar2017}
D{\"u}ndar, H. and Ak{\c{c}}ay{\i}r, M. (2017).
\newblock Tablet vs. paper: The effect on learners' reading performance.
\newblock {\em International Electronic Journal of Elementary Education}, 4(3):441--450.

\bibitem[Duran, 2013]{Duran2013}
Duran, E. (2013).
\newblock Investigation on views and attitudes of students in faculty of education about reading and writing on screen.
\newblock {\em Educational Research and Reviews}, 8(5):203.

\bibitem[Duval, 2011]{Duval2011}
Duval, E. (2011).
\newblock {Attention please!: learning analytics for visualization and recommendation}.
\newblock In {\em the 1st International Conference on Learning Analytics}.

\bibitem[Early and Saidy, 2014]{Early2014}
Early, J.~S. and Saidy, C. (2014).
\newblock A study of a multiple component feedback approach to substantive revision for secondary ell and multilingual writers.
\newblock {\em Reading and Writing}, 27(6):995--1014.

\bibitem[Echeverria et~al., 2018]{Echeverria2018}
Echeverria, V., Martinez-Maldonado, R., Granda, R., Chiluiza, K., Conati, C., and Shum, S.~B. (2018).
\newblock Driving data storytelling from learning design.
\newblock In {\em Proceedings of the 8th International Conference on Learning Analytics and Knowledge}, pages 131--140. ACM.

\bibitem[Edwards et~al., 2017]{Edwards2017}
Edwards, R.~L., Davis, S.~K., Hadwin, A.~F., and Milford, T.~M. (2017).
\newblock Using predictive analytics in a self-regulated learning university course to promote student success.
\newblock In {\em Proceedings of the Seventh International Learning Analytics \& Knowledge Conference}, LAK '17, pages 556--557, New York, NY, USA. ACM.

\bibitem[Endsley, 2016]{Endsley2016}
Endsley, M.~R. (2016).
\newblock {\em Designing for situation awareness: An approach to user-centered design}.
\newblock CRC press.

\bibitem[Fagen and Kamin, 2012]{Fagen2012}
Fagen, W. and Kamin, S. (2012).
\newblock Developing device-independent applications for active and collaborative learning with the slice framework.
\newblock In {\em Proceedings of world conference on educational multimedia, hypermedia and telecommunications}, pages 1565--1572.

\bibitem[Faigley and Witte, 1981]{Faigley1981}
Faigley, L. and Witte, S. (1981).
\newblock Analyzing revision.
\newblock {\em College Composition and Communication}, 32(4):400--414.

\bibitem[Farinosi et~al., 2016]{Farinosi2016}
Farinosi, M., Lim, C., and Roll, J. (2016).
\newblock Book or screen, pen or keyboard? a cross-cultural sociological analysis of writing and reading habits basing on germany, italy and the uk.
\newblock {\em Telematics and Informatics}, 33(2):410--421.

\bibitem[Fayyad et~al., 1996]{Fayyad1996}
Fayyad, U.~M., Piatetsky-Shapiro, G., Smyth, P., and Uthurusamy, R. (1996).
\newblock {\em Advances in knowledge discovery and data mining}, volume~21.
\newblock AAAI press Menlo Park.

\bibitem[Few, 2013]{Few2013}
Few, S. (2013).
\newblock {\em Information Dashboard Design: Displaying data for at-a-glance monitoring}.
\newblock Analytics Press.

\bibitem[Fitzgerald, 1987]{Fitzgerald1987}
Fitzgerald, J. (1987).
\newblock Research on revision in writing.
\newblock {\em Review of educational research}, 57(4):481--506.

\bibitem[Flesch, 1943]{Flesch1943}
Flesch, R.~F. (1943).
\newblock Marks of readable style : a study in adult education.
\newblock {\em Teachers College Contributions to Education}.

\bibitem[Flower et~al., 1986]{Flower1986}
Flower, L., Hayes, J.~R., Carey, L., Schriver, K., and Stratman, J. (1986).
\newblock Detection, diagnosis, and the strategies of revision.
\newblock {\em College composition and communication}, 37(1):16--55.

\bibitem[Fran\c{c}ois and Miltsakaki, 2012]{Francois2012}
Fran\c{c}ois, T. and Miltsakaki, E. (2012).
\newblock Do {NLP} and machine learning improve traditional readability formulas?
\newblock In {\em Proceedings of the First Workshop on Predicting and Improving Text Readability for Target Reader Populations}, PITR '12, pages 49--57, Stroudsburg, PA, USA. Association for Computational Linguistics.

\bibitem[Frawley et~al., 1992]{Frawley1992}
Frawley, W.~J., Piatetsky-Shapiro, G., and Matheus, C.~J. (1992).
\newblock Knowledge discovery in databases: An overview.
\newblock {\em AI magazine}, 13(3):57.

\bibitem[Ga{\v{s}}evi{\'c} et~al., 2015]{Gavsevic2015}
Ga{\v{s}}evi{\'c}, D., Dawson, S., and Siemens, G. (2015).
\newblock Let’s not forget: Learning analytics are about learning.
\newblock {\em TechTrends}, 59(1):64--71.

\bibitem[Geurts, 2010]{Geurts2010}
Geurts, J. (2010).
\newblock {\em A document engineering model and processing framework for multimedia documents}.
\newblock PhD thesis, Technische Universiteit Eindhoven.

\bibitem[Govaerts et~al., 2012]{Govaerts2012}
Govaerts, S., Verbert, K., Duval, E., and Pardo, A. (2012).
\newblock The student activity meter for awareness and self-reflection.
\newblock In {\em CHI'12 Extended Abstracts on Human Factors in Computing Systems}, pages 869--884. ACM.

\bibitem[Grabe, 2009]{Grabe2009}
Grabe, W. (2009).
\newblock {\em Reading in a second language: Moving from theory to practice}.
\newblock Ernst Klett Sprachen.

\bibitem[Graesser et~al., 2004]{Graesser2004}
Graesser, A.~C., McNamara, D.~S., Louwerse, M.~M., and Cai, Z. (2004).
\newblock Coh-metrix: Analysis of text on cohesion and language.
\newblock {\em Behavior research methods, instruments, \& computers}, 36(2):193--202.

\bibitem[Graham, 2009]{graham2009blended}
Graham, C.~R. (2009).
\newblock Blended learning models.
\newblock In {\em Encyclopedia of Information Science and Technology, Second Edition}, pages 375--382. IGI Global.

\bibitem[Gray, 1935]{Gray1935}
Gray, W.~S. (1935).
\newblock {\em What makes a book readable. With special reference to adults of limited reading ability. An initial study}.
\newblock University of Chicago Press.

\bibitem[Green et~al., 2010]{Green2010}
Green, T.~D., Perera, R.~A., Dance, L.~A., and Myers, E.~A. (2010).
\newblock Impact of presentation mode on recall of written text and numerical information: Hard copy versus electronic.
\newblock {\em North American Journal of Psychology}, 12(2).

\bibitem[Gros and Garc{\'\i}a-Pe{\~n}alvo, 2016]{Gros2016}
Gros, B. and Garc{\'\i}a-Pe{\~n}alvo, F.~J. (2016).
\newblock Future trends in the design strategies and technological affordances of e-learning.
\newblock {\em Learning, Design, and Technology: An International Compendium of Theory, Research, Practice, and Policy}, pages 1--23.

\bibitem[Gross, 2015]{Gross2015}
Gross, R. (2015).
\newblock {\em Psychology: The science of mind and behaviour 7th edition}.
\newblock Hodder Education.

\bibitem[Gunning, 1969]{Gunning1969}
Gunning, R. (1969).
\newblock The fog index after twenty years.
\newblock {\em Journal of Business Communication}, 6(2):3--13.

\bibitem[Gutierrez-Santos et~al., 2012]{Gutierrez2012}
Gutierrez-Santos, S., Geraniou, E., Pearce-Lazard, D., and Poulovassilis, A. (2012).
\newblock Design of teacher assistance tools in an exploratory learning environment for algebraic generalization.
\newblock {\em IEEE Transactions on Learning Technologies}, 5(4):366--376.

\bibitem[Guy, 2009]{Guy2009}
Guy, R. (2009).
\newblock {\em The evolution of mobile teaching and learning}.
\newblock Informing Science.

\bibitem[Gwizdka and Spence, 2007]{Gwizdka2007}
Gwizdka, J. and Spence, I. (2007).
\newblock Implicit measures of lostness and success in web navigation.
\newblock {\em Interacting with Computers}, 19(3):357--369.

\bibitem[Haenggi and Perfetti, 1992]{Haenggi1992}
Haenggi, D. and Perfetti, C.~A. (1992).
\newblock Individual differences in reprocessing of text.
\newblock {\em Journal of Educational Psychology}, 84(2):182.

\bibitem[Hauger et~al., 2011]{Hauger2011}
Hauger, D., Paramythis, A., and Weibelzahl, S. (2011).
\newblock Using browser interaction data to determine page reading behavior.
\newblock In Konstan, J.~A., Conejo, R., Marzo, J.~L., and Oliver, N., editors, {\em User Modeling, Adaption and Personalization}, pages 147--158, Berlin, Heidelberg. Springer Berlin Heidelberg.

\bibitem[Hayes, 2000]{Hayes2000}
Hayes, J.~R. (2000).
\newblock A new framework for understanding cognition and.
\newblock {\em Perspectives on writing: Research, theory, and practice}, page~6.

\bibitem[Hayes and Chenoweth, 2006]{Hayes2006}
Hayes, J.~R. and Chenoweth, N.~A. (2006).
\newblock Is working memory involved in the transcribing and editing of texts?
\newblock {\em Written Communication}, 23(2):135--149.

\bibitem[Hayes et~al., 1987]{Hayes1987}
Hayes, J.~R., Flower, L., Schriver, K.~A., Stratman, J., Carey, L., et~al. (1987).
\newblock Cognitive processes in revision.
\newblock {\em Advances in applied psycholinguistics}, 2:176--240.

\bibitem[Holstein et~al., 2017]{Holstein2017}
Holstein, K., McLaren, B.~M., and Aleven, V. (2017).
\newblock Intelligent tutors as teachers' aides: Exploring teacher needs for real-time analytics in blended classrooms.
\newblock In {\em Proceedings of the Seventh International Learning Analytics \&\#38; Knowledge Conference}, LAK '17, pages 257--266, New York, NY, USA. ACM.

\bibitem[Jabr, 2013]{Jabr2013}
Jabr, F. (2013).
\newblock The reading brain in the digital age: The science of paper versus screens.
\newblock {\em Scientific American}, 11.

\bibitem[Jourdan et~al., 1998]{Jourdan1998}
Jourdan, M., Laya{\"\i}da, N., Roisin, C., Sabry-Isma{\"\i}l, L., and Tardif, L. (1998).
\newblock Madeus, and authoring environment for interactive multimedia documents.
\newblock In {\em Proceedings of the sixth ACM international conference on Multimedia}, pages 267--272. ACM.

\bibitem[Kaplan and Haenlein, 2016]{Kaplan2016}
Kaplan, A.~M. and Haenlein, M. (2016).
\newblock Higher education and the digital revolution: About moocs, spocs, social media, and the cookie monster.
\newblock {\em Business Horizons}, 59(4):441--450.

\bibitem[Keim et~al., 2008]{Keim2008}
Keim, D., Andrienko, G., Fekete, J.-D., Gorg, C., Kohlhammer, J., and Melan{\c{c}}on, G. (2008).
\newblock Visual analytics: Definition, process, and challenges.
\newblock {\em Lecture notes in computer science}, 4950:154--176.

\bibitem[Kerly et~al., 2008]{Kerly2008}
Kerly, A., Ellis, R., and Bull, S. (2008).
\newblock Calmsystem: a conversational agent for learner modelling.
\newblock {\em Knowledge-Based Systems}, 21(3):238--246.

\bibitem[Khalil and Ebner, 2015]{Khalil2015}
Khalil, M. and Ebner, M. (2015).
\newblock Learning analytics: principles and constraints.
\newblock In {\em EdMedia: World Conference on Educational Media and Technology}, pages 1789--1799. Association for the Advancement of Computing in Education (AACE).

\bibitem[Kim et~al., 2015]{Kim2015}
Kim, J., Jo, I.~H., and Park, Y. (2015).
\newblock {Effects of learning analytics dashboard: analyzing the relations among dashboard utilization, satisfaction, and learning achievement}.
\newblock {\em Asia Pacific Education Review}, 17(1):13--24.

\bibitem[Kincaid et~al., 1975]{Kincaid1975}
Kincaid, J.~P., Fishburne~Jr, R.~P., Rogers, R.~L., and Chissom, B.~S. (1975).
\newblock Derivation of new readability formulas (automated readability index, fog count and flesch reading ease formula) for navy enlisted personnel.
\newblock Technical report, Naval Technical Training Command Millington TN Research Branch.

\bibitem[King and He, 2006]{King2006}
King, W.~R. and He, J. (2006).
\newblock A meta-analysis of the technology acceptance model.
\newblock {\em Information \& management}, 43(6):740--755.

\bibitem[Klerkx et~al., 2014]{Klerkx2014}
Klerkx, J., Verbert, K., and Duval, E. (2014).
\newblock Enhancing learning with visualization techniques.
\newblock In {\em Handbook of research on educational communications and technology}, pages 791--807. Springer.

\bibitem[Knowles, 1975]{Knowles1975}
Knowles, M.~S. (1975).
\newblock {\em Self-directed learning}.
\newblock New York: Association Press.

\bibitem[Kong et~al., 2018]{Kong2018}
Kong, Y., Seo, Y.~S., and Zhai, L. (2018).
\newblock Comparison of reading performance on screen and on paper: A meta-analysis.
\newblock {\em Computers \& Education}, 123:138--149.

\bibitem[Kosara, 2007]{Kosara2007}
Kosara, R. (2007).
\newblock Visualization criticism-the missing link between information visualization and art.
\newblock In {\em Information Visualization, 2007. IV'07. 11th International Conference}, pages 631--636. IEEE.

\bibitem[Kurata et~al., 2017]{Kurata2017}
Kurata, K., Ishita, E., Miyata, Y., and Minami, Y. (2017).
\newblock Print or digital? reading behavior and preferences in japan.
\newblock {\em Journal of the Association for Information Science and Technology}, 68(4):884--894.

\bibitem[Laha, 2010]{Laha2010}
Laha, A. (2010).
\newblock On the issues of building information warehouses.
\newblock In {\em Proceedings of the Third Annual ACM Bangalore Conference}, page~2. ACM.

\bibitem[Lee and Tedder, 2003]{Lee2003}
Lee, M.~J. and Tedder, M.~C. (2003).
\newblock The effects of three different computer texts on readers' recall: based on working memory capacity.
\newblock {\em Computers in Human Behavior}, 19(6):767--783.

\bibitem[Le{\'o}n and Carretero, 1995]{Leon1995}
Le{\'o}n, J.~A. and Carretero, M. (1995).
\newblock Intervention in comprehension and memory strategies: Knowledge and use of text structure.
\newblock {\em Learning and instruction}, 5(3):203--220.

\bibitem[Leony et~al., 2012]{Leony2012}
Leony, D., Pardo, A., de~la Fuente~Valent{\'\i}n, L., de~Castro, D.~S., and Kloos, C.~D. (2012).
\newblock Glass: a learning analytics visualization tool.
\newblock In {\em Proceedings of the 2nd international conference on learning analytics and knowledge}, pages 162--163. ACM.

\bibitem[Levine-Clark, 2015]{Levine2015}
Levine-Clark, M. (2015).
\newblock What do our users think about ebooks and print books?: 10 years of survey data at the university of denver.

\bibitem[Levy, 2016]{Levy2016}
Levy, D.~M. (2016).
\newblock {\em Scrolling forward: Making sense of documents in the digital age}.
\newblock Skyhorse Publishing, Inc.

\bibitem[Lin et~al., 2015]{Lin2015}
Lin, C.-L., Wang, M.-J.~J., and Kang, Y.-Y. (2015).
\newblock The evaluation of visuospatial performance between screen and paper.
\newblock {\em Displays}, 39:26--32.

\bibitem[Lin, 2003]{Lin2003}
Lin, D.-Y.~M. (2003).
\newblock Age differences in the performance of hypertext perusal as a function of text topology.
\newblock {\em Behaviour \& information technology}, 22(4):219--226.

\bibitem[Liu and Ram, 2011]{Liu2011}
Liu, J. and Ram, S. (2011).
\newblock Who does what: Collaboration patterns in the wikipedia and their impact on article quality.
\newblock {\em ACM Transactions on Management Information Systems (TMIS)}, 2(2):11.

\bibitem[Liu, 2005]{Liu2005}
Liu, Z. (2005).
\newblock Reading behavior in the digital environment: Changes in reading behavior over the past ten years.
\newblock {\em Journal of documentation}, 61(6):700--712.

\bibitem[Lockyer et~al., 2013]{Lockyer2013}
Lockyer, L., Heathcote, E., and Dawson, S. (2013).
\newblock Informing pedagogical action: Aligning learning analytics with learning design.
\newblock {\em American Behavioral Scientist}, 57(10):1439--1459.

\bibitem[Ma et~al., 2003]{Ma2003}
Ma, M., Schillings, V., Chen, T., and Meinel, C. (2003).
\newblock T-cube: A multimedia authoring system for elearning.
\newblock In {\em E-Learn: World Conference on E-Learning in Corporate, Government, Healthcare, and Higher Education}, pages 2289--2296. Association for the Advancement of Computing in Education (AACE).

\bibitem[Ma et~al., 2012]{Ma2012}
Ma, Y., Fosler-Lussier, E., and Lofthus, R. (2012).
\newblock Ranking-based readability assessment for early primary children's literature.
\newblock In {\em Proceedings of the 2012 Conference of the North American Chapter of the Association for Computational Linguistics: Human Language Technologies}, pages 548--552. Association for Computational Linguistics.

\bibitem[MacArthur and Graham, 2016]{Macarthur2016}
MacArthur, C. and Graham, S. (2016).
\newblock Writing research from a cognitive perspective.
\newblock {\em Handbook of writing research}, pages 24--40.

\bibitem[Mangen et~al., 2013]{Mangen2013}
Mangen, A., Walgermo, B.~R., and Br{\o}nnick, K. (2013).
\newblock Reading linear texts on paper versus computer screen: Effects on reading comprehension.
\newblock {\em International Journal of Educational Research}, 58:61--68.

\bibitem[Margolin et~al., 2013]{Margolin2013}
Margolin, S.~J., Driscoll, C., Toland, M.~J., and Kegler, J.~L. (2013).
\newblock E-readers, computer screens, or paper: Does reading comprehension change across media platforms?
\newblock {\em Applied Cognitive Psychology}, 27(4):512--519.

\bibitem[Marriott and Goyder, 2009]{Marriott2009}
Marriott, N. and Goyder, H. (2009).
\newblock {\em Manual for monitoring and evaluating education partnerships}.
\newblock International Institute for Educational Planning (IIEP).

\bibitem[Mart{\'\i}n~Fraile, 2007]{Martin2007}
Mart{\'\i}n~Fraile, L. (2007).
\newblock Monitoring and analysis tool for e-learning platforms.
\newblock {\em Final Degree Project directed by Zorrilla Pantale{\'o}n, M. University of Cantabria}.

\bibitem[Martinez~Maldonado et~al., 2012]{Martinez2012}
Martinez~Maldonado, R., Kay, J., Yacef, K., and Schwendimann, B. (2012).
\newblock An interactive teacher’s dashboard for monitoring groups in a multi-tabletop learning environment.
\newblock In {\em Intelligent Tutoring Systems}, pages 482--492. Springer.

\bibitem[Mathern, 2012]{Mathern2012}
Mathern, B. (2012).
\newblock {\em D{\'e}couverte interactive de connaissances {\`a} partir de traces d’activit{\'e}: Synth{\`e}se d’automates pour l’analyse et la mod{\'e}lisation de l’activit{\'e} de conduite automobile}.
\newblock PhD thesis, Universit{\'e} Claude Bernard-Lyon I.

\bibitem[Mayer, 2002]{Mayer2002}
Mayer, R.~E. (2002).
\newblock Multimedia learning.
\newblock {\em Psychology of learning and motivation}, 41:85--139.

\bibitem[Mc~Laughlin, 1969]{Laughlin1969}
Mc~Laughlin, G.~H. (1969).
\newblock Smog grading-a new readability formula.
\newblock {\em Journal of reading}, 12(8):639--646.

\bibitem[McNamara, 2012]{Mcnamara2012}
McNamara, D.~S. (2012).
\newblock {\em Reading comprehension strategies: Theories, interventions, and technologies}.
\newblock Psychology Press.

\bibitem[McNamara et~al., 2014]{McNamara2014}
McNamara, D.~S., Graesser, A.~C., McCarthy, P.~M., and Cai, Z. (2014).
\newblock {\em Automated Evaluation of Text and Discourse with Coh-Metrix}.
\newblock Cambridge University Press, New York, NY, USA.

\bibitem[McNamara and Magliano, 2009]{McNamara2009}
McNamara, D.~S. and Magliano, J. (2009).
\newblock Toward a comprehensive model of comprehension.
\newblock {\em Psychology of learning and motivation}, 51:297--384.

\bibitem[Millar and Schrier, 2015]{Millar2015}
Millar, M. and Schrier, T. (2015).
\newblock Digital or printed textbooks: which do students prefer and why?
\newblock {\em Journal of Teaching in Travel \& Tourism}, 15(2):166--185.

\bibitem[Mizrachi, 2014]{Mizrachi2014}
Mizrachi, D. (2014).
\newblock Online or print: Which do students prefer?
\newblock In {\em European Conference on Information Literacy}, pages 733--742. Springer.

\bibitem[Mobasher, 2007]{Mobasher2007}
Mobasher, B. (2007).
\newblock Data mining for web personalization.
\newblock In {\em The adaptive web}, pages 90--135. Springer.

\bibitem[Murphy et~al., 2003]{Murphy2003}
Murphy, P.~K., Long, J.~F., Holleran, T.~A., and Esterly, E. (2003).
\newblock Persuasion online or on paper: A new take on an old issue.
\newblock {\em Learning and Instruction}, 13(5):511--532.

\bibitem[Nakahara et~al., 2005]{Nakahara2005}
Nakahara, J., Hisamatsu, S., Yaegashi, K., and Yamauchi, Y. (2005).
\newblock itree: Does the mobile phone encourage learners to be more involved in collaborative learning?
\newblock In {\em Proceedings of th 2005 conference on Computer support for collaborative learning: learning 2005: the next 10 years!}, pages 470--478. International Society of the Learning Sciences.

\bibitem[Nelson et~al., 2012]{Nelson2012}
Nelson, J., Perfetti, C., Liben, D., and Liben, M. (2012).
\newblock Measures of text difficulty: Testing their predictive value for grade levels and student performance.
\newblock {\em Council of Chief State School Officers, Washington, DC}.

\bibitem[Nelson, 1965]{Nelson1965}
Nelson, T.~H. (1965).
\newblock Complex information processing: a file structure for the complex, the changing and the indeterminate.
\newblock In {\em Proceedings of the 1965 20th national conference}, ACM '65, pages 84--100, New York, NY, USA. ACM.

\bibitem[Nielsen, 2000]{Nielsen2000}
Nielsen, J. (2000).
\newblock Why you only need to test with 5 users.

\bibitem[Nielsen et~al., 1990]{Nielsen1990}
Nielsen, J. et~al. (1990).
\newblock {\em Hypertext and hypermedia}.
\newblock Academic Press.

\bibitem[Oh, 2013]{Oh2013}
Oh, K. (2013).
\newblock {\em Use of Reading Strategy to Assess Reading Medium Effectiveness: Application to Determine the Effects of Reading Medium and Generation in an Active Reading Task}.
\newblock PhD thesis, Virginia Polytechnic Institute and State University.

\bibitem[Olmanson et~al., 2016]{Olmanson2016}
Olmanson, J., Kennett, K., Magnifico, A., McCarthey, S., Searsmith, D., Cope, B., and Kalantzis, M. (2016).
\newblock Visualizing revision: Leveraging student-generated between-draft diagramming data in support of academic writing development.
\newblock {\em Technology, Knowledge and Learning}, 21(1):99--123.

\bibitem[Ortlieb et~al., 2014]{Ortlieb2014}
Ortlieb, E., Sargent, S., and Moreland, M. (2014).
\newblock Evaluating the efficacy of using a digital reading environment to improve reading comprehension within a reading clinic.
\newblock {\em Reading Psychology}, 35(5):397--421.

\bibitem[Otlet, 1934]{Otlet1934}
Otlet, P. (1934).
\newblock {\em Trait\'e de documentation. Le livre sur le livre. Th\'eorie et pratique}.
\newblock Number 197 in IIB Publication. Editiones Mundaneum.

\bibitem[Park and Jo, 2015]{Park2015}
Park, Y. and Jo, I.-H. (2015).
\newblock Development of the learning analytics dashboard to support students' learning performance.
\newblock {\em Journal of Universal Computer Science}, 21(1):110--133.

\bibitem[Patchan and Schunn, 2015]{Patchan2015}
Patchan, M.~M. and Schunn, C.~D. (2015).
\newblock Understanding the benefits of providing peer feedback: how students respond to peers’ texts of varying quality.
\newblock {\em Instructional Science}, 43(5):591--614.

\bibitem[Pattanasri et~al., 2012]{Pattanasri2012}
Pattanasri, N., Mukunoki, M., and Minoh, M. (2012).
\newblock Learning to estimate slide comprehension in classrooms with support vector machines.
\newblock {\em IEEE Transactions on Learning Technologies}, 5(1):52--61.

\bibitem[P{\'e}dauque, 2006]{Pedauque2006}
P{\'e}dauque, R.~T. (2006).
\newblock {\em Le Document {\`a} la lumi{\`e}re du num{\'e}rique: forme, texte, m{\'e}dium: comprendre le r{\^o}le du document num{\'e}rique dans l'{\'e}mergence d'une nouvelle modernit{\'e}}.
\newblock C \& F {\'E}ditions.

\bibitem[Peerani, 2013]{Peerani2013}
Peerani, N. (2013).
\newblock Barriers to distance learning: The educator's viewpoint.
\newblock {\em Distance Learning}, 10(2):29.

\bibitem[Philippakos, 2017]{Philippakos2017}
Philippakos, Z.~A. (2017).
\newblock Giving feedback: Preparing students for peer review and self-evaluation.
\newblock {\em The Reading Teacher}.

\bibitem[Pitler and Nenkova, 2008]{Pitler2008}
Pitler, E. and Nenkova, A. (2008).
\newblock Revisiting readability: A unified framework for predicting text quality.
\newblock In {\em Proceedings of the conference on empirical methods in natural language processing}, pages 186--195. Association for Computational Linguistics.

\bibitem[Podgorelec and Kuhar, 2011]{Podgorelec2011}
Podgorelec, V. and Kuhar, S. (2011).
\newblock Taking advantage of education data: Advanced data analysis and reporting in virtual learning environments.
\newblock {\em Elektronika ir Elektrotechnika}, 114(8):111--116.

\bibitem[Pohl et~al., 2012]{Pohl2012}
Pohl, A., Bry, F., Schwarz, J., and Gottstein, M. (2012).
\newblock Sensing the classroom: Improving awareness and self-awareness of students in backstage.
\newblock In {\em Interactive Collaborative Learning (ICL), 2012 15th International Conference on}, pages 1--8. IEEE.

\bibitem[Porion et~al., 2016]{Porion2016}
Porion, A., Aparicio, X., Megalakaki, O., Robert, A., and Baccino, T. (2016).
\newblock The impact of paper-based versus computerized presentation on text comprehension and memorization.
\newblock {\em Computers in Human Behavior}, 54:569--576.

\bibitem[Puchalski et~al., 1992]{Puchalski1992}
Puchalski, M.~M., Morra, M.~J., and Von~Wandruszka, R. (1992).
\newblock Fluorescence quenching of synthetic organic compounds by humic materials.
\newblock {\em Environmental science \& technology}, 26(9):1787--1792.

\bibitem[Ramos-Soto et~al., 2015]{Ramos2015}
Ramos-Soto, A., Lama, M., V{\'a}zquez-Barreiros, B., Bugar{\'\i}n, A., and Barro, M. M.~S. (2015).
\newblock Towards textual reporting in learning analytics dashboards.
\newblock In {\em 2015 IEEE 15th International Conference on Advanced Learning Technologies (ICALT)}, pages 260--264. IEEE.

\bibitem[Reimers and Neovesky, 2015]{Reimers2015}
Reimers, G. and Neovesky, A. (2015).
\newblock Student focused dashboards.
\newblock In {\em Proceedings of the 7th International Conference on Computer Supported Education-Volume 1}, pages 399--404. SCITEPRESS-Science and Technology Publications, Lda.

\bibitem[Rockinson-Szapkiw et~al., 2013]{Rockinson2013}
Rockinson-Szapkiw, A.~J., Courduff, J., Carter, K., and Bennett, D. (2013).
\newblock Electronic versus traditional print textbooks: A comparison study on the influence of university students' learning.
\newblock {\em Computers \& Education}, 63:259--266.

\bibitem[Rodr{\'\i}guez-Triana et~al., 2017]{Rodriguez2017}
Rodr{\'\i}guez-Triana, M.~J., Prieto, L.~P., Vozniuk, A., Boroujeni, M.~S., Schwendimann, B.~A., Holzer, A., and Gillet, D. (2017).
\newblock Monitoring, awareness and reflection in blended technology enhanced learning: a systematic review.
\newblock {\em International Journal of Technology Enhanced Learning}, 9(2-3):126--150.

\bibitem[Rom{\'a}n et~al., 2014]{Roman2014}
Rom{\'a}n, P.~E., Dell, R.~F., Vel{\'a}squez, J.~D., and Loyola, P.~S. (2014).
\newblock Identifying user sessions from web server logs with integer programming.
\newblock {\em Intelligent Data Analysis}, 18(1):43--61.

\bibitem[Romero and Ventura, 2010]{Romero2010}
Romero, C. and Ventura, S. (2010).
\newblock Educational data mining: a review of the state of the art.
\newblock {\em IEEE Transactions on Systems, Man, and Cybernetics, Part C (Applications and Reviews)}, 40(6):601--618.

\bibitem[Sadallah, 2019]{sadallah2019phd}
Sadallah, M. (2019).
\newblock {\em Models and Tools for Usage-based e-Learning Documents Reengineering}.
\newblock PhD thesis, Universit{\'e} Abderrahmane Mira de B{\'e}ja{\"\i}a (Alg{\'e}rie).

\bibitem[Sadallah, 2020]{sadallah2020coreada}
Sadallah, M. (2020).
\newblock Enhancing course revision: Introducing coreada-an advanced reading analytics dashboard.

\bibitem[Sadallah et~al., 2011]{Sadallah2011}
Sadallah, M., Aubert, O., and Pri{\'e}, Y. (2011).
\newblock Component-based hypervideo model: high-level operational specification of hypervideos.
\newblock In {\em Proceedings of the 11th ACM symposium on Document engineering}, pages 53--56. ACM.

\bibitem[Sadallah et~al., 2014]{Sadallah2014}
Sadallah, M., Aubert, O., and Pri{\'e}, Y. (2014).
\newblock Chm: an annotation-and component-based hypervideo model for the web.
\newblock {\em Multimedia tools and applications}, 70(2):869--903.

\bibitem[Sadallah et~al., 2013]{Sadallah2013}
Sadallah, M., Encelle, B., Maredj, A.-E., and Pri{\'e}, Y. (2013).
\newblock A framework for usage-based document reengineering.
\newblock In {\em Proceedings of the 2013 ACM Symposium on Document Engineering}, DocEng '13, pages 99--102, New York, NY, USA. ACM.

\bibitem[Sadallah et~al., 2015]{Sadallah2015}
Sadallah, M., Encelle, B., Maredj, A.-E., and Pri{\'e}, Y. (2015).
\newblock Towards reading session-based indicators in educational reading analytics.
\newblock In {\em Design for Teaching and Learning in a Networked World}, pages 297--310. Springer.

\bibitem[Sadallah et~al., 2020a]{sadallah2020leveraging}
Sadallah, M., Encelle, B., Maredj, A.-E., and Pri{\'e}, Y. (2020a).
\newblock Leveraging learners' activity logs for course reading analytics using session-based indicators.
\newblock {\em International journal of technology enhanced learning}, 12(1):53--78.

\bibitem[Sadallah et~al., 2020b]{sadallah2020towards}
Sadallah, M., Encelle, B., Maredj, A.-E., and Pri{\'e}, Y. (2020b).
\newblock Towards fine-grained reading dashboards for online course revision.
\newblock {\em Educational Technology Research and Development}, 68:3165--3186.

\bibitem[Salmer{\'o}n et~al., 2006]{Salmeron2006}
Salmer{\'o}n, L., Kintsch, W., and Ca{\~a}s, J.~J. (2006).
\newblock Reading strategies and prior knowledge in learning from hypertext.
\newblock {\em Memory \& Cognition}, 34(5):1157--1171.

\bibitem[Schriver, 1992]{Schriver1992}
Schriver, K.~A. (1992).
\newblock Teaching writers to anticipate readers' needs: A classroom-evaluated pedagogy.
\newblock {\em Written communication}, 9(2):179--208.

\bibitem[Schwendimann et~al., 2017]{Schwendimann2017}
Schwendimann, B.~A., Rodríguez-Triana, M.~J., Vozniuk, A., Prieto, L.~P., Boroujeni, M.~S., Holzer, A., Gillet, D., and Dillenbourg, P. (2017).
\newblock Perceiving learning at a glance: A systematic literature review of learning dashboard research.
\newblock {\em IEEE Transactions on Learning Technologies}, 10(1):30--41.

\bibitem[Siemens and Gasevic, 2012]{Siemens2012}
Siemens, G. and Gasevic, D. (2012).
\newblock Guest editorial-learning and knowledge analytics.
\newblock {\em Educational Technology \& Society}, 15(3):1.

\bibitem[Slater et~al., 2017]{Slater2017}
Slater, S., Joksimovi{\'c}, S., Kovanovic, V., Baker, R.~S., and Gasevic, D. (2017).
\newblock Tools for educational data mining: A review.
\newblock {\em Journal of Educational and Behavioral Statistics}, 42(1):85--106.

\bibitem[Slingsby et~al., 2011]{Slingsby2011}
Slingsby, A., Dykes, J., and Wood, J. (2011).
\newblock Exploring uncertainty in geodemographics with interactive graphics.
\newblock {\em IEEE Transactions on Visualization and Computer Graphics}, 17(12):2545--2554.

\bibitem[Staiger, 2012]{Staiger2012}
Staiger, J. (2012).
\newblock How e-books are used.
\newblock {\em Reference \& User Services Quarterly}, 51(4):355--365.

\bibitem[Stoop et~al., 2013a]{Stoop2013p2}
Stoop, J., Kreutzer, P., and G.~Kircz, J. (2013a).
\newblock Reading and learning from screens versus print: a study in changing habits: Part 2--comparing different text structures on paper and on screen.
\newblock {\em New Library World}, 114(9/10):371--383.

\bibitem[Stoop et~al., 2013b]{Stoop2013p1}
Stoop, J., Kreutzer, P., and Kircz, J. (2013b).
\newblock Reading and learning from screens versus print: a study in changing habits: Part 1--reading long information rich texts.
\newblock {\em New Library World}, 114(7/8):284--300.

\bibitem[Tabuenca et~al., 2014]{Tabuenca2014}
Tabuenca, B., Kalz, M., and Specht, M. (2014).
\newblock Binding daily physical environments to learning activities with mobile and sensor technology.
\newblock In {\em European Summit on Immersive Education}, pages 73--84. Springer.

\bibitem[Tanaka-Ishii et~al., 2010]{Tanaka2010}
Tanaka-Ishii, K., Tezuka, S., and Terada, H. (2010).
\newblock Sorting texts by readability.
\newblock {\em Computational Linguistics}, 36(2):203--227.

\bibitem[Teasley, 2017]{Teasley2017}
Teasley, S.~D. (2017).
\newblock Student facing dashboards: One size fits all?
\newblock {\em Technology, Knowledge and Learning}, 22(3):377--384.

\bibitem[Thompson, 2005]{Thompson2005}
Thompson, J.~B. (2005).
\newblock {\em Books in the digital age: The transformation of academic and higher education publishing in Britain and the United States}.
\newblock Polity.

\bibitem[Tuncer and Bahadir, 2014]{Tuncer2014}
Tuncer, M. and Bahadir, F. (2014).
\newblock Effect of screen reading and reading from printed out material on student success and permanency in introduction to computer lesson.
\newblock {\em TOJET: The Turkish Online Journal of Educational Technology}, 13(3).

\bibitem[van Ossenbruggen, 2001]{Van2001}
van Ossenbruggen, J. (2001).
\newblock {\em Processing Structured Hypermedia-A matter of style}.
\newblock PhD thesis, Amsterdam: Vrije Universiteit.

\bibitem[V{\'a}zquez et~al., 2006]{Vazquez2006}
V{\'a}zquez, A., Oliveira, J.~G., Dezs{\"o}, Z., Goh, K.-I., Kondor, I., and Barab{\'a}si, A.-L. (2006).
\newblock Modeling bursts and heavy tails in human dynamics.
\newblock {\em Physical Review E}, 73(3):036127.

\bibitem[Verbert et~al., 2013]{Verbert2013}
Verbert, K., Duval, E., Klerkx, J., Govaerts, S., and Santos, J.~L. (2013).
\newblock Learning analytics dashboard applications.
\newblock {\em American Behavioral Scientist}.

\bibitem[Verbert et~al., 2014]{Verbert2014}
Verbert, K., Govaerts, S., Duval, E., Santos, J.~L., Van~Assche, F., Parra, G., and Klerkx, J. (2014).
\newblock Learning dashboards: an overview and future research opportunities.
\newblock {\em Personal and Ubiquitous Computing}, 18(6):1499--1514.

\bibitem[V{\"o}lkel, 2007]{Volkel2007}
V{\"o}lkel, M. (2007).
\newblock From documents to knowledge models.
\newblock In {\em Proceedings of the 4th Conference on Professional Knowledge Management}, volume~2, pages 209--216.

\bibitem[Walsh, 2016]{Walsh2016}
Walsh, G. (2016).
\newblock Screen and paper reading research – a literature review.
\newblock {\em Australian Academic \& Research Libraries}, 47(3):160--173.

\bibitem[Wise et~al., 2014]{Wise2014a}
Wise, A., Zhao, Y., and Hausknecht, S. (2014).
\newblock Learning analytics for online discussions: Embedded and extracted approaches.
\newblock {\em Journal of Learning Analytics}, 1(2):48--71.

\bibitem[Wise et~al., 2016]{Wise2016}
Wise, A.~F., Vytasek, J.~M., Hausknecht, S., and Zhao, Y. (2016).
\newblock Developing learning analytics design knowledge in the “middle space”: The student tuning model and align design framework for learning analytics use.
\newblock {\em Online Learning}, 20(2).

\bibitem[Witte, 2013]{Witte2013}
Witte, S. (2013).
\newblock Preaching what we practice: A study of revision.
\newblock {\em Journal of Curriculum and Instruction}, 6(2):33--59.

\bibitem[Witten et~al., 2016]{Witten2016}
Witten, I.~H., Frank, E., Hall, M.~A., and Pal, C.~J. (2016).
\newblock {\em Data Mining: Practical machine learning tools and techniques}.
\newblock Morgan Kaufmann.

\bibitem[Xu and Ouyang, 2022]{xu2022application}
Xu, W. and Ouyang, F. (2022).
\newblock The application of ai technologies in stem education: a systematic review from 2011 to 2021.
\newblock {\em International Journal of STEM Education}, 9(1):59.

\bibitem[Yahiaoui et~al., 2011]{Yahiaoui2011}
Yahiaoui, L., Pri\'e, Y., Boufaida, Z., and Champin, P.-A. (2011).
\newblock Redocumenting computer-mediated activity from its traces: a model-based approach for narrative construction.
\newblock {\em Journal of Digital Information}, 12(3).

\bibitem[Yoo et~al., 2015]{Yoo2015}
Yoo, Y., Lee, H., Jo, I.-H., and Park, Y. (2015).
\newblock Educational dashboards for smart learning: Review of case studies.
\newblock In {\em Emerging Issues in Smart Learning}, pages 145--155. Springer.

\bibitem[Zhang et~al., 2007]{Zhang2007}
Zhang, H., Almeroth, K., Knight, A., Bulger, M., and Mayer, R. (2007).
\newblock Moodog: Tracking students' online learning activities.
\newblock In {\em EdMedia: World Conference on Educational Media and Technology}, pages 4415--4422. Association for the Advancement of Computing in Education (AACE).

\end{thebibliography}
\end{document}